%% file: eigsumr.tex
\documentclass{elsarticle}
\usepackage{amsmath,amssymb,graphics,epsfig,grffile,a4wide}
\usepackage{bbold}

\newcommand{\sign}{\text{sign}}

\graphicspath{{./figs/}}

\newcommand{\diag}[0]{\text{diag}}

\newcommand{\spa}[0]{\phantom{spa}}

\newcommand{\path}[1]{[#1]}

\usepackage{algorithm2e}

\def\Dslash{\mathchoice
    {D\hskip-0.62em\raise0.2ex\hbox{$\displaystyle/$}\hskip0.2em}%
    {D\hskip-0.62em\raise0.2ex\hbox{$\textstyle/$}\hskip0.2em}%
    {D\hskip-0.5em\raise0.15ex\hbox{$\scriptstyle/$}\hskip0.2em}%
    {D\hskip-0.5em\raise0.15ex\hbox{$\scriptscriptstyle/$}\hskip0.2em}}

\newcommand{\norm}[1]{\|#1\|}

\begin{document}
\begin{frontmatter}
\title{The eigSUMR inverter for overlap fermions}

\author[a]{Nigel Cundy}
\author[a]{Weonjong Lee}

\address[a]{   Lattice Gauge Theory Research Center, FPRD, and CTP, Department of Physics \&
    Astronomy, Seoul National University, Seoul, 151-747, South Korea\\
        E-mail:{ndcundy@phya.snu.ac.kr}}

\date{\today}
\begin{abstract}
We discuss the usage and applicability of deflation methods for the overlap lattice Dirac operator, focussing on calculating the eigenvalues using a method similar to the eigCG algorithm used for other Dirac operators. The overlap operator, which contains several theoretical advantages over other formulations of lattice Quantum Chromodynamics, is more computationally expensive because it requires the computation of the matrix sign function. The principle change made compared to deflation methods for other formulations of lattice QCD is that it is necessary for best performance to tune the accuracy of the matrix sign function as the computation proceeds. We present two possible relaxation strategies, one which provides a rigorous bound for the eigenvalues but seems to be too conservative in practice, and a second which is less conservative but, while its stability is not guaranteed, seems to work well in practice.

We adapt the original eigCG algorithm for two of the preferred inversion algorithms for overlap fermions, GMRESR(relCG) and GMRESR(relSUMR). Before deflation, the rate of convergence of these routines in terms of iterations is similar, but, since the Shifted Unitary Minimal Residual (SUMR) algorithm only requires one call to the matrix sign function compared to the two calls required for Conjugate Gradient (CG), SUMR is usually preferred for single inversions of the Dirac operator.  We construct bounds for the required accuracy of the matrix sign function during the eigenvalue calculation. For the SUMR algorithm, we use the standard Galerkin projection to perform the deflation; while for the CG algorithm, we are able to use a considerably superior spectral pre-conditioner. The superior performance of the spectral preconditioner, and its need for less accurate eigenvalues, almost erodes SUMR's advantage over CG as an inversion algorithm.

We see factor of three gains for the inversion algorithm from the deflation on our small test lattices; we expect larger gains over the undeflated algorithms in realistic simulations on larger lattices and with smaller masses. There is, however, a significant cost in the eigenvalue calculation because we cannot relax the accuracy of the matrix sign function as aggressively when calculating the eigenvalues as we do while performing the inversions. This set-up cost is, however, more than compensated for the gain in the deflation if enough right hand sides are required.
\end{abstract}
\end{frontmatter}
\section{Introduction}
The approximate and spontaneously broken chiral symmetry is important in Quantum Chromodynamics (QCD), the theory which describes the strong nuclear force, as it determines (to a large extent) the mass spectrum of the lightest hadrons. It also protects the effective Lagrangian describing the hadrons from picking up an additional mass renormalisation, which, in principle, will be sensitive to the ultra-violet cut-off. When simulating QCD numerically on a discrete space time lattice~\cite{Wilson:1974}, it is therefore advisable to break chiral symmetry as lightly as possible. While it is possible to remove the additive mass renormalisation by tuning the quark mass, this tuning is unlikely to be perfectly realised, creating a small error; the worse the chiral symmetry breaking, the harder it is to control this error. Equally, an action which violates chiral symmetry has many additional operators in the effective Lagrangian, making the analysis of some observables more challenging.

The difficulty is that simulating lattice QCD with an exact Ginsparg-Wilson chiral symmetry~\cite{Ginsparg:1982bj,Hasenfratz:1998ri,Luscher:1998pqa} (the closest to the continuum chiral symmetry achievable on the lattice given the Nielsen-Ninomoya theorem~\cite{Nishy-Ninny}) is expensive. The simplest practical and known lattice Dirac operator with exact\footnote{The overlap operator gives an exact chiral symmetry. However, in practice, we do not render the matrix sign function within the overlap operator to a perfect precision, which breaks chiral symmetry. It can, however, be approximated to an arbitrary precision controlled by the accuracy of the approximation to the matrix sign function and the precision of the floating point arithmetic on the computer. The amount of explicit chiral symmetry breaking can therefore be controlled, and the systematic error from this approximation reduced so that it is insignificant.} chiral symmetry is the overlap operator,~\cite{Narayanan:1993ss,Neuberger:1998fp,Neuberger:1997bg,Neuberger:1998my}
\begin{gather}
a D[\mu] = \frac{1}{2}(1+\mu + (1-\mu)\gamma_5 \sign(K)),
\end{gather}
where $\sign(K)$ is the matrix sign function, $\mu/(1-\mu)$ is proportional to the bare fermion mass, and we will call the Hermitian operator $K$ the kernel of the matrix sign function. There is a great deal of freedom while choosing $K$ -- the constraints are that we require that $D$ has the correct continuum limit (as the lattice spacing $a$ goes to zero), no fermion doublers, and is local. A convenient choice is the Wilson Dirac operator,
\begin{gather}
 aK_{xy} = \gamma_5 \left[\delta_{xy} - \kappa\sum_{\mu=1}^4((1-\gamma_\mu)\delta_{y,x+a\hat{\mu}}U_\mu(x) + (1+\gamma_\mu)\delta_{y,x-a\hat{\mu}}U^\dagger_\mu(y)) \right]
\end{gather}
with $\kappa = \frac{1}{8-2m}$ for $m$ a real parameter in the range $0\lesssim m < 2$, and where $U_\mu(x)$ is the gauge connection, an SU(3) matrix on every link of the lattice (we choose to use $\kappa = 0.19$, which has good locality properties for the Dirac operator). $\gamma_\mu$ ($\mu = 1,\ldots,5$) are the standard Hermitian form of the anti-commuting Dirac $\gamma$-matrices, which satisfy $\gamma_\mu^2 = 1$.

It is easy to show that the massless overlap Dirac operator satisfies the Ginsparg-Wilson relation
\begin{gather}
 D[0]\gamma_5(1-2aD[0]) + \gamma_5 D[0] = 0,
\end{gather}
which reduces to $\gamma_5 D[0] + D[0]\gamma_5=0$, the equation that permits chiral symmetry, in the naive continuum limit ($a \rightarrow 0$). The overlap Dirac operator is also $\gamma_5$-Hermitian, $D[\mu]^\dagger = \gamma_5 D[\mu]\gamma_5$, which guarantees that the eigenvalues of the Dirac operator are either real or come in complex conjugate pairs. This also means that $\gamma_5 D$ is Hermitian, and we shall call this the Hermitian Dirac operator. Given that $\gamma_5\sign(K)$ is unitary, the real eigenvalues of the massless operator are either at zero or 1, and the eigenvalues lie on a circle in the complex plane of radius $1/2$ centred at $1/2$.  Furthermore, since the squared Hermitian Dirac operator commutes with $\gamma_5$, $[\gamma_5,D^\dagger D]=0$, we see the eigenvalues of this operator must be degenerate, and can be expressed as exact eigenvectors of $\gamma_5$ which we may denote as $\psi_{+i}$ and $\psi_{-i}$, and, except when the corresponding eigenvalues are at $0$ and $\pm 1$, the eigenvalues of $\gamma_5 D$ occur in $\pm$ pairs which are linear combinations of  $\psi_{+i}$ and $\psi_{-i}$. The eigenvectors of $D^\dagger D$ are independent of the mass. Finally, since $[D,D^\dagger] = 0$, the paired eigenvectors of $D$ (with complex conjugate eigenvalues) are also linear combinations of  $\psi_{+i}$ and $\psi_{-i}$. As $D$ is a normal operator (being shifted unitary), its left and right eigenvectors are the same, and its eigenvectors are orthogonal.

The difficulties with using overlap fermions are due to the matrix sign function, and these are both algorithmic and concern the large computational cost required. The matrix sign function is defined in terms of a spectral decomposition,
\begin{gather}
 \sign(K) = \sum_i {\tilde{\psi}_i}{\tilde{\psi}_i}^\dagger \sign(\tilde{\lambda}_i),
\end{gather}
where ${\tilde{\psi}_i}$ and $\tilde{\lambda}_i$ are the eigenvectors and eigenvalues of $K$. In practice, one simulates the matrix sign function by deflating the smallest eigenvectors of $K$, and then use some approximation to cover the rest of the eigenvalue spectrum. Various methods have been proposed for the approximation: polynomial approximations~\cite{Hernandez:1999cu}, rational approximations~\cite{vandenEshof:2002ms}, an approach based on the Lanczos algorithm~\cite{Borici:2007bp} and continued fractions, which are most easily expressed in terms of a five dimensional representation~\cite{Narayanan:2000qx,Borici:2001ua,Edwards:2005an,Kennedy:2006ax}. One may also use the approximation for the bulk of the eigenvalue spectrum without the deflation of the small eigenvectors giving a continuous function of $K$ where the matrix sign function breaks down for small eigenvalues of $K$~\cite{Shamir:1993zy,Kaplan:1992bt,Chiu:2002ir,Cundy:2010uq} which, although it loses the exact chiral symmetry, still maintains a very good chiral symmetry, and tends to be quicker and removes various algorithmic difficulties associated with the discontinuity in the matrix sign function~\cite{Fodor:2003bh,DeGrand:2004nq,DeGrand:2005vb,Cundy:2005pi,Cundy:2005mr,Cundy:2008zc,Cundy:2007la}. In this work, we shall use deflation and the optimum Zolotarev rational approximation~\cite{vandenEshof:2002ms,Zolotarev}, allowing us to control the accuracy of the matrix sign function, and therefore the breaking of chiral symmetry, with the only restriction on the accuracy being the numerical precision of the computer. In practical applications, since the kernel eigenvectors only need to be calculated once for each gauge field configuration, the cost of calculating these eigenvectors (for example by using a polynomial preconditioned restarted Arnoldi algorithm~\cite{Sorensen97acceleratingthe,IRAM}) is negligible.  However, we need not always use the full accuracy matrix sign function during the computation, and, since low accuracy estimations of the matrix sign function are considerably faster to compute, it is advantageous to relax the accuracy for almost all calls to the overlap operator while still maintaining the full accuracy of the final result.

For physical applications, we need to invert the overlap operator and, for some observables, to calculate its eigenvectors. There are two general strategies used to invert the overlap operator. The first is to invert the five dimensional representation of the matrix sign function~\cite{Hashimoto:2007vv}; the second is to use a nested inversion~\cite{Arnold:2003sx,Cundy:2004pza}; with an inner inversion to calculate the matrix sign function and an outer inversion of the overlap operator. It is not yet clear which of these approaches is superior. In this work, we concentrate on the nested approach.

Two of the most efficient algorithms to invert the overlap operator are Conjugate Gradient (CG) and Shifted Unitary Minimal Residual (SUMR). The first of these algorithms, which is used for the Hermitian squared operator $D^\dagger D$, is well known and understood. The second, less well known option, was originally proposed by Jagels and Reichel~\cite{Jagels,Arnold:2003sx}, and it is specifically designed for operators such as the overlap operator. In principle, they should require roughly the same number of iterations to converge (although in practice without preconditioning the convergence of CG tends to be a little faster, though not by any significant amount). However, since SUMR requires one call to the matrix sign function per iteration rather than CG's two, for a single inversion of the overlap operator SUMR should be the preferred method (although CG tends to be better than two SUMRs for inverting $D^\dagger D$). The relative performance of CG and SUMR is shown for one typical configuration in figure \ref{fig:2a}.

Both of these methods can be improved significantly from the naive inversion algorithm. The first key idea is to use as low an accuracy as possible for the matrix sign function for the bulk of the calculation. This can be done firstly by relaxing the accuracy of the matrix sign function as the inversion progresses~\cite{Arnold:2003sx}, and secondly by using a low accuracy inversion of the overlap operator as a preconditioner for a high accuracy inversion of the overlap operator (giving us three inversions: the inner calculation of the matrix sign function; the middle inversion -- the CG or SUMR preconditioner, and the outer inversion of the full-accuracy overlap operator)~\cite{Cundy:2004pza}. By further incorporating mixed precision methods, calling a single precision matrix sign function when its required accuracy is sufficiently low, this approach can give an order of magnitude gain over the naive CG and SUMR algorithms.

Further improvements can be made by preconditioning the CG or SUMR algorithm used in the middle inversion. In general, adding an additional nested inverter does not help much, but we have found that deflating the lowest overlap eigenvalues can gain up to an additional factor of five, with a higher gain at smaller quark mass~\cite{Cundy:2005mn} (our results in this work, on a small lattice and relatively large mass, give a factor of three improvement). Our previous approach was to calculate the overlap eigenvectors ${\psi_i}$ with eigenvalues $\lambda_i$ in advance, and then to use them to construct a preconditioner for the CG inversion. This approach only worked for the CG algorithm, as the preconditioner spoils the shifted unitary structure of the overlap operator and thus means that the SUMR algorithm no longer works, and there is an additional set-up cost of calculating the overlap eigenvalues. However, we do not need to  to calculate the overlap eigenvectors to a high precision for the preconditioning to be effective, and there are other ways, albeit less robust, for adapting deflation to SUMR.

Of course, a better approach would be to combine the inversion and eigenvalue calculation. It has been known how to do this for a general operator for some time (see, for example,~\cite{Stathopoulos:2007zi}), and the resulting eigCG algorithm is widely used. The purpose of this work is to report on our efforts to adapt the eigCG approach to overlap fermions; both to create an eigCG inverter and an eigSUMR inverter. The principle novelty compared to other, simpler, operators is that we need to control the accuracy of the matrix sign function during the inversion. The optimum relaxation strategy controlling the accuracy of the matrix sign function during the inversion algorithm differs from that of the optimum strategy for the eigenvalue calculation. Our main result is to provide bounds to control the accuracy of the matrix sign function as the eigenvalue calculation proceeds. Our second result is to confirm that these methods are superior to the undeflated inversions if enough right hand sides are required. Thirdly, we study the effectiveness and limitations of eigSUMR as an eigenvalue routine.

Unlike simpler Dirac operators, the additional cost of eigCG or eigSUMR over the deflated CG or SUMR algorithm is not negligible, because we cannot relax the accuracy of the matrix sign function as vigorously as we would prefer during the inversion. We are instead forced to use a relatively high accuracy matrix sign function during the inversions used to calculate the eigenvectors. However, we do not need a high accuracy estimate of the eigenvectors for the deflation to be effective, and in general two or three inversions are sufficient to calculate the eigenvectors to a sufficient accuracy. Although there is some significant set-up cost, the gain from the deflation more than compensates for this if enough inversions are required (where `enough' depends on the lattice volume and the quark mass).

This article is arranged as follows. In section \ref{sec:2} we review the SUMR inversion algorithm, while in section \ref{sec:3} we describe the relaxation and preconditioning techniques which are currently used to accelerate the inversion of the overlap operator. Section \ref{sec:4} discusses deflation of the overlap inversion, both the methods we have been using to deflate the CG algorithm, and our proposal to apply deflation to the SUMR algorithm. In section \ref{sec:5} we discuss how the Krylov subspace used to perform the inversions can also be used to calculate eigenvalues, and we present numerical results in section \ref{sec:6}. We conclude in section \ref{sec:7}.

\section{The SUMR routine}\label{sec:2}
It has been shown in~\cite{Jagels} and introduced to the lattice community in~\cite{Arnold:2003sx} that it is possible to use a short recurrence to build up the Arnoldi vectors for a Unitary or shifted Unitary system. Both of these are normal matrices, so the left and right eigenvectors are the same and the eigenvectors are orthogonal to each other.  We consider a a matrix of the form $A = \rho + U$, where $U$ is an unitary $N\times N$ matrix and $\rho$ is a real number (implicitly multiplied by the identity operator). The overlap operator can be reduced to this form with a simple scaling. Now it is clear that the Krylov subspace for $A$ generated from some vector $b$ is the same as the Krylov subspace for $U$ starting from the same point. We describe the theory behind the generation of the Krylov subspace for the SUMR algorithm in appendix \ref{app:A}. The resulting algorithm, which can be used to generate an orthonormal basis $q_i$, $i= 1,2,\ldots n$ which spans the Krylov subspace $\{b,Ab,A^2b,\ldots , A^{n-1}b\}$ is given in algorithm \ref{alg:sumrbasis}. The algorithm requires a second vector, $\tilde{q}$, alongside $q$ which also spans the Krylov subspace.

\begin{algorithm}[t]
\begin{align*}
&q_0 = \tilde{q}_0 = b/\norm{b}\nonumber\\
&\text{for }j\text{ in }0,1,2,3,4, \ldots;\text{ do}\nonumber\\
&\spa u = U q_j \nonumber\\
&\spa \gamma_j =- (\tilde{q}_j,u)\nonumber\\
&\spa \sigma_j = \sqrt{1-|\gamma_j|^2}\nonumber\\
&\spa q_{j+1} = \frac{1}{\sigma_j} (u + \gamma_j \tilde{q}_j)\nonumber\\
&\spa \tilde{q}_{j+1} = \sigma_{j} \tilde{q}_j + \gamma_j^* q_{j+1}\nonumber\\
&\text{done}.
\end{align*}
\caption{The modified Arnoldi algorithm to generate an orthonormal Krylov basis $\mathbf{q}$ for a unitary operator $U$.}\label{alg:sumrbasis}
\end{algorithm}
It is straightforward to show by induction that in perfect arithmetic, $q$ and $\tilde{q}$ should remain normalised to $1$. However, in practice, we have found $q$ and $\tilde{q}$ can lose their normalisation after a number of iterations, particularly when the matrix sign function is rendered only approximately, and this affects our eigenvalue routine. It is therefore necessary to monitor the normalisation of $q$, and to restart the algorithm (by normalising $q$, setting $\tilde{q} = q$ and resetting $\sigma$, $\gamma$ and the other parameters needed in the SUMR inversion routine in algorithm \ref{alg:sumr}) if $\norm{q}$ starts growing significantly larger or smaller than one.

This construction of the Krylov subspace can be used to create a short-recurrence inversion algorithm for a shifted unitary matrix which is effectively an efficient representation of the GMRES algorithm, with the important improvement that there is no need to store all the $q$ vectors~\cite{Jagels}; this approach has been called SUMR in~\cite{Arnold:2003sx}. Experience has taught us that SUMR requires only slightly more iterations than the unpreconditioned conjugate gradient (CG) algorithm to converge; but it has the advantage that it only requires one call to the matrix sign function per iteration rather than two for CGNE, so that unpreconditioned SUMR is almost twice as fast as unpreconditioned CGNE for a single inversion of the overlap operator (i.e. a computation of the propagator).

The full SUMR algorithm for an initial guess $x_0$ is given in algorithm \ref{alg:sumr}.
\begin{algorithm}
\begin{align*}
 &\tilde{r} = b-Ax_0; \delta = \norm{\tilde{r}};x=x_0; \hat{\varphi} = \frac{1}{\delta} ; \hat{\tau} = \frac{\delta}{\rho}\nonumber\\
 &\omega = 0; q_{old} = 0;p = 0;\varphi = 0;s = 0 ;q = \frac{\tilde{r}}{\delta}; \tilde{q} = q\nonumber\\
 &\lambda = 0; r' = 0; r = 1;\gamma=1;c=1;\sigma=1\nonumber\\
& \text{for }j\text{ in }0,1,2,3,\ldots\nonumber\\
 &\spa r'' = r\nonumber\\
 &\spa u = Uq\nonumber\\
 &\spa \gamma = (\tilde{q},u)\nonumber\\
 &\spa \sigma = \sqrt{1-|\gamma|^2}\nonumber\\
 &\spa \alpha = \gamma\delta\nonumber\\
 &\spa \gamma = - \gamma\nonumber\\
 &\spa r' = \alpha \varphi + {s}\zeta/{\rho}\nonumber\\
 &\spa \hat{r} = \alpha \hat{\varphi} + {c^\dagger}\zeta/{\rho}\nonumber\\
 &\spa c^\dagger = \frac{\hat{r}}{\sqrt{|\hat{r}|^2 + \sigma^2}}\nonumber\\
 &\spa s = - \frac{\sigma}{\sqrt{|\hat{r}|^2+\sigma^2}} \nonumber\\
 &\spa r = s \sigma - c \hat{r}\nonumber\\
 &\spa \tau = -c\hat{\tau}\nonumber\\
 &\spa \hat{\tau} = s\hat{\tau}\nonumber\\
 &\spa \eta = {\tau}/{r} \nonumber\\
&\spa \kappa = {r'}/{r''}\nonumber\\
&\spa\omega_{old} = \omega \nonumber\\
&\spa \omega = \alpha p + \kappa(q_{old} - \omega)\nonumber\\
&\spa p = p + \lambda (q_{old} - \omega_{old})\nonumber\\
&\spa x = x+ \eta(q-\omega)\nonumber\\
&\spa \delta = \sigma \delta\nonumber\\
&\spa \varphi = {s\gamma^\dagger}/{\delta} - c \hat{\varphi}\nonumber\\
&\spa \lambda = \frac{\varphi}{r}\nonumber\\
&\spa\hat{\varphi} = s \hat{\varphi} + {c^\dagger\gamma^\dagger}/{\delta}\nonumber\\
&\spa q_{old} = q\nonumber\\
&\spa q = (\gamma \tilde{q} + u)/{\sigma}\nonumber\\
&\spa \tilde{q} = \sigma \tilde{q} + \gamma^\dagger q\nonumber\\
& \text{done}.
 \end{align*}

 \caption{The SUMR inversion routine. $(\tilde{q},u) \equiv \tilde{q}^\dagger u$ represents the inner product of two complex vectors.}\label{alg:sumr}

\end{algorithm}

As with CG, the SUMR algorithm can be trivially be adapted to give a multi-shift solver. However, given that it is impossible to combine this with the various preconditioning methods discussed below, it is not clear that it would be beneficial to do so unless a large number of inversions at various distinct low masses are required.
\section{Relaxation and GMRESR preconditioning for overlap fermions}\label{sec:3}
There are two principle approaches for inverting the overlap operator; firstly to use a nested four dimensional (4D) approach, with an inner and outer inversion~\cite{Arnold:2003sx,Cundy:2004pza,Cundy:2005mn}, and secondly to express the overlap operator in terms of a five dimensional (5D) operator, and invert that 5D operator~\cite{Narayanan:2000qx,Borici:2001ua,Edwards:2005an,Kennedy:2006ax,Hashimoto:2007vv}. To our knowledge, the optimum 4D approach and the optimum 5D approach have not been compared. In~\cite{Hashimoto:2007vv}, a 5D inversion routine was compared against a sub-optimum 4D method (which used relaxation, but not GMRESR or deflation), and while the 5D inverter was shown to be superior to the relaxed 4D CG inverter by a factor of 3-4; the relaxed 4D deflated GMRESR routine shows an even bigger gain over that routine~\cite{Cundy:2004pza,Cundy:2005mn}. We therefore still consider it an open question whether the nested 4D or 5D approach is superior. It may well depend on the computer architecture used.

The previous state of the art method for the nested inversion routine was developed in~\cite{Arnold:2003sx,Cundy:2004pza,Cundy:2005mn}. There are three steps: relaxation, GMRESR preconditioning, and deflation of the inversion (which should not be confused with the deflation of the matrix sign function which accelerates the calculation of the sign function: they are separate steps using separate sets of eigenvectors), which is an entirely. The key is to ensure that we do not have to evaluate the matrix sign function to a high accuracy during the inversion. A low  accuracy approximation to the matrix sign function is considerably faster to evaluate than a high accuracy approximation to the matrix sign function. The accuracy of the matrix sign function can be measured in various ways. The accuracy of the approximation to the sign function is defined as
\begin{gather}
 \eta'' = \max_{b \in \norm{b}=1} {{b^\dagger \tilde{s}^2 b} - 1} ,
\end{gather}
where $\tilde{s}$ is our approximation to the matrix sign function and the maximum is over all vectors $b$ normalised to 1. This is impossible to measure practically, so one usually estimates it using
\begin{gather}
 \eta' = \frac{b^\dagger \tilde{s}^2 b}{b^\dagger b} - 1,
\end{gather}
for one particular choice of $b$.
The accuracy can also be controlled by tuning the approximation used to represent the matrix sign function. For example, the Zolotarev rational approximation for a real variable $x$,
\begin{gather}
 \epsilon(x) = \sum_{i = 1}^{N_Z} \frac{\omega^Z_i x}{x^2 + \sigma_i^Z},
\end{gather}
produces an $|\epsilon|$ which oscillates between $1 + \Delta$ and $1-\Delta$ for a given range $\mathfrak{a}<|x|<\mathfrak{b}$. $\mathfrak{a}$ and $\mathfrak{b}$ are fixed as soon as we have selected $K$: $\mathfrak{b}^2$ is the largest eigenvalue of $K^2$, while $\mathfrak{a}^2$ is the largest of the small eigenvalues of $K^2$ used in the deflation of the matrix sign function. $N_Z$ is the order of the rational approximation, and the coefficients $\omega^Z$ and $\sigma^Z$, as well as the error $\Delta$, can be expressed as functions of  $N_Z$, $a$ and $b$ in terms of Jacobi elliptic functions~\cite{vandenEshof:2002ms,Zolotarev,Kennedy:2004tk}. The accuracy of the matrix sign function, $\eta$, is therefore $\Delta$ or the error in the inversion of $x^2 + \sigma_i^Z$, whichever is larger. $\Delta$ can be controlled by varying the order of $N_Z$. In general, $N_Z \sim 20$ allows the matrix sign function to be calculated to double precision accuracy. The accuracy of the matrix sign function can be controlled accurately by varying $N_Z$ and the accuracy of the inversion of $x^2 + \sigma_i^Z$.

We define the residual vector for the inversion $x = A^{-1} b$ as
\begin{gather}
 r = Ax - b,
\end{gather}
and write $\xi = \norm{r}$. Here $A$ represents either the Dirac operator for SUMR or $D^\dagger D$ for CG. The question then becomes how accurate do we need $\eta$ in order that $\xi$ is smaller than our desired tolerance (i.e. so that the errors from the inaccuracy of the calculation of the propagator are negligible compared to the other errors affecting our result)? We do not need to maintain a high $\eta_0$ for the whole outer iteration to invert $A$. We need to start with an accurate matrix sign function, but can gradually reduce the accuracy as the inversion progresses. The optimal relaxation strategy depends on the inversion routine.
In a Krylov subspace method, an approximation to the residual is calculated at every step. If the matrix $A$ is not exactly calculated, then this residual will diverge from the true residual by an increasing amount. The true residual $||b - A x_k||$ can be compared against the computed residual $||r_k||$, leading to the inequality
\begin{gather}
||b-Ax_k|| \le ||r_k - (b-Ax_k)|| + ||r_k||.
\end{gather}
The goal is to control the residual gap, $||r_k - (b-Ax_k)||$, so that it remains less than the required accuracy of the inversion; otherwise the true residual will stabilise at a value close to the gap. At the $j$th step of the inversion, we compute the matrix sign function to an accuracy $\eta_j$.

For the CG algorithm, with the iteration step,
\begin{gather}
q_{j-1} = A p_{j-1}\nonumber
r_j = r_{j-1} - \alpha_{j-1} q_{j-1}\nonumber\\
x_j = x_{j-1} + \alpha_{j-1} p_{j-1}\nonumber\\
p_j = r_j + \frac{\gamma_j}{\gamma_{j-1}} p^{j-1},\label{eq:cgrel}
\end{gather}
 a sensible relaxation strategy is to compute the sign function in the CG algorithm to an accuracy $\epsilon_A \norm{b} \sqrt{\zeta}$, where $\zeta = \sum_i \norm {r_i}^{-2}$. This means that we want to choose $\eta_j = \epsilon_A \norm{b} \sqrt{\zeta}$, where $\epsilon_A$ is the desired relative accuracy of the outer inversion~\cite{Arnold:2003sx,Cundy:2004pza}.

In minimal residual algorithms, including SUMR, the relaxation strategy is $\eta_j = \epsilon_A \norm{b}/\norm{r_j}$~\cite{Arnold:2003sx,Cundy:2004pza}. We call the relaxed algorithms relCG and relSUMR respectively.

We note that our numerical experience is that it is not possible to completely relax the accuracy of the matrix sign function since if it is allowed to become very inaccurate then the eigenvalue solver no longer converges; but instead we should ensure that the approximation to the matrix sign function remains better than some minimal accuracy. This minimal bound ought to be tuned for each ensemble; we found that an accuracy of $10^{-2}$ was effective on the ensembles we used.

The second strategy is to use a low accuracy inversion of the overlap operator as a pre-conditioner for a high accuracy inversion of the overlap operator. Because the required accuracy of the matrix sign function depends on the required accuracy for the outer inversion, this means that the bulk of the time, which is spent in the pre-conditioner, only requires a very low accuracy matrix sign function. We therefore now have three inversions in the system: An outer inversion of $A$, with a middle inversion of $A$ used as a pre-conditioner for the outer inversion, and the inner inversion required for the matrix sign function. The preferred routine for the outer inversion is GMRES, since this does not require a constant pre-conditioner, and since we only require three or four iteration steps for this routine, the cost of storing all the Arnoldi vectors is relatively small. We therefore use algorithm \ref{alg:gmresr}~\cite{Cundy:2004pza}:
\begin{algorithm}
\begin{align}
&x=0;r=b;C=0;U=0;i=0\nonumber\\
&\text{while }||r||> \epsilon ||b||;\text{ do}\nonumber\\
&\spa \text{solve }Au=r\text{ to relative accuracy } \tilde{\eta}\nonumber\\
&\spa \text {compute }c\text{ with }||Au-c|| < \epsilon ||b|| ||u||/ ||r||\nonumber\\
&\spa \text{for }j\text{ in }1,2,\ldots,i; \text{ do}\nonumber\\
&\spa\spa \beta = (C[j],c)\nonumber\\
&\spa\spa c = c- \beta C[j]\nonumber\\
&\spa\spa u = u - \beta U[j] \nonumber\\
&\spa \text{done}\nonumber\\
&\spa c = c/||c||\nonumber\\
&\spa u = u/||c|| \nonumber\\
&\spa C[i+1] = c  \nonumber\\
&\spa U[i+1] = u \nonumber\\
&\spa \alpha = (c,r)\nonumber\\
&\spa x = x + \alpha u\nonumber\\
&\spa r = r - \alpha c\nonumber\\
&\spa i = i+1\nonumber\\
&\text{done}
\end{align}
\caption{The GMRESR algorithm for overlap fermions}\label{alg:gmresr}
\end{algorithm}
$C$ and $U$ are here arrays of vectors; $c$, $r$, $u$ and $x$ vectors, $\beta$ and $\alpha$ scalars, $i$ an integer. $A$ represents the overlap operator. $\epsilon$ is the accuracy to which we require the inversion, and $\tilde{\eta}$ is the accuracy of the pre-conditioner, which can (and should) be tuned to optimise the inversion. This can be used whether we use CG or SUMR to invert the overlap operator, and we call the resultant algorithms GMRESR(relCG) and GMRESR(relSUMR)\footnote{GMRESR refers to an algorithm suggested in~\cite{vandervosrt:1994}, where a GMRES inversion was used as a pre-conditioner for a second GMRES algorithm as an alternative to restarting. We have replaced the preconditioner with either a CG or SUMR inversion.}. In our numerical tests, we have neglected the additional application of the overlap operator needed to convert the CG algorithm into a CGNE algorithm directly comparable with SUMR.

This approach can easily be combined with a mixed precision calculation: if the matrix sign function is required to a worse accuracy than $\sim 2^{-24}\sim 10^{-7}$ (in practice, the tolerance needs to be a little larger than this, and to find the optimal value requires some tuning), one can use a single precision representation of the matrix sign function; while if it needs to be better than this one uses the double precision matrix sign function. This allows the bulk of the work to be performed in single precision, giving approximately a factor of two gain over the fully double precision algorithm.

The combination of GMRESR preconditioning, relaxation and mixed precision tends to give a factor of ten-fifteen gain over the original unimproved algorithm.

\section{Deflation}\label{sec:4}
Further improvements are possible by preconditioning the relCG or relSUMR pre-conditioners. In~\cite{Cundy:2005mn}, it was proposed to do so by building a pre-conditioner from the eigenvectors of $H = \gamma_5 D$ (see also~\cite{DuffBeer,Erhel:2000:ACG:354353.354400,NLA:NLA1680020105} for variations on this approach).

Suppose that we have calculated approximations to the $n$ smallest eigenvectors of $H^2$, $\psi_1,\psi_2,\ldots,\psi_n$, with eigenvalues $\lambda_1,\lambda_2,\ldots, \lambda_n$. The accuracy of the eigenvectors is measured using the residual
\begin{gather}
r_i = H^2 \psi - \frac{(\psi_i,H^2\psi_i)}{(\psi_i,\psi_i)} \psi_i.
\end{gather}
We label the Ritz estimate of the eigenvalue as $(\psi_i,H^2\psi_i)/(\psi_i,\psi_i) = \mu_i$, where $\psi$ now represents a guess of the eigenvector. It is possible to construct a preconditioner for the CG algorithm using eigenvectors calculated with a fairly low accuracy (how low an accuracy depends on how the eigenvectors are calculated and the simulation parameters). We build up a spectral pre-conditioner
\begin{gather}
\hat{H}^{-1} = 1 + \sum_i |\psi_i\rangle\langle \psi_i | \left(\frac{c}{\sqrt{\mu_i}} - 1\right),
\end{gather}
used on both the left and the right of $H^2$, where $c$ is a constant chosen to be somewhere in the bulk of the eigenvalue spectrum of $H$ (we use $c = \frac{\sqrt{3}}{2}$). If the eigenvectors were exact, this would project the smallest eigenvalues of $\hat{H}^{-1} H^2 \hat{H}^{-1}$ to $c$, and it would be equivalent to a full deflation algorithm, albeit with a little more cost as we have to repeatedly apply the eigenvector projection. There is no need to calculate the eigenvectors accurately, and our experience is that a relative residual, $\norm{r_i}/\mu_i$, of around $10^{-1}$ for the eigenvectors is usually good enough to obtain the optimal convergence rate (although our results in this study seem to suggest that it does not even need to be as good as this).  If there are a number of small eigenvalues of $H$ separated from the rest of the spectrum, this simple trick may accelerate the inversion a great deal, and improvements in the inversion algorithm of a factor of three or four are not uncommon, with better results with smaller masses; of course, this has to be offset against the cost of calculating the eigenvectors in the first place. The disadvantage with deflation methods occur on larger lattices, where since the density of small eigenvalues increases, the number of eigenvectors needed to have the same effect also increases. This is as much of a problem for the required memory to store the additional eigenvectors as computational time.

The deflated CGNE algorithm can be constructed in the standard way. For a zero initial guess, algorithm \ref{alg:CGNE} is used to find the solution $x$ to $Hx-b = r_{\infty} =0$.
\begin{algorithm}
\begin{align}
& r_0 = p_0 = \hat{H}^{-1}b; x_0 = 0\nonumber\\
&\text{for } j\text{ in } 1,2,3,\ldots \text{until } \norm{r_j} < \norm{b}\epsilon\text{; do}\nonumber\\
&\spa z = \hat{H}^{-1} H^2 \hat{H}^{-1} p_{j-1}\nonumber\\
&\spa \alpha_j = \frac{(r_{j-1},r_{j-1})}{(r_{j-1},z)}\nonumber\\
&\spa r_j =r_{j-1} -\alpha_j z\nonumber\\
&\spa x_j =x_{j-1} +\alpha_j p_{j-1}\nonumber\\
&\spa \beta_j =\frac{(r_j,r_j)}{(r_{j-1},r_{j-1})}\nonumber\\
&\spa p_j = r_j + \beta_j p_{j-1}\nonumber\\
&\text{done}\nonumber\\
&x = H\hat{H}^{-1}x_j
\end{align}
\caption{The preconditioned CGNE algorithm.}\label{alg:CGNE}
\end{algorithm}
This is more stable and robust than the Galerkin deflation algorithm which is commonly used~\cite{Stathopoulos:2007zi,Giusti:2002sm,Saad:2003:IMS:829576}. In particular, the Galerkin algorithm tends to only improve the convergence rate of the algorithm up to the point where the inversion residual is comparable to the accuracy of the eigenvectors, while this method maintains a result as good as full deflation even for very low accuracy eigenvectors. The reason that this algorithm is not generally used in QCD applications is because of the cost of applying the preconditioner, which needs to be applied on every application of the operator $H^2$; however, for overlap fermions, the additional cost of the pre-conditioner is negligible compared to that of the matrix sign function, so we may as well use the more robust method. This also has a disadvantage when using the same Krylov subspace to calculate or improve the eigenvectors, as we shall do below, since the Krylov subspace generated is that of the preconditioned operator rather than the operator we want. We found that this was not significant in practice if we use the eigenvectors for deflation, since only a very low accuracy of eigenvectors is required for the full effect of the deflation. However, it does mean that it is impractical to combine this inversion routine with a precise computation  of the eigenvalues if they are needed for purposes other than deflation.

However, the spectral preconditioner cannot be used for SUMR since it breaks the unitary structure of the operator. We must therefore find some alternative. The Galerkin algorithm uses some orthonormal subspace constructed from $N$ (when $N$ is much smaller than the size of the matrix $A$) vectors ${v}_i$ which are bundled together in the matrix $V$ to project the initial guess of the inversion out of the subspace. So given an initial guess $\tilde{x}_0$, the new initial vector used for the inversion algorithm is given by
\begin{gather}
x_0 = \tilde{x_0} + V \frac{1}{V^\dagger A V} V^\dagger (b - A\tilde{x}_0).
\end{gather}
This has been applied to the SUMR and other routines for overlap fermions in~\cite{Chiarappa:2006hz}, though without using the optimum relaxation and preconditioning strategies.

For SUMR, we have used a slightly modified form of this approach. In this case, we are interested in the inversion of the Dirac operator $D$, however it is more convenient to use the eigenvectors of the Hermitian operator $\gamma_5 D$.  The non-zero eigenvalues of the overlap Dirac operator come in complex conjugate pairs. Given that $D^\dagger D=H^2$ commutes with $\gamma_5$ and $[D,D^\dagger] = 0$, the eigenvectors of $D^\dagger D$ come in pairs of opposite chirality, the eigenvectors of $\gamma_5 D$ come in pairs with eigenvalues of opposite sign. Each of these sets of eigenvector pairs are linear combinations of each other; so if an eigenvector pair of $\gamma_5D$ is $\{\psi_+,\psi_-\}$, then the the corresponding eigenvector pair of $D$ is a linear combination of these two vectors. There is no difference, then, in constructing the starting vector from a linear combination of the $n$ lowest eigenvectors of $\gamma_5 D$ rather than a linear combination of the eigenvectors of $D$. Given an orthonormal set of approximate eigenvectors $\psi_i$ of $\gamma_5 D$, we first of all construct a new basis $\psi'= Y \psi$ for unitary $Y$ where $(\psi'_i,\gamma_5 D\psi'_j)$ is diagonal. This is always possible for a Hermitian operator, since $(\psi_i,\gamma_5 D\psi_j)$ would also be Hermitian and $Y$ would just contain the eigenvectors of this $n\times n$ matrix.

We then construct the starting vector for the SUMR algorithm
\begin{gather}
 x_0 = \sum_j \alpha_j \psi'_j,
\end{gather}
where $\alpha_j$ is chosen to minimise the norm of the residual
\begin{gather}
 r' = b-\sum_{j=1}^N \alpha_j D \psi'_j,
\end{gather}
which is the same as minimising the norm of
\begin{gather}
 r = \gamma_5 b - \sum_j \alpha_j \gamma_5 D \psi'_j
\end{gather}
so
\begin{gather}
 \norm{r}^2 = (b,b) + \sum_j \left[|\alpha_j|^2 (D\psi'_j,D\psi'_j) - (b,D\psi'_j)\alpha_j - (D\psi'_j,b)\alpha_j^\dagger\right],
\end{gather}
where $\dagger$ applied to a scalar indicates complex conjugation. The residual is minimised for
\begin{gather}
 \alpha_j = \frac{(D\psi'_j,b)}{(D\psi'_j,D\psi'_j)}.
\end{gather}
This procedure will obviously work for any set of vectors $\psi'_j$, not just the eigenvectors.

If $\psi_j'$ were an exact eigenvector of $D$, this is equivalent to exact deflation; we project the low lying eigenvectors out of the initial residual vector. However, for approximate eigenvectors of $D$, we may still obtain a similar speed up of the inversion up to the accuracy of the eigenvectors. This is similar to the Galerkin projection (for an initial guess $\tilde{x} = 0$, it is equivalent to the Galerkin projection where $A = D^\dagger D$ with a right hand side $D^\dagger b$), but avoids the need to invert $V^\dagger A V$ for each right hand side.

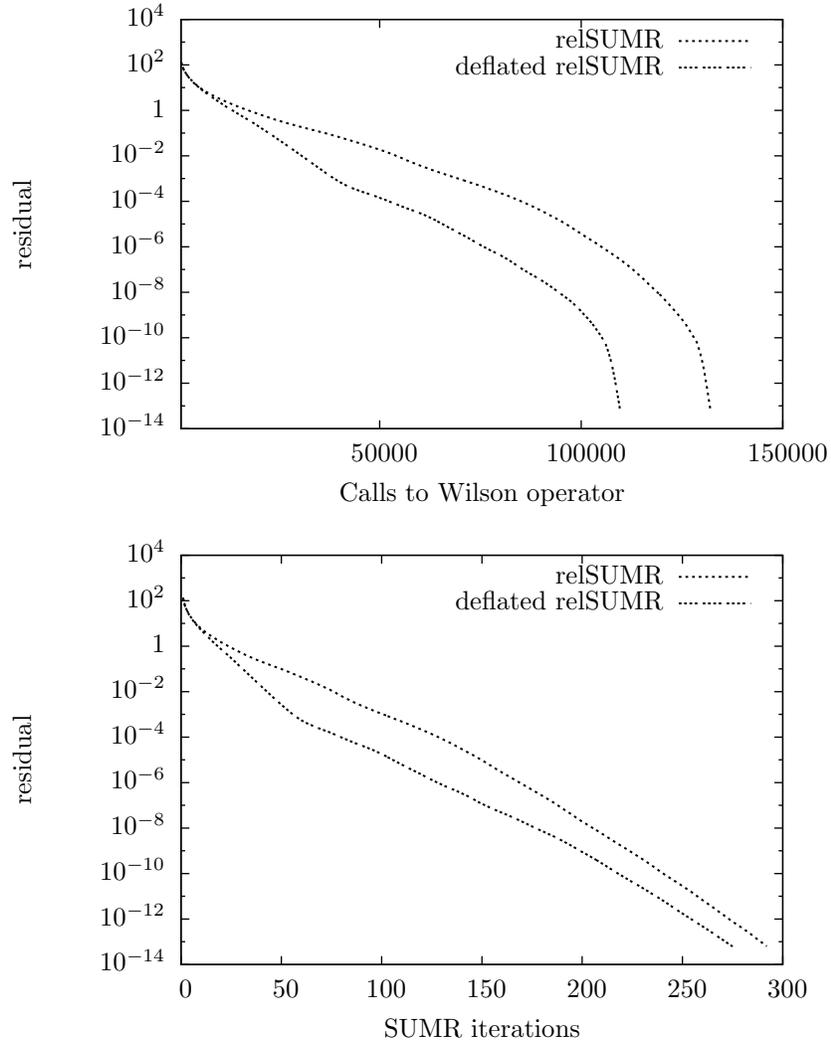
\begin{figure}
 \begin{center}
 \begin{tabular}{c}
  \input{figs/s8t32m0.03_c55_n30SUMRWilsonCalls2.tex}\\
  \input{figs/s8t32m0.03_c55_n30SUMRiterations2.tex}
  \end{tabular}
 \end{center}
\caption{The convergence in terms of the number of iterations of the inversion and the number of Wilson operator calls (counting a double precision Wilson operator as two calls) of the a) undeflated SUMR routine; b) deflated SUMR routine, with 30 overlap eigenvalues calculated with residuals below $10^{-3}$ on a $8^3\times 32$ dynamical overlap lattice.}\label{fig:1}
\end{figure}

Unlike the spectral decomposition, this method only improves the convergence of the SUMR algorithm up to the accuracy of the eigenvectors. For example, if we switch off  GMRESR preconditioning, the residual is plotted against the iteration count and number of calls to the Wilson operator in figure \ref{fig:1}, on a $8^3\times32$ lattice, both for the original undeflated algorithm and the deflated algorithm. It can be seen that the rate of convergence with respect to the number of iterations rate of the algorithm initially improves, but then reverts to the convergence of the undeflated system. In terms of the iteration count, there is no real difference whether we use deflation or not. However, there is a small improvement in the number of calls to the kernel operator. This is because the initial steps of the inversion are when the most accurate matrix sign function is required. Fewer accurate calls to the matrix sign function are required, so the overall cost in the routine is reduced a little. However, the gain is still much smaller than we would usually desire from a deflation algorithm. This picture can be improved by improving the accuracy of the eigenvectors.

This problem is averted when we use GMRESR preconditioning. We only need solve the preconditioner for the GMRES algorithm (the middle inversion) to a low accuracy. This means that while calculating the preconditioner, we remain in the regime where the projection is advantageous.

\section{Calculation of Eigenvalues}\label{sec:5}
The main challenge with deflation is the calculation of the eigenvalues. One can, of course, use a previous calculation of the eigenvalues, using, for example, an implicitly restarted Lanczos method, or a CG minimisation of the Ritz functional. An alternative is to use the Krylov subspace generated for an inversion to calculate the eigenvalues. This has previously been adapted for the CG algorithm for non-overlap fermions in~\cite{Stathopoulos:2007zi}; the resulting algorithm was called eigCG.

Suppose that we want to construct $n$ eigenvalues $\{\psi_1,\psi_2,\psi_3,\ldots\}$. The basic idea is to construct a basis of vectors $v$ from the CG residuals, $\{v_i,v_j,\ldots\}$ and the matrix $M_{ij} = (Dv_i,Dv_j)$. After $m$ iterations, one diagonalises $M$ using some unitary operator $U$, so $U^\dagger M U = \Lambda$ where $\Lambda$ is diagonal and sorted so that its smallest element has the smallest index. The eigenvalues of $M$ form an estimate of the eigenvalues of $D^\dagger D$ while the vectors $\mathbf{v}' = U \mathbf{v}$ form estimates of the eigenvectors. We then construct a new set of $v$ vectors with the first $n$ elements being the first $n$ elements of $\textbf{v}'$, and the remaining $m-n$ elements being built up from the next residuals of the CG inversion, after re-orthogonalising $\textbf{v}$ if necessary (although in practice it is quicker and more robust to initially calculate $M$ with the unorthogonalised $\textbf{v}$ and manipulate $M$ to simulate the orthogonalising of the eigenvectors). Until the first eigenvector-restart (when we calculate the eigenvectors of $M$), the method is equivalent to the Lanczos method (given that $D^\dagger D$ is Hermitian); and, as long as the orthogonality of the CG residuals does not break down the convergence of the eigenvectors continues to follow that of the un-restarted Lanczos behaviour throughout the CG inversion. When we complete the initial inversion, we resume the inversion against the next input vector, deflating those eigenvectors already calculated to the desired accuracy, with the first $n$ elements of the new $v$ taken from our best estimates of the eigenvectors. Once all the eigenvalues are calculated to the desired accuracy, one switches off the eigenvalue part of the routine, and continues with a simple deflated CG.

There is, of course, a small cost for eigCG over a standard CG inversion, but it is not large (neglecting, for the moment, the effects of relaxation and preconditioning). Most of the manipulations of the small matrix $M$, including finding its eigenvectors and eigenvalues, are negligible compared with the computations involving the full-sized matrix (such as the Dirac operator and spinor algebra). One needs a few additional matrix vector products to build up $M$ and to re-orthogonalise $\textbf{v}$, and a few vector manipulations to reconstruct $\textbf{v}'$. For overlap fermions, this cost (if we exclude relaxation) is entirely negligible compared against the cost of the matrix sign function. We are required to store the vectors $\mathbf{v}^D=\{D v_1,Dv_2,Dv_3\ldots\}$ as well as $\mathbf{v} = \{v_1,v_2,v_3\ldots\}$ during the simulation. This allows us to easily measure the residual of the eigenvector, $r^v = \gamma_5 D v - (v,\gamma_5 D v) v$, without having to apply any additional matrix vector products.

For overlap fermions, since the eigenvalues of $D^\dagger D$ come in degenerate pairs, it is useful to separate these pairs to improve the stability of the algorithm. We do this by in place of $M_{ij} = (D v_i,D v_j)$ using $M_{ij} = (D v_i,D v_j) + 2\mu \delta  (v_i,\gamma_5 D v_j)$, where $0<\delta \ll 1$ is a tuned small constant. This $M_{ij}$ remains Hermitian and positive definite (if the eigenvalues of $D^\dagger D$ are $\lambda_i^2$, then $\lambda_i^2 > 4\mu^2$ so the smallest eigenvalues of $D^\dagger D + 2\mu \delta \gamma_5 D$ are $4\mu^2 (1-\delta)$, but its eigenvectors will be the eigenvectors of $\gamma_5 D$ rather than an random mixture of the eigenvector pairs of $D^\dagger D$. The eigenvalue pairs of $M$ will be separated by $4\mu \delta \epsilon |\lambda|$; which avoids the danger of the Lanczos algorithm missing one of a pair of degenerate eigenvectors (This, of course, does not help resolving degenerate zero eigenvalues). In a CG algorithm, one applies $\gamma_5 D \gamma_5 D p$; to allow a splitting of the eigenvalue pairs requires that we store $\gamma_5 D p$ instead. Indeed, we have used both $p$ and $\gamma_5 D p$ as separate $v$ vectors. This, of course, destroys the natural orthogonality of the residual vectors.

Our principle interest is to create an eigSUMR algorithm for overlap fermions. The construction follows the same principles as the eigCG algorithm, and the crucial part of the algorithm, the generation of the subspace, can be summarised in algorithm \ref{alg:eigsumr} for the first inversion. We seek to find the first $n$ eigenvalues to an accuracy $\epsilon_v$ and use a maximum subspace size $m > n$.  Algorithm \ref{alg:eigsumr} can be used either as a stand-alone eigenvalue routine or incorporated into a SUMR inverter. As long as we do not need to recalculate $v^D = Dv$ (which may be necessary if a low accuracy sign function is used) or relax the accuracy of the matrix sign function the additional overhead on top of the SUMR inversion is negligible.
\begin{algorithm}
\begin{align*}
&q_0 = \tilde{q}_0 = b/\norm{b}\nonumber\\
&\mathbf{v} = 0; \mathbf{v}^D = 0; k = 0\nonumber\\
&\text{for }j\text{ in }0 ,1 ,2 ,3, 4, \ldots;\text{ do}\nonumber\\
&\spa u = U q_j ; \nonumber\\
&\spa
v_k = q_j ; \nonumber\\
&\spa
v^D_k = \frac{(1-\mu)}{2} u + \frac{(1+\mu)}{2}q_j\nonumber\\
&\spa k = k+1 \nonumber\\
& \spa\text{if }(k == m)\text{; then}\nonumber\\
&\spa\spa \text{Recalculate the eigenvalues following algorithm }\ref{alg:eigsumr2}.\nonumber\\
&\spa \text{end if}\nonumber\\
&\spa \gamma_j =- (\tilde{q}_j,u) ; 
\sigma_j = \sqrt{1-|\gamma_j|^2}\nonumber\\
&\spa q_{j+1} = \frac{1}{\sigma_j} (u + \gamma_j \tilde{q}_j)\nonumber\\
&\spa \tilde{q}_{j+1} = \sigma_{j} \tilde{q}_j + \gamma_j^* q_{j+1}\nonumber\\
&\text{done}\nonumber\\
& \text{Recalculate the eigenvalues following algorithm }\ref{alg:eigsumr2}.\nonumber\\
&\text{done}
\end{align*}
\caption{The eigSUMR eigenvalue routine, where $U = \gamma_5 \sign(K)$.}\label{alg:eigsumr}
\end{algorithm}
\begin{algorithm}
 \begin{align*}
& \text{Construct }E_{ij} = (v_i,v_j)\nonumber\\
& \text{Perform the LDU decomposition on }E\text{, giving an upper triangular matrix }U\nonumber\\
&\spa\spa\spa\spa\text{ and a lower triangular matrix } L = U^\dagger\nonumber\\
& \text{Construct }M_{ij} = (v^D_i,v^D_j) + 2\mu\delta (v_i,\gamma_5 v^D_j)\nonumber\\
& \text{Find the unitary matrix }V\text{ which diagonalises }L^{-1} M U^{-1}\nonumber\\
& {v}_{0,\ldots n-1} = (VU^T \mathbf{v})_{0,\ldots n-1}\nonumber\\
& {v}^D_{0,\ldots n-1} = (VU^T \mathbf{v}^D)_{0,\ldots n-1}\nonumber\\
& k = n\nonumber\\
& \text{for }i\text{ in }0,1,\ldots, n-1\text{; do}\nonumber\\
& \spa r_{v,i} = v^D_i - v_i (v_i,\gamma_5 v^D_i)\nonumber\\
& \text{done}
 \end{align*}
\caption{The recalculation of the eigenvalues in the eigSUMR routine.}\label{alg:eigsumr2}
\end{algorithm}

 We have explicitly re-orthonormalised the $v$ vectors for each calculation of the eigenvectors. This is necessary because we find that the orthogonalisation of the vectors in the SUMR can be quickly lost, especially when we use a low accuracy computation of the matrix sign function. To improve stability and speed, we have done so using a LDU decomposition. This allows us to avoid necessary additional spinor operations, which are both costly and a source for the propagation of rounding errors. Our method is equivalent to the Gram-Schmidt algorithm. The matrix $E_{ij} = (v_i,v_j)$ is decomposed as $E = L\tilde{D}U$, where $L$ is lower triangular, $\tilde{D}$ is diagonal and $U$ is upper triangular.\footnote{For the general LDU decomposition, we use the convention that $\tilde{D}_{ii}$, the elements of $\tilde{D}$ satisfy $|\tilde{D}_{ii}| = 1$ while $L_{ii} = U_{ii}^\dagger$.}  In our case, since $E$ is Hermitian and positive definite, $\tilde{D}$ is just the identity matrix and $L = U^\dagger$ (the LDU decomposition reduces to the Cholesky decomposition).

After each restart of the Krylov subspace (i.e. each new right hand side), we repeated the diagonalisation procedure in algorithm \ref{alg:eigsumr2} using a basis constructed from the previous three sets of computed eigenvectors (i.e. the eigenvectors as they were calculated at the end of the inversion, the eigenvectors from the inversion before that, and the eigenvectors outputed from the inversion before that). In principle, the largest contributions of the errors to the approximate eigenvectors come from the next highest eigenvectors. Each guess of the eigenvectors thus contains those vectors we want, plus the next highest eigenvectors plus some additional noise. By performing this diagonalisation, we hoped to remove the next highest eigenvectors from our present guesses. This gave a small performance gain in the computation of the eigenvectors at a negligible additional cost.

\subsection{Relaxation}\label{sec:5.1}
The question remains as to what is the optimal strategy for controlling the accuracy of the matrix sign function. The accuracy is determined by how good the approximation of $v^D = D v$ is. In the routine above, we assumed that the matrix sign function was calculated to infinite accuracy. This is in practice impossible. Instead, we use an approximate matrix sign function $\tilde{s}$ (with $s$ the exact sign function) which leads to an approximate Dirac operator $\tilde{D}$ and an approximate $\tilde{v}^D$. We can write,
\begin{gather}
 \tilde{v}^D = v^D + \delta,
\end{gather}
and our goal is to keep $\norm{\delta}$ sufficiently small so that it has no significant effect on the estimate of the eigenvalue or the residual. Our experience is that using too low an accuracy for $v^D$ can be disastrous: the estimate of the eigenvalues of $M$ rapidly diverges from the true eigenvalue spectrum, and the residual of the eigenvectors correspondingly grows worse with each restart. While we wish to relax the matrix sign function as much as possible, we cannot relax it too much.

The first question is how we can measure $\norm{\delta}$; and unfortunately a direct measurement requires an application of the matrix sign function and is therefore expensive. It is, however, possible to get a quick order of magnitude estimate by using
\begin{gather}
 (v,\gamma_5\tilde{v}^D) = (v,\gamma_5 v^D) + (v,\gamma_5 \delta).
\end{gather}
$(v,\gamma_5 v^D)$ is an estimate of the eigenvalue of $\gamma_5 D$, and this quantity is real. Therefore, the imaginary part of $(v,\gamma_5\tilde{v}^D)$,  only comes from the imaginary part of $(v,\gamma_5 \delta)$, $\Im(v,\gamma_5 \delta)$. We can write $(v,\gamma_5\delta) = \norm{v}\norm{\delta} \cos\theta e^{i\phi}$, in which case
\begin{gather}
|\Im(v,\gamma_5\tilde{v}^D)|  \norm{v}\norm{\delta}|\cos\theta \sin\phi| = \norm{\delta}|\cos\theta\sin\phi|,\label{eq:bound1}
\end{gather}
given that $\norm{v} = 1$. 

An additional estimate of the error in $v^D$ is to consider the vector $\hat{v}' = 2\frac{v^D - \frac{1+\mu}{2}v}{1-\mu}$. In exact arithmetic, this should be $\gamma_5 \sign v$, so $\norm{\hat{v}'}^2$ should be one. The inaccuracy of $v^D$ can be thus be estimated by the deviation
\begin{gather}
\norm{\hat{v}'}^2 - 1=  \frac{4\delta^2}{(1-\mu)^2} - ((\delta,v) + (v,\delta)) \frac{1+\mu}{1-\mu} .\label{eq:bound2}
\end{gather}
$(\delta,v^D) = 0$, and given that $v^D  - \mu v = r^v$, where $\mu$ is the Ritz estimate of the eigenvalue and $r^v$ the residual of the eigenvector, we find that $(\delta,v) \sim (\delta,r^v)/|\mu| \sim 2 \norm{\delta}\norm{r^v} \cos \theta'\cos\phi'/\mu$, where $\theta'$ and $\phi'$ parametrise the angle between $\delta$ and $r^v$. We write
\begin{align}
 \gamma =&  \frac{1-\mu^2}{4} (\norm{\hat{v}'}^2 - 1)\nonumber\\
 \beta = & \frac{(1+\mu^2)r^v}{2|\mu|}\nonumber\\
 \alpha = & \cos \theta'\cos\phi',
\end{align}
so
\begin{gather}
 \gamma = \norm{\delta}^2 - \norm{\delta} \beta \alpha,
\end{gather}
which gives
\begin{gather}
 \norm{\delta} = \frac{1}{2}\beta\alpha \pm \frac{1}{2}\sqrt{\beta^2 \alpha^2 + 4\gamma},
\end{gather}
where we can cheaply measure $\gamma$ and $\beta$ (at least to a reasonable accuracy), while $\alpha$ is in the range $-1\le\alpha\le 1$. Note that we require $\norm{\delta} > 0$ and $\norm{\delta} = \sqrt{\gamma}$ when $\beta = 0$, which means we should take the positive root of the solution when $\gamma > 0$. This means that we can establish bounds on $\norm{\delta}$. 
\begin{gather}
 \max(0,-\frac{1}{2} \beta  + \frac{1}{2} \sqrt{\beta^2 + 4 \gamma})\le \norm{\delta} \le \frac{1}{2} \beta  + \frac{1}{2} \sqrt{\beta^2 + 4 \gamma})
\end{gather}
These results will therefore give two different estimates of $\norm{\delta}$. Neither of these quantities are ideal; the first just gives a minimum bound on $\delta$, while the second bound depends on our estimate of $\norm{r^v}$, which can deviate significantly from the true value if $\norm{\delta}$ is too large. However, they each give us some estimate of the error without having to continually apply a high accuracy matrix sign function. 
We may thus keep track of these estimates of $\delta$, and recalculate $v^D$ to a high accuracy when one of them indicates that $\norm{\delta}$ might have risen sufficiently close to $\epsilon_\delta$, our maximum allowed tolerance for $\delta$ (where the precise meaning of `sufficiently' has to be judged by experience).
 In an eigSUMR routine, we may also measure the accuracy of $\tilde{v}^D$ more indirectly through the difference between the computed and exact residuals from the inversion (the exact residual is calculated when we restart the pre-conditioner in an GMRESR routine).

The residual of the estimate of the eigenvector is defined (in exact arithmetic) as
\begin{gather}
 r_{\text{true}}^v = \gamma_5 D v - v (v,\gamma_5 D v) = \gamma_5 v^D - v(v,\gamma_5 v^D).
\end{gather}
In practice in inexact arithmetic, we will have
\begin{gather}
r^v = r_{\text{true}}^v + \gamma_5 \delta - v(v,\gamma_5 \delta),
\end{gather}
which gives
\begin{gather}
 \norm{r^v}^2 = \norm{r^v_\text{true}}^2 + (r^v_{\text{true}},(1-v v^\dagger)\gamma_5 \delta) + ((1-v v^\dagger)\gamma_5 \delta,r^v_{\text{true}}) + (\delta,(1-vv^\dagger) \delta).
\end{gather}
The residual gap, $g$, is the difference between $\norm{r^v}^2$ and $\norm{r^v_\text{true}}^2$, and we want to keep this below $\epsilon^2$, where $\epsilon$ is the desired accuracy for the eigenvector. We have
\begin{gather}
 g \le 2\norm{r^v_{\text{true}}}\norm{\delta} + \norm{\delta}^2 < \epsilon^2.
\end{gather}
This bound gives
\begin{gather}
 \norm{\delta} < \sqrt{\epsilon^2 + \norm{r^v_{\text{true}}}^2} - \norm{r^v_{\text{true}}}. \label{eq:targetdelta}
\end{gather}
During each update of the eigenvectors, we know that
\begin{align}
 v_i \rightarrow & (UV)_{ij} v_j \nonumber\\
 \tilde{v}^D_i \rightarrow & (UV)_{ij} \tilde{v}^D_j = (UV)_{ij} {v}^D_j + (UV)_{ij}\delta^D_j,
\end{align}
where $\delta^D_j$ is either the previously calculated error from the previous eigenvector-restart if $v_j$ is one of the vectors held back from the previous iteration, or due to the inaccuracy of the matrix sign function as we generate the new Krylov subspace.
Thus, neglecting any error on $(UV)$ due to the fact that they have been calculated using $\tilde{v}^D$ rather than $v^D$, the new $\delta_i$ satisfies the bound
\begin{gather}
 \norm{\delta_i} < \sum_{j=1}^m |(UV)_{ij}| \norm{ \delta^D_j}.
\end{gather}
This means that to avoid having to recalculate $v^D$ during the iteration, we may compute the matrix sign function to an accuracy
\begin{gather}
 \norm{ \delta^D_j} < \frac{\sqrt{\epsilon^2 + \norm{r^v_{\text{true},0}}^2} - \norm{r^v_{\text{true},0}}}{(m-n)k \max_{n<j\le m, i <n'}|(UV)_{ij}|},\label{eq:nearlybound}
\end{gather}
where $k$ is the expected number of iterations required for the eigenvector routine to converge, or between high accuracy recalculations of $v^D$, and $\norm{r^v_{\text{true},0}}$ is the residual of the best converged eigenvector (we may use the computed residual rather than the true residual here, as the difference between the two will not be significant enough to affect anything).  $\max_{ij}|(UV)_{ij}|$ may be estimated from the previous eigenvector-restarts; the general trend in the long-term is that once the eigenvectors start to converge this quantity decreases as the iteration proceeds, as may be expected since it provides the correction to the ever-improving eigenvector. However, it can fluctuate significantly from one iteration to the next, and for this reason we took the average value from the previous ten eigenvalue calculations, which worked well in practice. Of course, this estimate is conservative, and in practice we relaxed it by assuming that the additional errors $\delta^D_j$ will not all add together coherently. It is more likely that the effects of $\delta^D_j$ will partially cancel each other out. We assume that the effects of these vectors on the error on $v^D$ will resemble a random walk, suggesting that need only scale the accuracy of the matrix sign function according to the inverse square root of the number of vectors rather than the inverse of the number of vectors. After the initial few calculations, even this proved to be too conservative since most values of $|UV|$ were several orders of magnitude smaller than the maximum value which only affected a different eigenvector each time. Therefore, we scaled the bound by the square root of the number of eigenvectors to simulate that only one of these eigenvectors was affected.
\begin{align}
 \norm{ \delta^D_j} < & \frac{\chi \sqrt{\epsilon^2 + \norm{r^v_{\text{true},0}}^2} - \norm{r^v_{\text{true},0}}}{\sqrt{(m-n)k} \max_{n'<j\le m, i <n}|(UV)_{ij}|}\label{eq:bound}\\
 \chi =&\left\{\begin{array}{lr} 1&\text{  First two diagonalisations}\\
                \sqrt{n}&\text{  Subsequent diagonalisations}
               \end{array}\right.\nonumber .
\end{align}
This is the expression we have used in our simulations. While, unlike (\ref{eq:nearlybound}), it is not guaranteed to avoid large errors in the eigenvectors, it nonetheless seems to work well. When using SUMR, it is advisable to recalculate $v^D$ after the completion of the eigenvector calculation to avoid errors entering the calculation of the residual during deflation.

It is also possible to reduce the accuracy of $\epsilon$ as the algorithm proceeds; starting with a high value of $\epsilon$ and gradually reducing it to the desired precision. We need to explicitly recalculate $v^D$ every $k$ iterations to maintain accuracy, and there is no cost if we adjust $\epsilon$ before this recalculation.

Finally, we note that $\delta^D_j$ may be measured when we apply the matrix sign function to $v_j$. If $u = \tilde{s} v$, then
\begin{align}
 u =& s v + \gamma_5 \delta^D\nonumber\\
 \frac{\norm{u}^2}{\norm{v}^2} - 1 =& \frac{\norm{\delta^D}^2 + (v,s\gamma_5 \delta^D) + (\delta^D,\gamma_5 s v)}{\norm{v}^2} \lesssim 2 \norm{\delta^D},
\end{align}
where in the last step we have used the normalisation $\norm{v} = 1$.

When using the algorithm as a stand-alone eigenvalue solver, we may freely use the bound (\ref{eq:bound}), adjusting the accuracy of the matrix sign function after each eigenvector-restart. When incorporating this into a relaxed GMRESR(SUMR) routine, there are problems because this bound is not the same bound as used in the inversion. There are several possibilities:
\begin{itemize}
 \item Only incorporate those $q$ vectors where the calculation of the sign function satisfies the bound (\ref{eq:bound}) into the Krylov subspace, at the cost of a slower calculation of the eigenvectors;
 \item Relax the inversion routine as normal, including all the eigenvectors, at the cost of having to recalculate $v^D$ to a high accuracy when we construct the eigenvectors;
 \item Relax the inversion routine according to the maximum of the two bounds for the accuracy of the matrix sign function, at the cost of having a slower inversion routine.
\end{itemize}

In the numerical simulation presented here, we applied the third of these strategies. This significantly increases the cost of the initial inversions which are used to calculate the eigenvalues; however it means that we are able to use larger subspaces to calculate the eigenvalues so that they converge better; and our experience is that only three or four inversions are sufficient to have the smallest eigenvectors converge to the accuracy which gives the optimum rate of improvement for the inversions. Our experience of the first strategy on smaller test lattices showed that the eigenvectors did not converge, as only three or four additional vectors were used for each call to the GMRESR preconditioner to improve the already estimated eigenvectors: not enough to make any serious progress. This approach therefore offered no real gain over the undeflated GMRES(SUMR) routine as the eigenvectors were never of a sufficient accuracy to substantially accelerate the inversion. The second of these strategies is obviously inferior, as we would be calculating the matrix sign function twice, once to a high accuracy, per eigenvector.  However, whichever strategy we use, there is a cost to the initial inversions used to calculate the eigenvectors. However, once we have calculated enough eigenvectors to a good enough accuracy, we can switch off the eigenvector calculation and use the optimum inversion relaxation strategy. Given a typical number of right hand sides, the gain from the deflation far outweighs these additional costs. It is necessary to recalculate $v^D$ after $k$ calls to the eigenvector routine, where $k$ can be tuned to give the best performance.

\section{Numerical Results}\label{sec:6}
We tested the algorithm on $8^3\times 32$ dynamical overlap configurations, generated with a Luscher-Weisz gauge action~\cite{TILW,TILW2,TILW5} at $\beta = 8.15$ and a quark mass of $\mu = 0.03$ corresponding to a pion mass $m_\pi a= 0.27(4)$ (measured via the axial correlator, $\sim 460 $~MeV) and a lattice spacing of $a =0.118(1)$~fm  (measured from a string tension of $(420~\text{MeV})^2$). We used one sweep of stout smearing on the gauge configuration, matching the set up used to generate the ensemble. The simulations were run on a desktop computer at Seoul National university. We only present results from a single configuration with zero topological charge, since we did not observe any significant difference from configuration to configuration; except that the improvement from the deflation was up to a factor of three better if the Dirac operator had an exact zero mode. We will compare the eigSUMR, eigCG, GMRESR(relCG) and GMRESR(relSUMR) algorithms. We set $k = 25$, which allows us to run five inversions without having to recalculate $v^D$ and and the desired accuracy of the eigenvectors to $\epsilon = 10^{-4}$. The residual of the GMRESR preconditioner was tuned so that the GMRESR routine should converge in three steps after the eigenvectors were calculated and two steps while the eigenvectors are being calculated (to allow us to make use of a larger subspace). We deflated the eleven lowest eigenvectors of the Kernel operator to accelerate the computation of the matrix sign function (the Wilson operator tends to be well conditioned on a smeared dynamical overlap configuration, so this was all that was necessary). We tested the algorithm with 15, 30 and 45 overlap eigenvectors. In all the inversions we used a $Z_2$ source vector, $b$, where each component of the spinor was equally likely to be $\pm 1$. We used different choices of this source vector on our smaller test lattices, and it made no difference to the results. In all the plots for the inversions, we show the residual calculated during the CG or eigSUMR routine. For all the data, baring the issue mentioned in the discussion of \ref{fig:2a} below, this does not differ from the true residual by any noticeable amount. For the GMRESR outer inversion, the true and calculated residuals did not differ significantly for any of our results.

\subsection{GMRESR and relaxation}
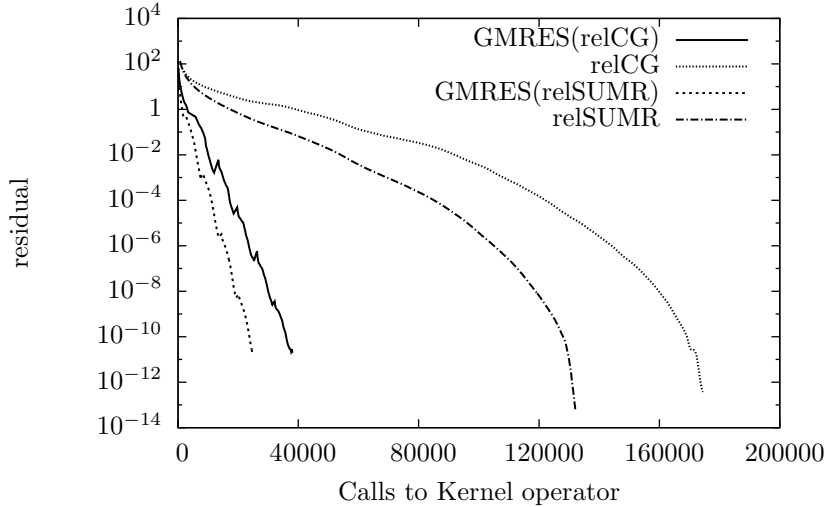
\begin{figure}
 \begin{center}
 \begin{tabular}{c}
  \input{figs/s8t32m0.03_c55_n30cgvsumr.tex}
  \end{tabular}
 \end{center}
\caption{The convergence in terms of the number of Wilson operator calls (counting a double precision Wilson operator as two calls) of a) the relaxed SUMR inversion; b) the GMRESR(SUMR) inversion without deflation; c) the relaxed CG inversion, and d) the GMRESR(CG) inversion algorithm without deflation.}\label{fig:2a}
\end{figure}

Figure \ref{fig:2a} provides a comparison between the relaxed and GMRES preconditioned CG and SUMR routines, showing the gain of both the GMRES preconditioning and SUMR over CG. We generally will use the number of calls to the Wilson operator as a measure of the time for the inversion. Since the bulk of the cost of each algorithm is in the evaluation of the matrix sign function, and the cost of that is proportional to the number of times the Kernel operator is called, this is a reasonable measure to use, and it avoids various machine and run dependent fluctuations which would occur if we measured the total time. In all our figures, if the kernel operator was calculated in double precision, it was counted as two calls to the operator; while if it was calculated in single precision it was counted as one call to the operator. This compensates for the observation that a double precision Wilson operator is roughly half the speed of the single precision operator.
Figure \ref{fig:2a} merely reproduces the results from earlier works. The curvature on the convergence of the relaxed CG and SUMR iterations is due to the relaxation. If we excluded the relaxation, the plot would be roughly a straight line, with the same gradient at seen on our plot in the early stages of the inversion. It can be seen that relaxation gives a vastly superior performance to the unrelaxed algorithm, and the GMRES algorithm (which we set to use five calls to the preconditioner) is about a factor of four superior to the relaxed algorithm. Thus, in total, these algorithms already give an order of magnitude improvement over the naive CG or SUMR algorithms. The small spikes seen in the GMRES(CG) curve are at the end of of the preconditioning steps. The plot shows the measured residual; however, because of our aggressive relaxation strategy (we sacrificed a small amount of accuracy in the middle inversion for speed, since the GMRES inversion will correct for any small error), the true residual deviates from this by a small amount at the end of the inversion. The GMRES algorithm then resets the residual to its true value for the next call to the preconditioner. It can also be seen that SUMR is faster than CG for a single inversion, though by considerably less than the factor of two we might naively expect. This is partly caused by SUMR requiring a few more iterations to converge, and partly from the different relaxation strategies for CG and SUMR.

\subsection{Deflation}
\begin{figure}
 \begin{center}
 \begin{tabular}{c}
  \input{figs/s8t32m0.03_c55_n30deflate.tex}
  \end{tabular}
 \end{center}
\caption{The convergence in terms of the number of Wilson operator calls (counting a double precision Wilson operator as two calls) of the a) undeflated GMRESR(relSUMR) routine; b) deflated GMRESR(deflated relSUMR) routine, with 30 overlap eigenvalues previously calculated to a high accuracy; c) undeflated GMRESR(relCG) routine, and d) the deflated GMRESR(deflated relCG) routine with 30 eigenvalues previously calculated.}\label{fig:2b}
\end{figure}
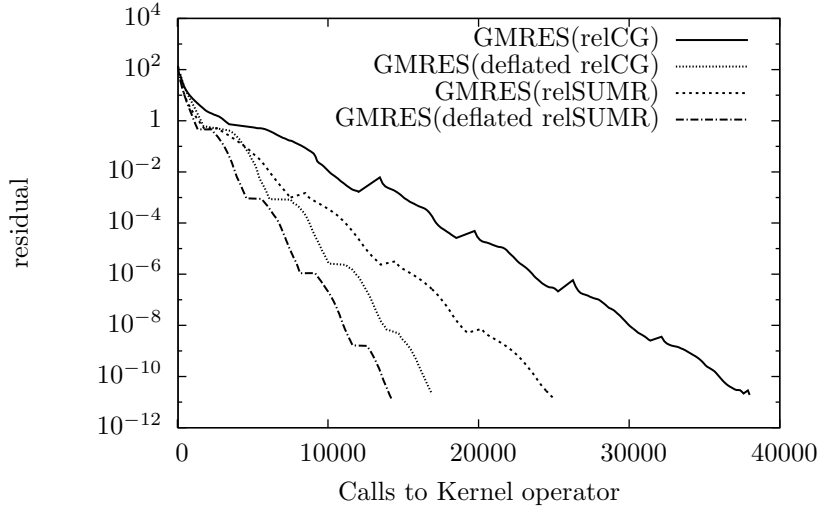

We now turn to the effects of deflation. For these plots, we calculate and use thirty overlap eigenvectors during the deflation. We consider what occurs when we vary the number of eigenvectors later. We used five inversions to calculate the eigenvalues, which we label as \textit{calculating relCG} or \textit{calculating relSUMR}, and then ran several inversions without improving the eigenvectors, and we label these runs as \textit{deflated relCG} or \textit{deflated relSUMR}.
Figure \ref{fig:2b} compares the convergence of the deflated GMRESR(deflated relSUMR) and GMRESR(deflated relCG) algorithms with the undeflated algorithms. The plot is of the residual of the inversion against the number of calls to the kernel of the matrix sign function (given that almost all of the time required for the calculation is due to the computation of the matrix sign function, and the cost of the matrix sign function is proportional to the number of times the kernel matrix is called, this will be a good measurement of the overall cost of the calculation).
Figure \ref{fig:3} shows the cost of the initial eigenvector calculation, plotting the number of kernel operator calls against the residual for those GMRESR(calculating relSUMR) routines used to calculate the eigenvectors, and comparing it against the undeflated and fully deflated GMRESR(relSUMR) routines. In figure \ref{fig:3b}, we do the same thing for the iteration count of the inversion.
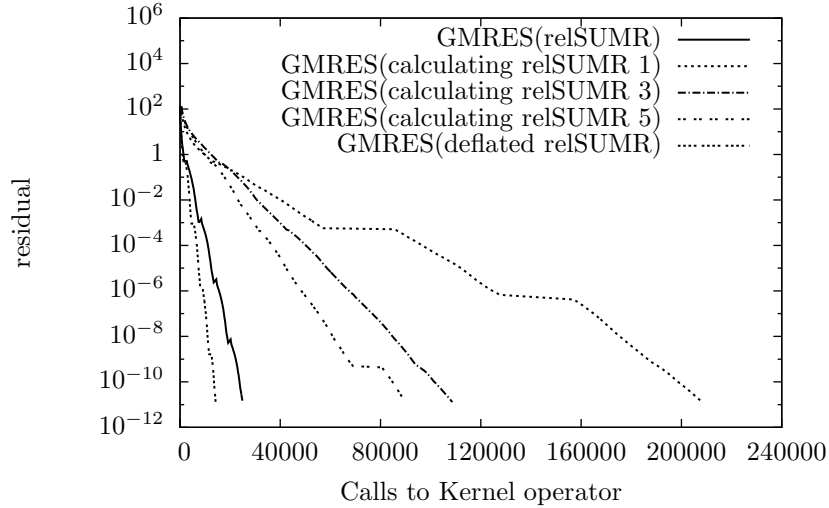
\begin{figure}
 \begin{center}
 \begin{tabular}{c}
  \input{figs/s8t32m0.03_c55_n30GMRESSUMRWilsonCalls.tex}
  \end{tabular}
 \end{center}
\caption{The convergence in terms of the number of Wilson operator calls (counting a double precision Wilson operator as two calls) of a) the undeflated GMRESR(SUMR) routine; b) the deflated GMRESR(eigSUMR) routine, with 30 overlap eigenvalues previously calculated to a high accuracy; c) the first, third and fifth of the GMRESR(eigSUMR) routines where the eigenvectors where calculated. The long plateaus for calculating relSUMR 1 indicates places where we needed to directly recalculate $v^D$ because of a poor estimate of $|UV|$.}\label{fig:3}
\end{figure}
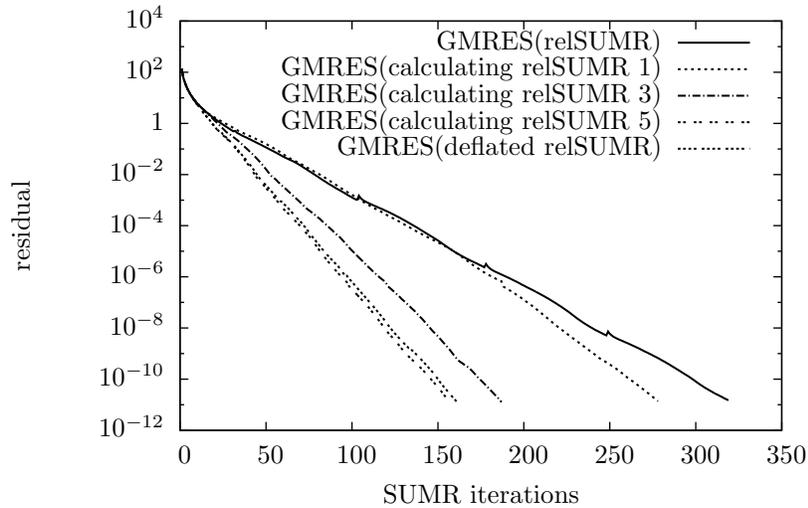
\begin{figure}
 \begin{center}
 \begin{tabular}{c}
  \input{figs/s8t32m0.03_c55_n30GMRESSUMRiterations.tex}
  \end{tabular}
 \end{center}
\caption{The convergence in terms of the number of calls to the matrix sign function of  a) the undeflated GMRESR(SUMR) routine; b) the deflated GMRESR(eigSUMR) routine, with 30 overlap eigenvalues previously calculated; c) the first, third and fifth of the GMRESR(eigSUMR) routines where the eigenvectors where calculated. }\label{fig:3b}
\end{figure}
Figures \ref{fig:3c} and \ref{fig:3d} repeat these results for the CG algorithm.
\begin{figure}
 \begin{center}
 \begin{tabular}{c}
  \input{figs/s8t32m0.03_c55_n30GMRESCGWilsonCalls.tex}
  \end{tabular}
 \end{center}
\caption{The convergence in terms of the number of Wilson operator calls (counting a double precision Wilson operator as two calls) of the a) undeflated GMRESR(CG) routine; b) deflated GMRESR(eigCG) routine, with 30 overlap eigenvalues previously calculated; c) the deflated  GMRESR(eigCG) routine at various stages during the calculation of the eigenvalues. }\label{fig:3c}
\end{figure}
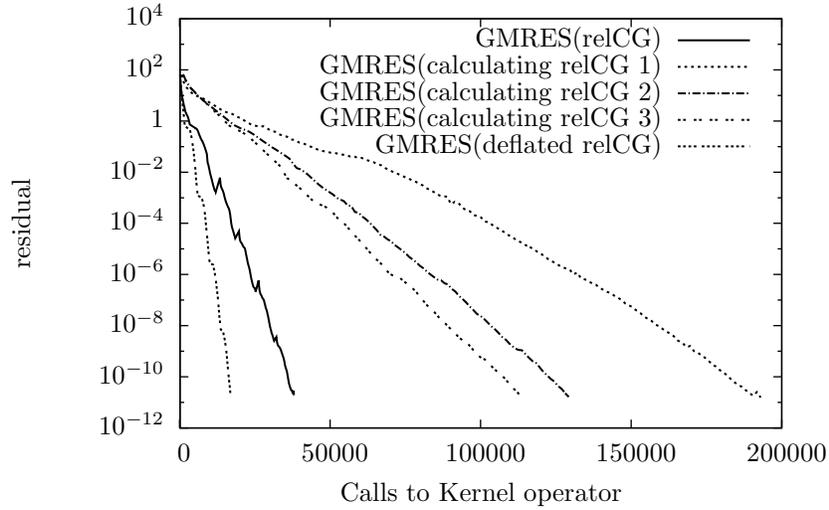
\begin{figure}
 \begin{center}
 \begin{tabular}{c}
  \input{figs/s8t32m0.03_c55_n30GMRESCGiterations.tex}
  \end{tabular}
 \end{center}
\caption{The convergence in terms of the number of calls to the squared Dirac operator (containing two applications of the matrix sign function) of the a) undeflated GMRESR(CG) routine; b) deflated GMRESR(eigCG) routine, with 30 overlap eigenvalues previously calculated; c) the deflated  GMRESR(eigCG) routine during the calculation of the eigenvalues.}\label{fig:3d}
\end{figure}
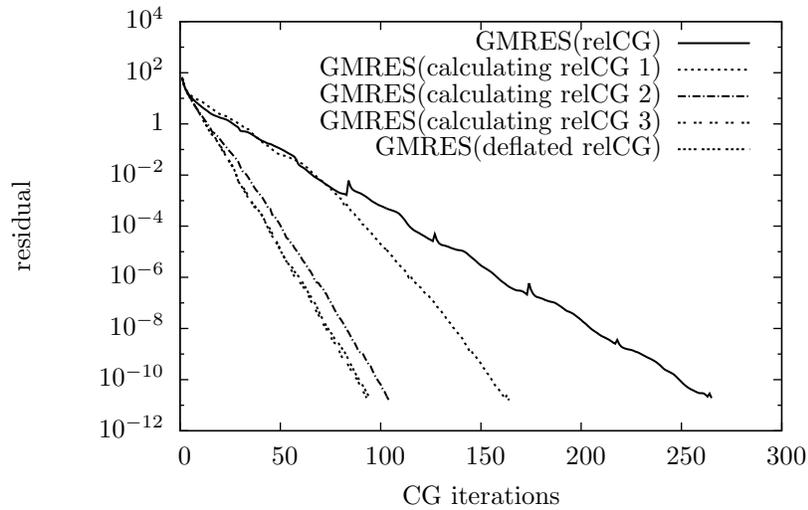
There is some initial cost for the SUMR algorithm during the calculation routines before the eigenvalue computation. This is to improve the accuracy of $v^D$, which we require to a reasonably accuracy while constructing the Galerkin projection otherwise the inversion algorithm will not converge to the correct result. This could have been achieved by keeping the accuracy of $v^D$ higher during the computation of the matrix sign function; however this will slow down the inversion algorithm considerably as more expensive matrix sign functions would be needed. We found that a better balance was achieved by keeping the accuracy of $v^D$ at the level needed to improve the eigenvectors, and have a small additional cost at the start of each inversion.

On the configuration we used to generate these plots, the lowest eigenvectors of $\gamma_5 D$ had eigenvalues of $\pm 0.114$, while the twenty ninth and thirtieth eigenvalues were $\pm 0.302$. We therefore expect an approximately $0.302/0.114 =  2.65$ gain in the iteration number for the inversion, if the deflation is successful. In practice, for the CG algorithm we see a gain of $265/98 = 2.70$ in the iteration count (which is close enough to the optimum value), while SUMR gives a gain of $329/165 = 1.99$ (which is still far from the expected efficiency). The eigenvector residuals after five applications of the eigCG algorithm and used to deflate the CG inversions ranged from $0.01$ to $0.2$, while the eigenvectors used in the deflated SUMR algorithm (also after five eigSUMR routines to calculate the eigenvalues) had residuals ranging from $0.0003$ to $0.01$. This tells us that firstly the eigSUMR algorithm is far more efficient at calculating the eigenvectors (which we expected, given the wrong Krylov subspace is used in the CG algorithm), and secondly that the deflation algorithm used in the eigCG routine works to its theoretical maximum even when the eigenvectors are exceptionally poor, while the projection/GMRES deflation used in SUMR did not work at full efficiently even with more accurate eigenvectors. We would require the eigenvectors to a greater accuracy, i.e. more inversions calculating the eigenvectors, to achieve the optimum improvement. Additionally, the full improvement of deflation is visible for CG after just the second eigenvalue calculation, while for SUMR the required iteration count gradually improves as the eigenvalues are computed. Nonetheless, despite the deflation method used for the CG algorithm's greater efficiency, we still see that eigSUMR is more efficient than eigCG.

We do not see quite the same gain when we consider the number of calls to the Wilson operator (in terms of the number of calls to the Kernel operator, the gain for CG was only a factor of 2.2). This is mostly due to the higher accuracy applications to the matrix sign function in the outer GMRES inversion. The deflation only affects the middle pre-conditioner, while the GMRES part of the algorithm remains untouched -- a fixed cost regardless of whether we deflate. The deflation reduces the calls to the Wilson operator in the pre-conditioner at a similar rate to the reduction of the iteration count. The remaining inefficiency when converting iteration count into calls of the Wilson operator is due to the relaxation.

Finally, we can see that the cost of calculating the eigenvectors is substantial, particularly for the SUMR routine where more calculating inversions are required before the gain from the deflation starts to become significant. The cost is usually largest for the first inversions, and even for the more mature calculations around a factor of 2-4 for CG and 4-5 for SUMR (i.e. one calculating inversion is twice to five times more expensive than a undeflated inversion). This is solely because we cannot relax the accuracy of the matrix sign function efficiently while calculating the eigenvectors. However, in a real-life simulation, only three or four of these steps are required (at least on the small lattices used in this study), while we will be able to use the deflated algorithm far more times. The gain from deflation will far outweigh the cost of the calculation of the eigenvectors.

We also note that the deflation will become more effective as the quark mass is reduced, and on configurations with a non-zero topological charge (and thus exact zero eigenvectors of the massless operator).
\subsection{Eigenvector calculation}
We now consider the efficiency of the eigenvalue calculation, and whether our proposed relaxation schema is successful. For this calculation, we used a standalone eigenvalue solver, starting from a $Z_2$ source, and continuing the eigenvalue calculation after the inversion had converged. We restarted the calculation after every twenty four diagonalisations of the lanczos vectors (i.e. when we were due to recalculate $v^D$), or when the eigenvectors reached the required precision for that restart. At each restart, we performed an additional  diagonalisation of the eigenvectors, using the computed eigenvectors and those from the previous two restarts.  We used the final $q$ vector from the previous run to restart the simulation, after re-orthogonalising it against the already calculated eigenvectors. This allowed us to further adjust the desired accuracy of the eigenvectors as the calculation progressed, so we could always set the tolerance for the eigenvector run to be suitably lower than the expected residual at the next restart. We also restarted the eigenvector calculation when $\norm{q} - 1$  grew larger than $0.1$ (which indicates a breakdown in the algorithm due to the imprecision of the matrix sign function. If $\norm{q}$ grows too much then the algorithm can become unstable). This therefore does not fully replicate the situation in an eigSUMR inverter; however the rate of convergence is similar whether we use eigSUMR or this stand alone eigenvalue solver. We compare the residual for the eigenvector against the number of calls to the Kernel operator and the number of calls to the matrix sign function for both when we relax the accuracy of the solver according to the prescription described in section \ref{sec:5.1} and when we use a full accuracy matrix sign function for the whole calculation. In addition to the thirty eigenvectors, we used an additional 80 Arnoldi vectors when updating the eigenvectors. In an eigSUMR or eigCG calculation, this number would be restricted by the number of iterations needed for the inversion to converge.

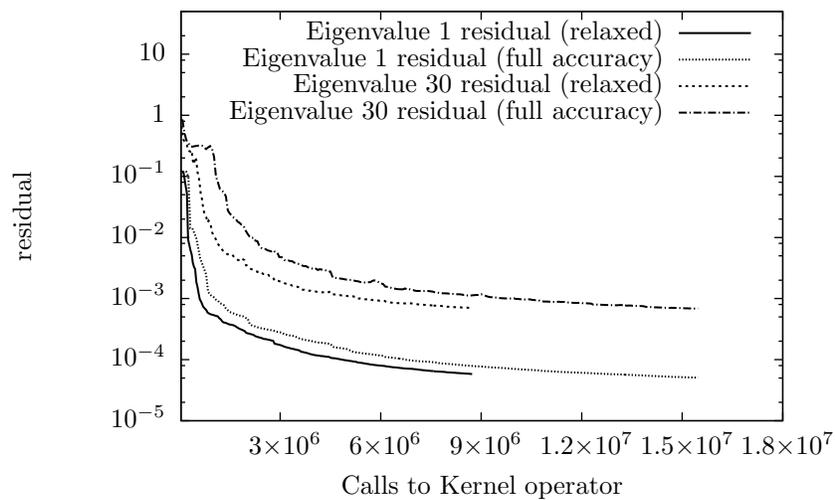
\begin{figure}
 \begin{center}
 \begin{tabular}{c}
  \input{figs/s8t32m0.03_c55_n30UnitaryResidual_0_29.tex}
  \end{tabular}
 \end{center}
\caption{The convergence of the first and 30th eigenvectors in terms of the number of calls to the kernel operator, comparing the relaxed and full accuracy eigenvalue solvers.}\label{fig:4}
\end{figure}

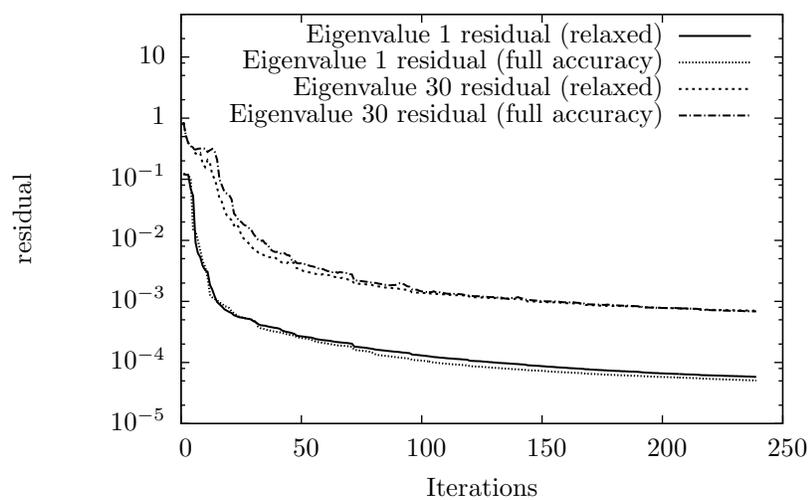
\begin{figure}
 \begin{center}
 \begin{tabular}{c}
  \input{figs/s8t32m0.03_c55_n30UnitaryResidualIter_0_29.tex}
  \end{tabular}
 \end{center}
\caption{The convergence of the first and 30th eigenvectors in terms of the number of calls to the matrix sign function, comparing the relaxed and full accuracy eigenvalue solvers.}\label{fig:4b}
\end{figure}

Figure \ref{fig:4} shows the residual of the eigenvectors plotted against the number of calls to the kernel operator for the first and last eigenvalues calculated, while figure \ref{fig:4b} shows the residuals plotted against the number of recalculations of the eigenvectors. We see that, at least for the lowest eigenvector, the convergence is similar compared to the number of calls to the recalculation routine for both the relaxed and full accuracy eigenvalue solver. This means that our relaxation has not added to the number of iterations required to solve for the eigenvectors: the error in the eigenvectors is under control. When comparing the residue against the number of calls to the Kernel operator, we see a significant gain for the relaxed algorithm at the early stages of the calculation.

However, while our eigenvalue routine works well to obtain the eigenvalues and eigenvectors to a low accuracy, it slows down considerably beyond a certain point, which depends on the number of eigenvectors calculated. If high accuracy eigenvectors are required in addition to the inversion, it may be advantageous to use eigSUMR to calculate the eigenvectors to a low precision, and then some other method such as inverse iteration or the Jacobi-Davidson algorithm to polish the eigenvectors to the required accuracy. The Jacobi-Davidson algorithm~\cite{Jacobi-Davidson}, a more robust, faster and efficient generalisation of inverse iteration, generally converges well once the eigenvalues are known to a moderate accuracy, but less well if the eigenvalues are not known, although the cost of Jacobi-Davidson is also proportional to the number of eigenvectors required. The expensive part of the Jacobi-Davidson algorithm is a low accuracy inversion of the operator, which we already know how to do efficiently for overlap fermions. This is the opposite to what we have found for the Arnoldi process used in the eigSUMR routine. If high accuracy eigenvectors are required, we therefore suggest using eigSUMR to obtain an initial estimate of the eigenvectors and eigenvalues which can be used both as an input for the Jacobi Davividson routine and to construct a spectral preconditioner to accelerate it.

\begin{figure}
 \begin{center}
 \begin{tabular}{c}
  \input{figs/s8t32m0.03_c55_n30UnitaryAccuracy_29.tex}
  \end{tabular}
 \end{center}
\caption{The accuracy of $v^D$ when the accuracy of the matrix sign function was tuned by equation \ref{eq:bound}, compared to the target accuracy given in equation \ref{eq:targetdelta}.}\label{fig:6}
\end{figure}
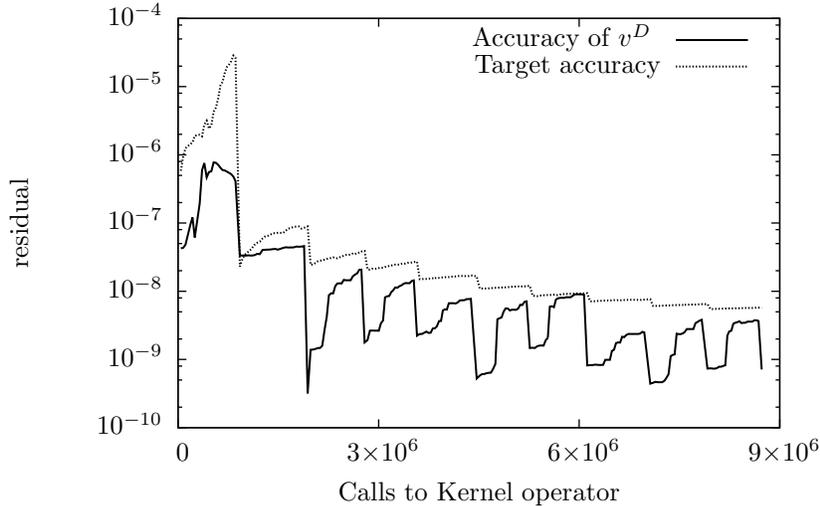

We show the error $\norm{v^D - D v}$ plotted for the thirtieth overlap eigenvector in figure \ref{fig:6}. We see that the accuracy of $v^D$ remains slightly below the target accuracy, which suggests that our relaxation strategy works reasonably well. We calculate $v^D$ accurately after every twenty four diagonalisations of the Lanczos vectors, which causes the troughs seen in the plot.
\subsection{Varying the number of eigenvectors}
Finally, we consider the effect of varying the number of calculated eigenvectors, both on the convergence of the inversion and on the convergence of the eigenvector calculation. We consider 15, 30 and 45 eigenvectors. We expect that the inversions will will improve as we deflate more eigenvectors, while it is less clear how the eigenvector calculation will fare. Figure \ref{fig:7} shows the convergence of the deflated CG inversion as different numbers of eigenvectors are calculated, and Figure \ref{fig:9} shows the same thing for SUMR. Once again, the CG algorithm rapidly, after only a few iterations to calculate the eigenvectors, reaches the optimum convergence rate, while SUMR requires more than the five inversions we used to reach it. The ratio of the SUMR against CG improvement factor becomes worse as we deflate more eigenvalues, which means that the more eigenvalues we calculate, the more inversions are needed to have the eigenvalues calculated to a high enough accuracy for the deflation to take effect. For CG, this is not the case, and no matter how many eigenvalues we calculate (at least up to the numbers we study here) only two or three inversions are required to calculate the eigenvectors to the required accuracy.

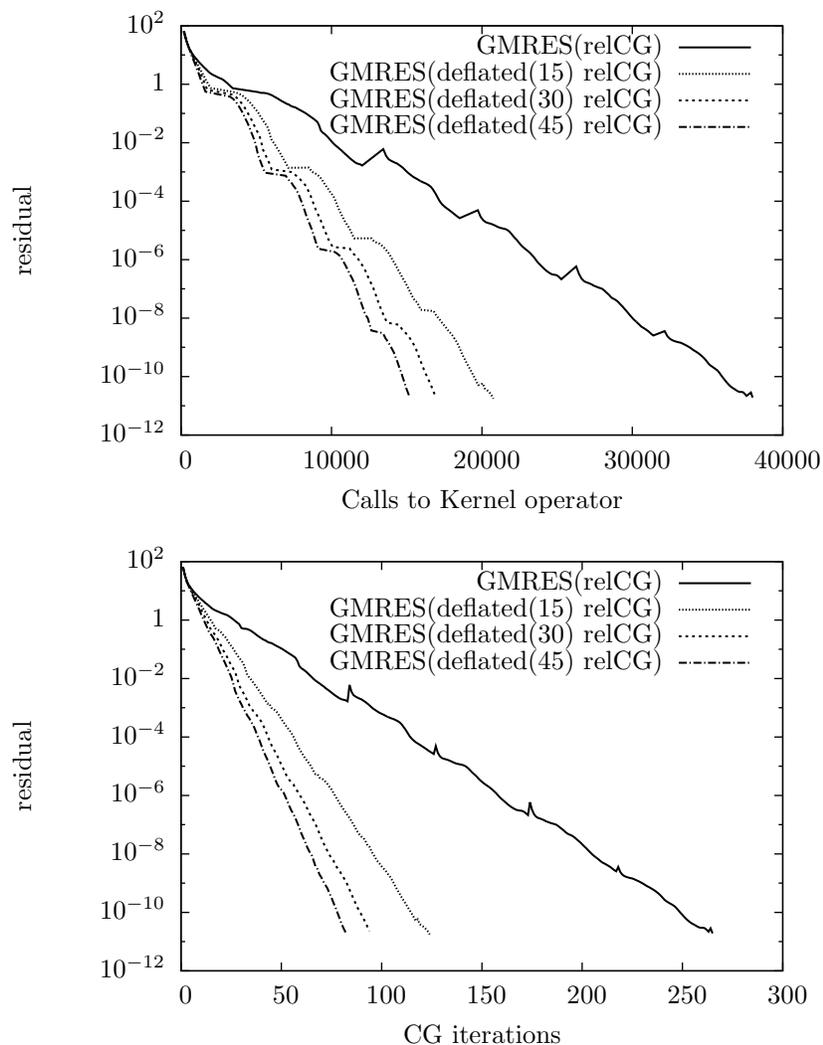
\begin{figure}
 \begin{center}
 \begin{tabular}{c}
  \input{figs/s8t32m0.03_c55_n30GMRESCGComparison1.tex}\\
  \input{figs/s8t32m0.03_c55_n30GMRESCGiterationsComparison1.tex}
  \end{tabular}
 \end{center}
\caption{The convergence of the CG inverter after deflating 0, 15, 30 and 45 eigenvectors. The maximum possible improvement over the original routine is given by the ratio of the lowest eigenvalue of $\gamma_5 D$ not included in the deflation to the lowest eigenvalue of the operator. The ratio of the fifteenth to the first eigenvalue of $\gamma_5 D$ is $2.1$, of the thirtieth to the first eigenvalue $2.6$ and of the 45th to the first eigenvalue is $3.1$. The improvement in the number of iterations after deflation is a factor of $2.1$ for 15 eigenvalues, $2.7$ for 30 eigenvalues, and $3.1$ for 45 eigenvalues.}\label{fig:7}
\end{figure}
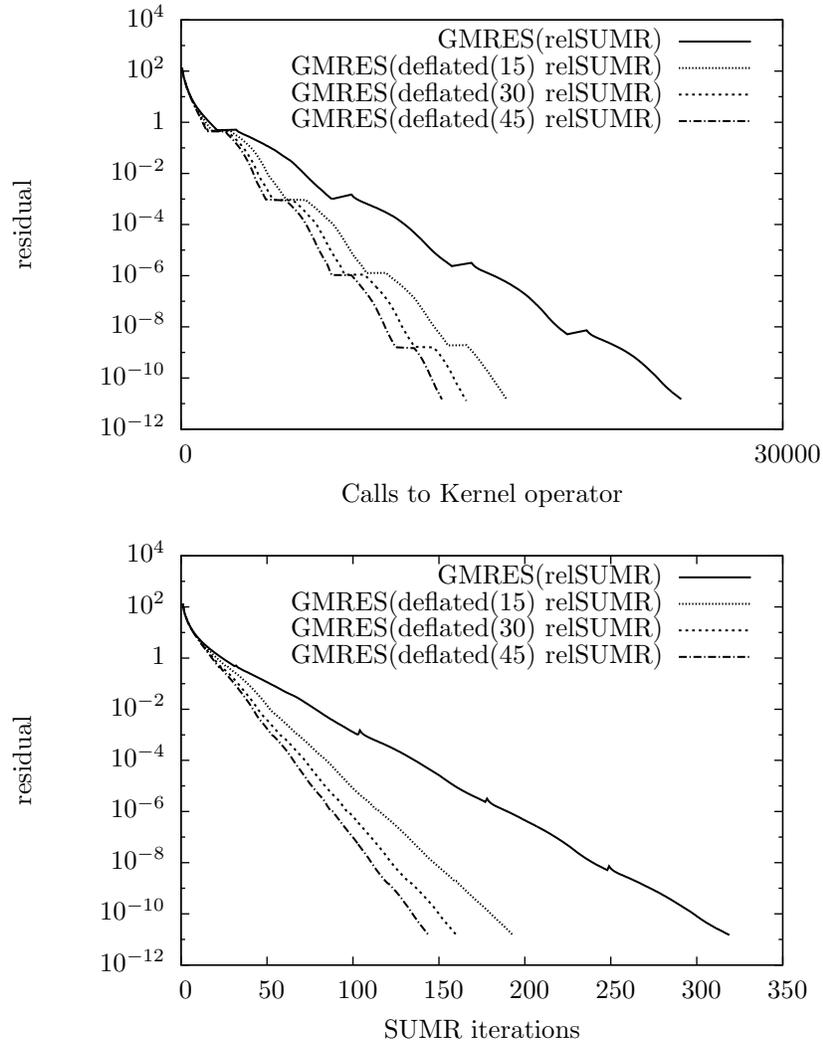
\begin{figure}
 \begin{center}
 \begin{tabular}{c}
  \input{figs/s8t32m0.03_c55_n30GMRESSUMRComparison1.tex}\\
  \input{figs/s8t32m0.03_c55_n30GMRESSUMRiterationsComparison1.tex}
  \end{tabular}
 \end{center}
\caption{The convergence of the SUMR inverter after deflating 0, 15, 30 and 45 eigenvectors. The improvement in the number of iterations for the deflated routines is $1.6$ for 15 eigenvalues, $2.0$ for 30 eigenvalues, and $2.2$ for 45 eigenvalues.}\label{fig:9}
\end{figure}

\begin{figure}
 \begin{center}
 \begin{tabular}{c}
  \input{figs/s8t32m0.03_c55_n30GMRESSUMRComparison3.tex}\\
  \input{figs/s8t32m0.03_c55_n30GMRESSUMRiterationsComparison3.tex}
  \end{tabular}
 \end{center}
\caption{The convergence of the SUMR inverter while calculating 0, 15, 30 and 45 eigenvectors.}\label{fig:10a}
\end{figure}
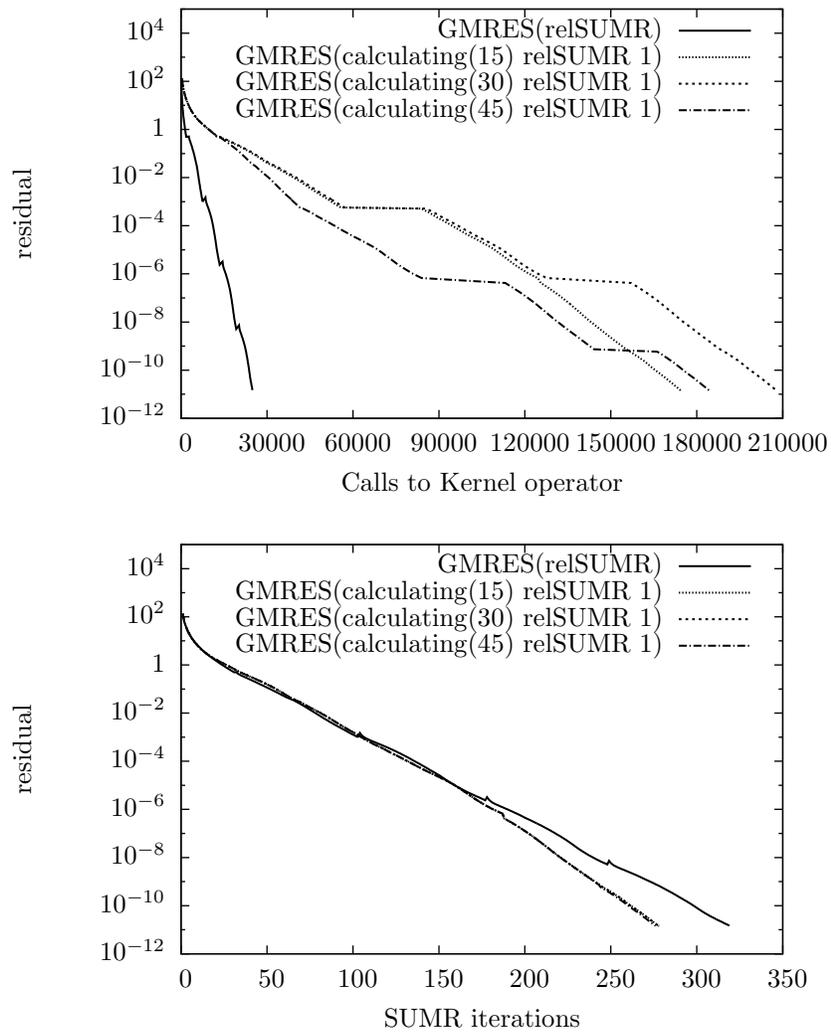

Of course, the number of inversions required for the full improvement factor is one part of the cost of calculating the eigenvalues; the second is whether the cost changes per inversion for the CG and SUMR inversions used to calculate the eigenvectors. We display this in figures \ref{fig:8} and \ref{fig:10}, for the third calculating CG or calculating SUMR inversion, and in figures \ref{fig:8a} and \ref{fig:10a} for the initial calculating CG or calculating SUMR inversion. It can be seen that the additional cost per inversion required to invert extra eigenvalues is negligible.

\begin{figure}
 \begin{center}
 \begin{tabular}{c}
  \input{figs/s8t32m0.03_c55_n30GMRESCGComparison2.tex}\\
  \input{figs/s8t32m0.03_c55_n30GMRESCGiterationsComparison2.tex}
  \end{tabular}
 \end{center}
\caption{The convergence of the CG inverter while calculating 0, 15, 30 and 45 eigenvectors.}\label{fig:8}
\end{figure}
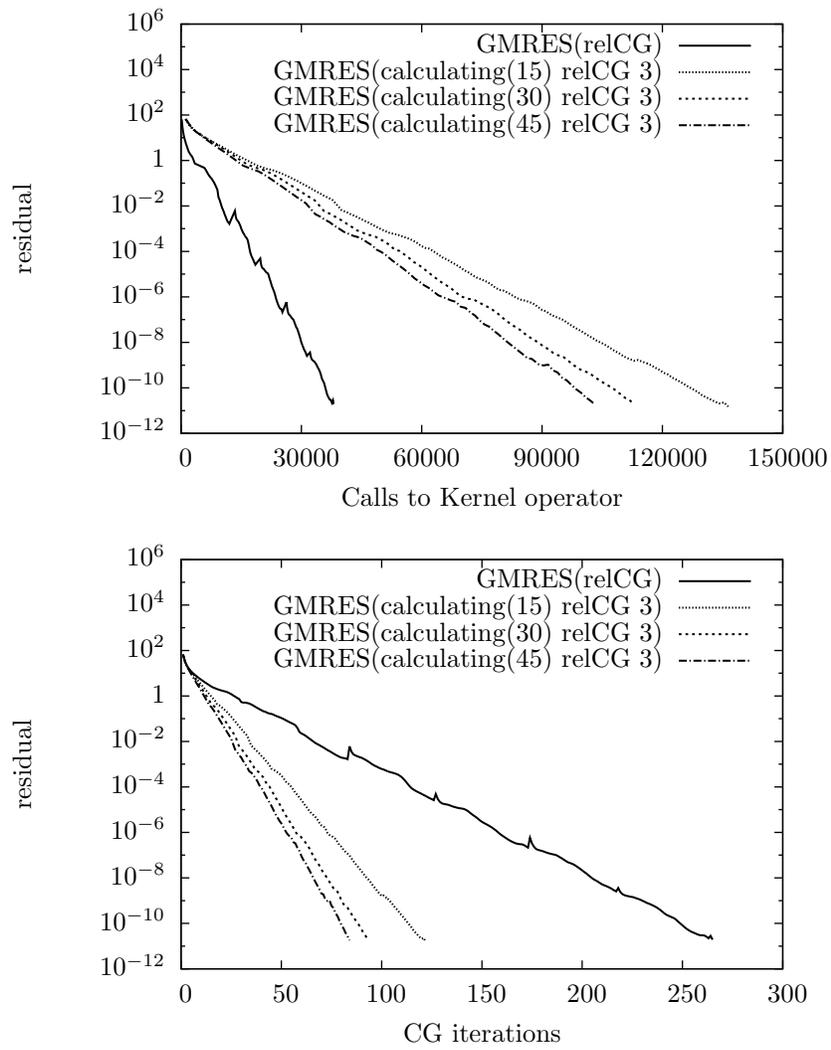

\begin{figure}
 \begin{center}
 \begin{tabular}{c}
  \input{figs/s8t32m0.03_c55_n30GMRESSUMRComparison2.tex}\\
  \input{figs/s8t32m0.03_c55_n30GMRESSUMRiterationsComparison2.tex}
  \end{tabular}
 \end{center}
\caption{The convergence of the SUMR inverter while calculating 0, 15, 30 and 45 eigenvectors.}\label{fig:10}
\end{figure}
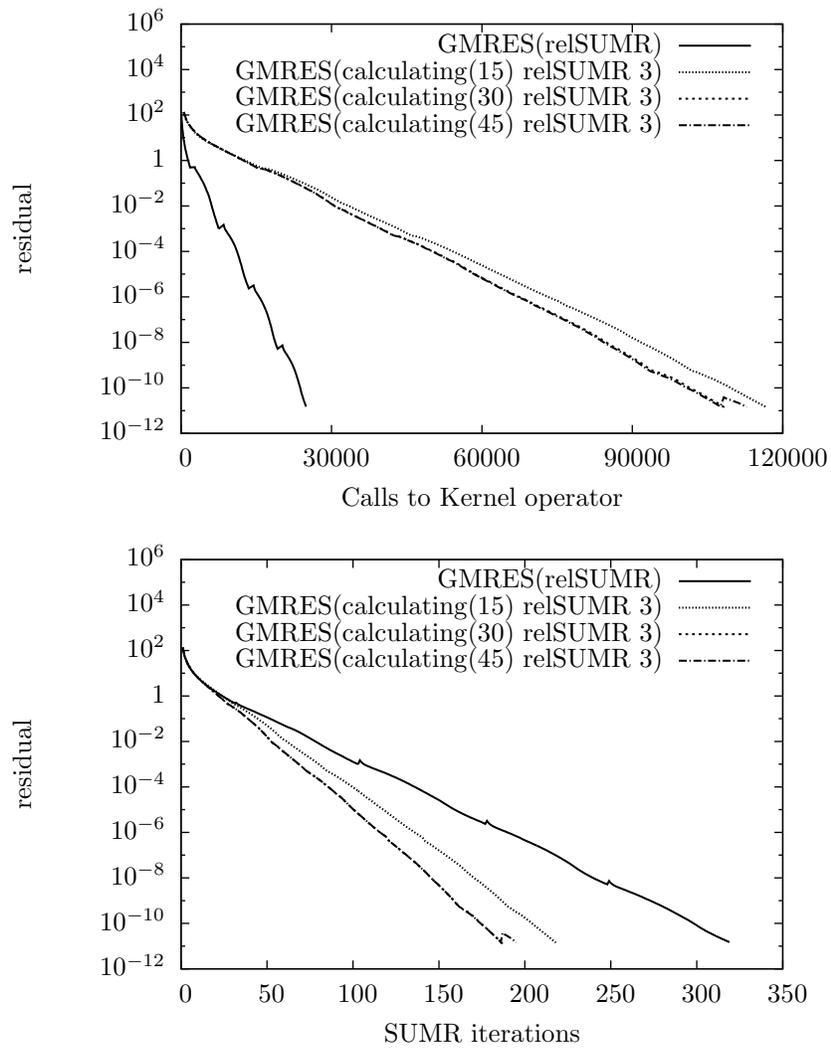
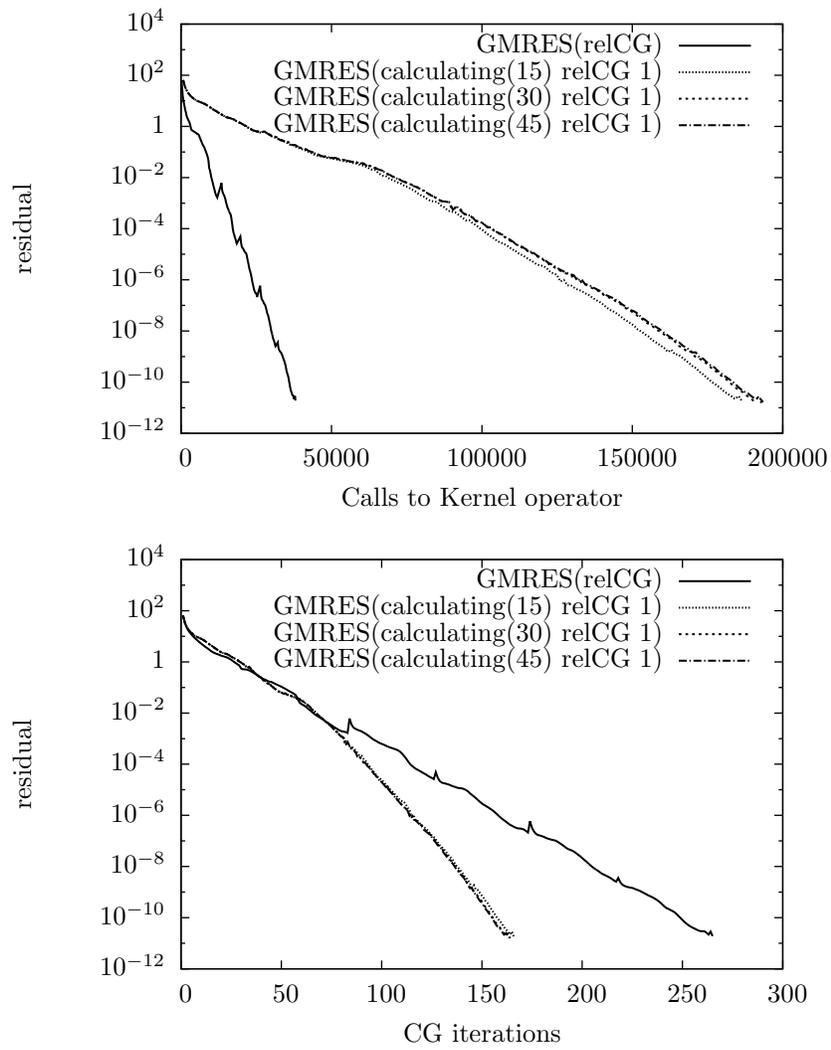
\begin{figure}
 \begin{center}
 \begin{tabular}{c}
  \input{figs/s8t32m0.03_c55_n30GMRESCGComparison3.tex}\\
  \input{figs/s8t32m0.03_c55_n30GMRESCGiterationsComparison3.tex}
  \end{tabular}
 \end{center}
\caption{The convergence of the CG inverter while calculating 0, 15, 30 and 45 eigenvectors.}\label{fig:8a}
\end{figure}

We conclude that in terms of computer time, the cost for the eigenvalue calculation for the CG algorithm is roughly independent of the number of eigenvalues calculated, at least when the size of the Krylov subspace used in each of the CG inversions is smaller than the number of eigenvalues needed. For SUMR, the computational cost to get the maximum gain for the deflation will increase as the number of eigenvalues increases. This difference is due to the improved preconditioning method used for the CG algorithm, which does not need the eigenvectors to be calculated to any significant accuracy to be effective; while the Galerkin projection method requires that the accuracy of the eigenvectors is comparable to the accuracy needed for the SUMR preconditioner in the GMRESR algorithm. With the number of eigenvalues we tested, deflated SUMR is still more efficient than deflated CG. That might change if more eigenvectors are included, or at lower masses, or fewer right hand sides are needed. Of course, for both CG and SUMR the required memory increases as the number of eigenvectors increases, and this may well be a limiting factor on some machines.

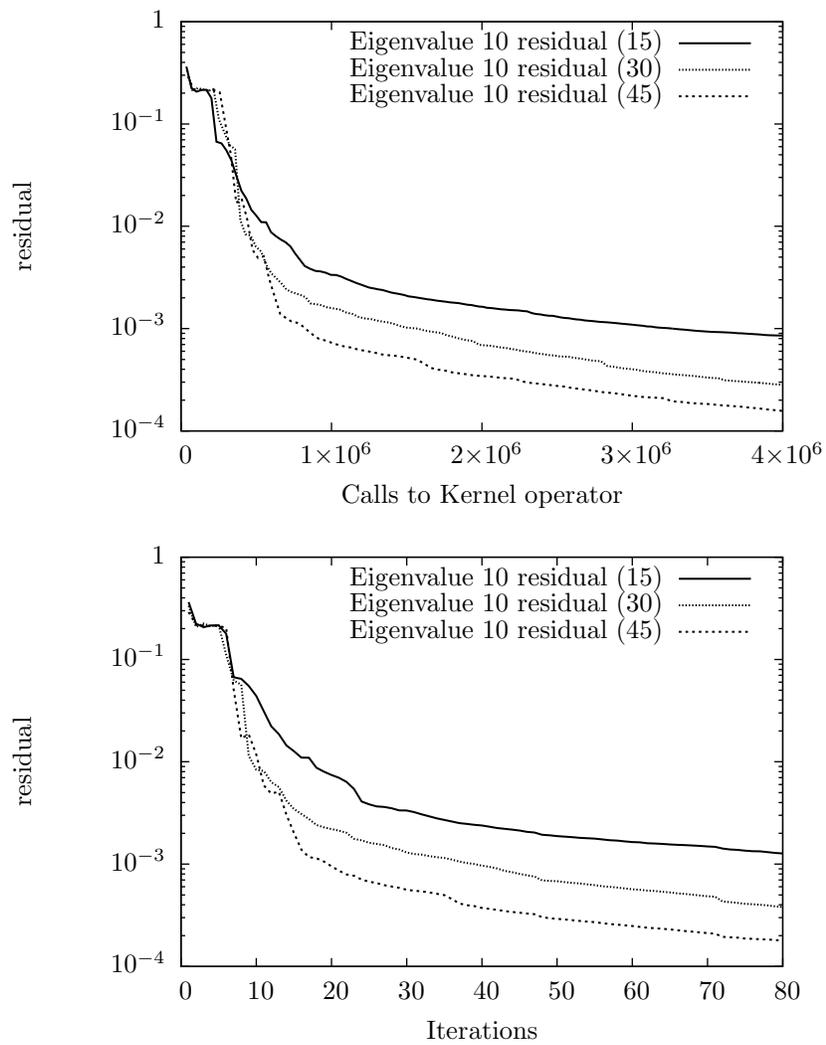
\begin{figure}
 \begin{center}
 \begin{tabular}{c}
  \input{figs/s8t32m0.03_c55_n30UnitaryResidual_10Comparison.tex}\\
    \input{figs/s8t32m0.03_c55_n30UnitaryResidual_10ComparisonIterations.tex}
  \end{tabular}
 \end{center}
\caption{The convergence of the tenth eigenvector while calculating 15, 30 and 45 eigenvectors with the relaxed eigenvalue solver.}\label{fig:11}
\end{figure}

\begin{figure}
 \begin{center}
 \begin{tabular}{c}
  \input{figs/s8t32m0.03_c55_n30UnitaryResidual_LastComparison.tex}\\
    \input{figs/s8t32m0.03_c55_n30UnitaryResidual_LastComparisonIterations.tex}
  \end{tabular}
 \end{center}
\caption{The convergence of the last calculated eigenvector while calculating  15, 30 and 45 eigenvectors with the relaxed eigenvalue solver.}\label{fig:12}
\end{figure}
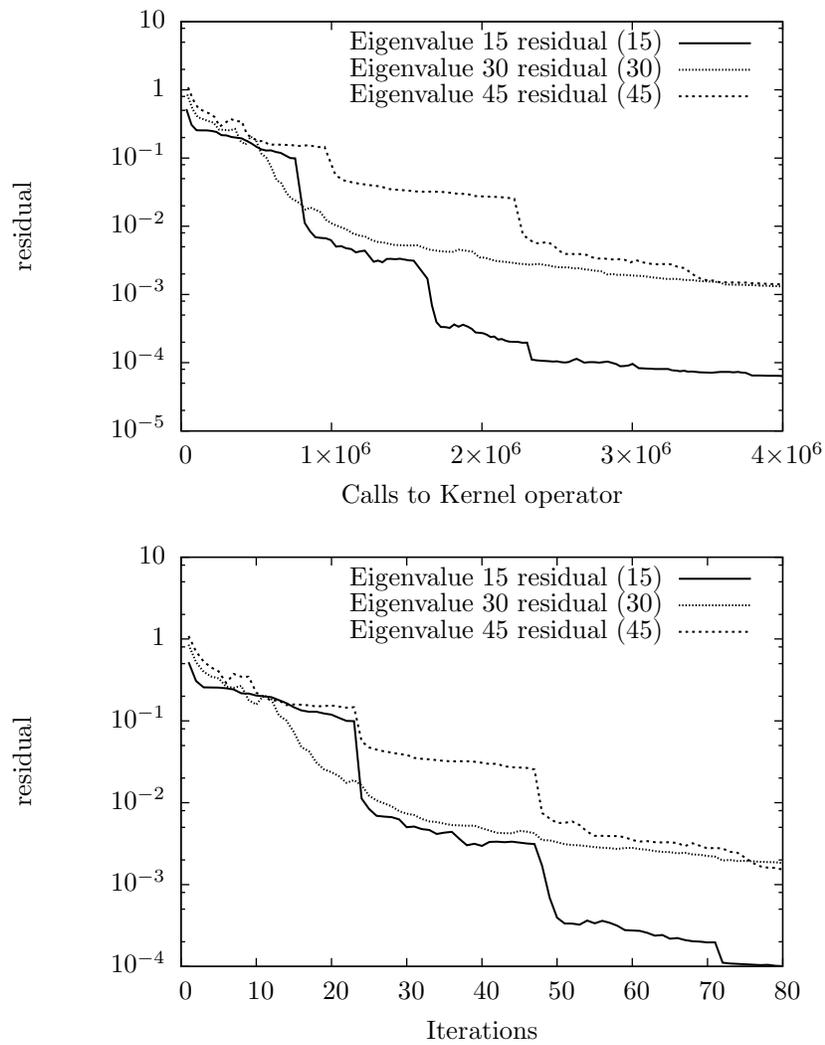

In figures \ref{fig:11} and \ref{fig:12} we consider the convergence of the eigenvector calculation as more eigenvectors are calculated. We consider the tenth eigenvector, which is calculated in all our runs, and the last eigenvector calculated in each run. It can be seen that the convergence of the tenth eigenvector improves as we increase the number of computed eigenvectors, while the final eigenvector generally converges slower the more eigenvectors are calculated. The sharp drops in the eigenvector residual in figure \ref{fig:12} occur when we reset the computation. They are partially caused by our diagonalisation of the eigenvectors against the eigenvalues computed from the previous iterations, which seems to particularly benefit the last few eigenvectors. If the eigenvectors are required in themselves, and not just for the deflation, then it may be advantageous to calculate more eigenvectors than necessary so that the convergence of the wanted eigenvectors improves.

\section{Conclusions}\label{sec:7}
We have tested the application of deflation methods to overlap fermions; focussing on the eigCG and eigSUMR algorithms. We construct a rigorous (though conservative) bound for the accuracy of the matrix sign function, and suggest a less rigorous and less conservative bound which seems to work well in practice. We see no loss in accuracy for the eigenvectors during the calculation.

Our deflated inversion algorithms show the usual level of improvement. In particular, the spectral decomposition method we have used for the CG algorithm accelerates the inversions according to the theoretical maximum even for exceptionally low quality approximations of the eigenvectors. This pre-conditioner is impractical for other fermion formulations because of its cost, but for overlap fermions the cost is negligible. A method such as this should be preferred over the Galerkin projection for overlap inversions of $D^\dagger D$. For SUMR, we have to use a variation of the Galerkin projection, which performs less well until the eigenvectors are calculated to a high enough accuracy.

As an eigenvector routine, our proposed eigCG algorithm performs particularly poorly, which is probably because the pre-conditioned inverter generates a different Krylov subspace than the optimal one for the calculation of the eigenvectors. However, it was still able to calculate the eigenvectors to a good enough accuracy to achieve the maximum possible acceleration from the deflation after only one or two right hand sides. The eigSUMR algorithm works well up to a certain accuracy, after which it shows a significant slowing down. To obtain high accuracy eigenvectors, it is beneficial to combine eigSUMR with some other routine such as inverse iteration or the Jacobi-Davidson method.

There is a considerable cost in using eigSUMR or eigCG to calculate the eigenvectors over a straight inversion algorithm, because we are unable to relax the accuracy of the matrix sign function as aggressively as is possible in an inversion. This set-up cost will, however, be negligible compared to the gain if enough right hand sides are required.

Because the deflation for the CG algorithm is more efficient than for SUMR, it remains an open question about which will be better in practical applications (if we do not also require high accuracy eigenvectors as well as the inverse -- we expect that if the eigenvectors are calculated accurately enough then the Galerkin projection will work as well as the spectral preconditioner). Which of these methods is superior may vary from simulation to simulation.

Obviously, deflation methods become less efficient on larger lattice volumes, since the increased density of small eigenvalues requires a larger number of eigenvectors to be calculated, leading to greater costs for the computational time and memory. However, given that our spectral pre-conditioner for the CG algorithm works with the maximum efficiency with even very low accuracy eigenvectors, and there does not seem to be an additional overhead for calculating more eigenvalues, we do not expect that there should be a significant increase in the computational cost at larger volumes (although, obviously, this needs to be confirmed). The memory requirement is likely to be more of a bottleneck, with the memory costs for the same degree of improvement likely to increase as O($V^2$), where $V$ is the lattice volume, since the number of eigenvectors required for the same improvement is likely to increase with the volume. This should not be a problem on the relatively small (compared to other fermion actions) lattices which are currently accessible to overlap fermions, but it is something which will need to be addressed in the future.

\section*{Acknowledgements}
\input{Acknowledgements}
Numerical simulations were carried out on servers and Desktop computers at Seoul National University.

\appendix
\section{Construction of the SUMR algorithm}\label{app:A}
The Arnoldi equation is $ U Q = Q B$, where $B$ is an $n\times n$ matrix, and $Q$ is an $N\times n$ matrix containing the $n$ orthonormal Arnoldi vectors $q_i$ in its columns. This gives $Q^{\dagger} U^{\dagger} U Q = B^{\dagger} Q^{\dagger} Q B$, Or $B^{\dagger} B = 1$, i.e $B$ is also unitary. This means that $B$ can be decomposed in terms of Given's matrices and a $U(1)^n$ factor, which can be absorbed into the phase of the $q$ vectors. Given that $B$ is also upper Hessenburg (like all Arnoldi matrices), the only Given's matrices which can contribute to it are those with only diagonal and sub-diagonal terms. We can therefore decompose,
\begin{gather}
B_n = G_0(\gamma_0) G_1(\gamma_1) \ldots G_{n-1} (\gamma_{n-1}) \tilde{G}_n (\tilde{\gamma}_n),\label{eq:sumr0}
\end{gather}
where
\begin{multline}
G_m(\gamma_m) =\\ \left( \begin{array}{c c c c}
1_{(m-1)\times (m-1)} &0_{ (m-1)\times 1}&0_{ (m-1)\times 1}&0_{(n-m-1)\times(m-1)}\\
0_{1\times (m-1)}&- \gamma_m&\sigma_m&0_{(n-m-1)\times 1}\\
0_{1\times (m-1)}&\sigma_m&\gamma_m^\dagger &0_{(n-m-1)\times 1}\\
0_{(n-m-1)\times (m-1)}&0_{ (n-m-1)\times 1}&0_{ (n-m-1)\times 1}&1_{(n-m-1)\times(n-m-1)}
\end{array} \right),
\end{multline}
where the $\dagger$ indicates the complex conjugate and $\sigma_m = \sqrt{1-|\gamma_m|^2}$, so that the $G_m$ matrices are unitary.
The final matrix is
\begin{gather}
\tilde{G}_n (\tilde{\gamma}_n) = \diag(1,1,\ldots,\tilde{\gamma_n}).
\end{gather}
By comparing the $m$th row, we can see that
\begin{gather}
U Q_m = Q_{m+1} \hat{B}_m, \label{eq:sumr2}
\end{gather}
where $\hat{B}_m$ is the $(m+1)\times (m)$ matrix defined in terms of the $(m+1) \times(m+1)$ Given's matrices $G_i(\gamma_i)$ and the $(m+1) \times m$ matrix $\hat{G}$,
\begin{align}
\hat{B}_m =& G_0(\gamma_0) G_1(\gamma_1) \ldots \hat{G}_m(\gamma_m)\nonumber\\
\hat{G}_m(\gamma_m)= & \left(\begin{array}{c c c}
1_{(m-1)\times (m-1)}& 0&0\\
0&1&-\gamma_m\\
0&0&\sigma_m
\end{array} \right).
\end{align}
We also know that
\begin{gather}
B = \left( \begin{array} {c c c c}
(q_0,U q_0)&(q_0,U q_1)&(q_0,U q_2)&(q_0,U q_3)\\
(q_1,U q_0)&(q_1,U q_1)&(q_1,U q_2)&(q_1,U q_3)\\
0&(q_2,Uq_1)&(q_2,U q_2)&(q_2,U q_3)\\
0&0&(q_3,U q_2)&(q_3,U q_3)
\end{array} \right),\label{eq:sumr1}
\end{gather}
where $(,)$ indicates the scalar product of two vectors.

Our intention is to build up the Arnoldi vectors $q_i$ via a short recurrence. We will first illustrate the process using an example when $n = 4$, and later generalise to arbitrary $n$. Equating equations (\ref{eq:sumr0}) and (\ref{eq:sumr1}) gives
\begin{align}
B = & \left(\begin{array}{c c c c}
-\gamma_0&\sigma_0&0&0\\
\sigma_0&\gamma_0^\dagger&0&0\\
0&0&1&0\\
0&0&0&1
\end{array}\right)\left(\begin{array}{c c c c}
1&0&0&0\\
0&-\gamma_1&\sigma_1&0\\
0&\sigma_1&\gamma_1^\dagger&0\\
0&0&0&1
\end{array}\right)\nonumber\\
&\phantom{spacespace}
\left(\begin{array}{c c c c}
1&0&0&0\\
0&1&0&0\\
0&0&-\gamma_2&\sigma_2\\
0&0&\sigma_2&\gamma_2^\dagger
\end{array}\right)\left(\begin{array}{c c c c}
1&0&0&0\\
0&1&0&0\\
0&0&1&0\\
0&0&0&-\tilde{\gamma}_3
\end{array}\right)\nonumber\\
=&\left(\begin{array}{c c c c}
-\gamma_0&\sigma_0&0&0\\
\sigma_0&\gamma_0^\dagger&0&0\\
0&0&1&0\\
0&0&0&1
\end{array}\right) \left(\begin{array}{c c c c}
1&0&0&0\\
0&-\gamma_1& -\sigma_1\gamma_2& -\sigma_1\sigma_2\tilde{\gamma}^3\\
0&\sigma_1&-\gamma_1^\dagger\gamma_2&-\sigma_2\gamma_1^\dagger \tilde{\gamma}_3\\
0&0&\sigma_2&-\tilde{\gamma}_3\gamma_2^\dagger
\end{array}\right)\nonumber\\
=&\left( \begin{array} {c c c c }
(q_0,U q_0)&(q_0,U q_1)&(q_0,U q_2)&(q_0,U q_3)\\
(q_1,U q_0)&(q_1,U q_1)&(q_1,U q_2)&(q_1,U q_3)\\
0&(q_2,Uq_1)&(q_2,U q_2)&(q_2,U q_3)\\
0&0&(q_3,U q_2)&(q_3,U q_3)
\end{array} \right).
\end{align}
A quick manipulation gives us
\begin{align}
&\left(\begin{array}{c c c c}
1&0&0&0\\
0&-\gamma_1& -\sigma_1\gamma_2& -\sigma_1\sigma_2\tilde{\gamma}^3\\
0&\sigma_1&-\gamma_1^\dagger\gamma_2&-\sigma_2\gamma_1^\dagger \tilde{\gamma}_3\\
0&0&\sigma_2&-\tilde{\gamma}_3\gamma_2^\dagger
\end{array}\right) =  \left(\begin{array}{c c c c}
-\gamma_0^\dagger&\sigma_0&0&0\\
\sigma_0&\gamma_0&0&0\\
0&0&1&0\\
0&0&0&1
\end{array}\right)\nonumber\\
&\phantom{spacespace}
\left( \begin{array} {c c c c }
(q_0,U q_0)&(q_0,U q_1)&(q_0,U q_2)&(q_0,U q_3)\\
(q_1,U q_0)&(q_1,U q_1)&(q_1,U q_2)&(q_1,U q_3)\\
0&(q_2,Uq_1)&(q_2,U q_2)&(q_2,U q_3)\\
0&0&(q_3,U q_2)&(q_3,U q_3)\end{array}\right)
\nonumber\\
&\phantom{space}=\left( \begin{array} {c c c c }
(\tilde{q}'_0,U q_0)&(\tilde{q}'_0,U q_1)&(\tilde{q}'_0,U q_2)&(\tilde{q}'_0,U q_3)\\
(\tilde{q}_1,U q_0)&(\tilde{q}_1,U q_1)&(\tilde{q}_1,U q_2)&(\tilde{q}_1,U q_3)\\
0&(q_2,Uq_1)&(q_2,U q_2)&(q_2,U q_3)\\
0&0&(q_3,U q_2)&(q_3,U q_3)\end{array}\right),
\end{align}
with
\begin{align}
\tilde{q}_1^{\dagger} =&  \sigma_0 q_0^{\dagger} + \gamma_0 q_1^{\dagger}\\
(\tilde{q}'_0)^{\dagger} = &  - \gamma_0^\dagger q_0^\dagger + \sigma_0 q_1^\dagger.
\end{align}
This gives,
\begin{gather}
\gamma_1 = - (\tilde{q}_1,U q_1).
\end{gather}
The general form of this process is~\cite{Jagels}
\begin{align}
\gamma_n = & - (\tilde{q}_n,U q_n)\nonumber\\
\sigma_n = & \sqrt(1- |\gamma_n|^2)\nonumber\\
\tilde{q}_{n+1} = & \sigma_{n} \tilde{q}_n + \gamma_n^\dagger q_{n+1}
\end{align}
with $\tilde{q}_0 = q_0$.
We can construct the next $q$ vector from equation (\ref{eq:sumr2}).
\begin{gather}
U q_m = (Q_{m+1})G_0(\gamma_0) G_1(\gamma_1) \ldots \hat{G}_m.
\end{gather}
With $Q_{m+1} = (q_0,q_1,q_2,q_3,\ldots)$, applying $Q_{m+1} G_0$ gives
\begin{align}
 Q_{m+1} G_0 =& (-\gamma_0 q_0 + \sigma_0 q_1,\sigma_0 q_0 + \gamma_0^\dagger q_1,q_2,q_3,\ldots)\nonumber\\
 =&(\tilde{q}'_0,\tilde{q}_1,q_2,q_3,q_4,\ldots).
\end{align}
Similarly,
\begin{gather}
 Q_{m+1} G_0 G_1 G_2 \ldots G_{n-1} = ((\tilde{q}'_0,\tilde{q}'_1,\ldots,\tilde{q}'_{n-1},\tilde{q}_n,q_{n+1},q_{n+2},\ldots).
\end{gather}
So we finish up with
\begin{gather}
 U q_m =(\ldots,\tilde{q}'_{m-1},\tilde{q}_m,q_{m+1} )\hat{G}_m.
\end{gather}
We thus get
\begin{gather}
U q_m = \sigma_m q_{m+1} - \gamma_m \tilde{q}_m
\end{gather}
or
\begin{gather}
q_{m+1} = \frac{1}{\sigma_m} (U q_m + \gamma_m \tilde{q}_m).
\end{gather}
This leads to the modified Arnoldi algorithm shown in algorithm \ref{alg:sumrbasis}.

\bibliographystyle{elsarticle-num}
\bibliography{weyl}
\end{document}

%% file: figs/s8t32m0.03_c55_n30SUMRWilsonCalls2.tex
\begingroup
  \makeatletter
  \providecommand\color[2][]{%
    \GenericError{(gnuplot) \space\space\space\@spaces}{%
      Package color not loaded in conjunction with
      terminal option `colourtext'%
    }{See the gnuplot documentation for explanation.%
    }{Either use 'blacktext' in gnuplot or load the package
      color.sty in LaTeX.}%
    \renewcommand\color[2][]{}%
  }%
  \providecommand\includegraphics[2][]{%
    \GenericError{(gnuplot) \space\space\space\@spaces}{%
      Package graphicx or graphics not loaded%
    }{See the gnuplot documentation for explanation.%
    }{The gnuplot epslatex terminal needs graphicx.sty or graphics.sty.}%
    \renewcommand\includegraphics[2][]{}%
  }%
  \providecommand\rotatebox[2]{#2}%
  \@ifundefined{ifGPcolor}{%
    \newif\ifGPcolor
    \GPcolorfalse
  }{}%
  \@ifundefined{ifGPblacktext}{%
    \newif\ifGPblacktext
    \GPblacktexttrue
  }{}%
  \let\gplgaddtomacro\g@addto@macro
  \gdef\gplbacktext{}%
  \gdef\gplfronttext{}%
  \makeatother
  \ifGPblacktext
    \def\colorrgb#1{}%
    \def\colorgray#1{}%
  \else
    \ifGPcolor
      \def\colorrgb#1{\color[rgb]{#1}}%
      \def\colorgray#1{\color[gray]{#1}}%
      \expandafter\def\csname LTw\endcsname{\color{white}}%
      \expandafter\def\csname LTb\endcsname{\color{black}}%
      \expandafter\def\csname LTa\endcsname{\color{black}}%
      \expandafter\def\csname LT0\endcsname{\color[rgb]{1,0,0}}%
      \expandafter\def\csname LT1\endcsname{\color[rgb]{0,1,0}}%
      \expandafter\def\csname LT2\endcsname{\color[rgb]{0,0,1}}%
      \expandafter\def\csname LT3\endcsname{\color[rgb]{1,0,1}}%
      \expandafter\def\csname LT4\endcsname{\color[rgb]{0,1,1}}%
      \expandafter\def\csname LT5\endcsname{\color[rgb]{1,1,0}}%
      \expandafter\def\csname LT6\endcsname{\color[rgb]{0,0,0}}%
      \expandafter\def\csname LT7\endcsname{\color[rgb]{1,0.3,0}}%
      \expandafter\def\csname LT8\endcsname{\color[rgb]{0.5,0.5,0.5}}%
    \else
      \def\colorrgb#1{\color{black}}%
      \def\colorgray#1{\color[gray]{#1}}%
      \expandafter\def\csname LTw\endcsname{\color{white}}%
      \expandafter\def\csname LTb\endcsname{\color{black}}%
      \expandafter\def\csname LTa\endcsname{\color{black}}%
      \expandafter\def\csname LT0\endcsname{\color{black}}%
      \expandafter\def\csname LT1\endcsname{\color{black}}%
      \expandafter\def\csname LT2\endcsname{\color{black}}%
      \expandafter\def\csname LT3\endcsname{\color{black}}%
      \expandafter\def\csname LT4\endcsname{\color{black}}%
      \expandafter\def\csname LT5\endcsname{\color{black}}%
      \expandafter\def\csname LT6\endcsname{\color{black}}%
      \expandafter\def\csname LT7\endcsname{\color{black}}%
      \expandafter\def\csname LT8\endcsname{\color{black}}%
    \fi
  \fi
  \setlength{\unitlength}{0.0500bp}%
  \begin{picture}(6236.00,3968.00)%
    \gplgaddtomacro\gplbacktext{%
      \csname LTb\endcsname%
      \put(1220,640){\makebox(0,0)[r]{\strut{}$10^{-14}$}}%
      \put(1220,983){\makebox(0,0)[r]{\strut{}$10^{-12}$}}%
      \put(1220,1326){\makebox(0,0)[r]{\strut{}$10^{-10}$}}%
      \put(1220,1669){\makebox(0,0)[r]{\strut{}$10^{-8}$}}%
      \put(1220,2012){\makebox(0,0)[r]{\strut{}$10^{-6}$}}%
      \put(1220,2355){\makebox(0,0)[r]{\strut{}$10^{-4}$}}%
      \put(1220,2698){\makebox(0,0)[r]{\strut{}$10^{-2}$}}%
      \put(1220,3041){\makebox(0,0)[r]{\strut{}$1$}}%
      \put(1220,3384){\makebox(0,0)[r]{\strut{}$10^{2}$}}%
      \put(1220,3727){\makebox(0,0)[r]{\strut{}$10^{4}$}}%
      \put(2837,440){\makebox(0,0){\strut{} 50000}}%
      \put(4356,440){\makebox(0,0){\strut{} 100000}}%
      \put(5875,440){\makebox(0,0){\strut{} 150000}}%
      \put(160,2183){\rotatebox{-270}{\makebox(0,0){\strut{}residual}}}%
      \put(3607,140){\makebox(0,0){\strut{}Calls to Wilson operator}}%
    }%
    \gplgaddtomacro\gplfronttext{%
      \csname LTb\endcsname%
      \put(4972,3564){\makebox(0,0)[r]{\strut{}relSUMR}}%
      \csname LTb\endcsname%
      \put(4972,3364){\makebox(0,0)[r]{\strut{}deflated relSUMR}}%
    }%
    \gplbacktext
    \put(0,0){\includegraphics{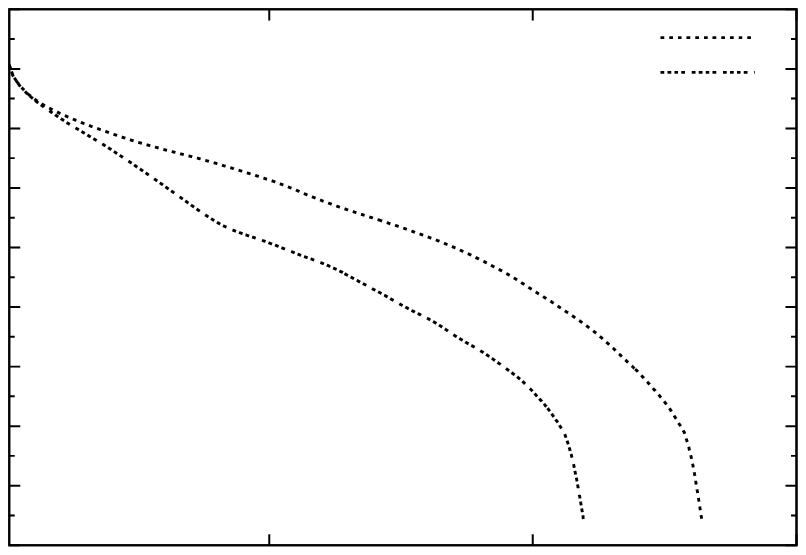}}%
    \gplfronttext
  \end{picture}%
\endgroup

%% file: figs/s8t32m0.03_c55_n30SUMRiterations2.tex
\begingroup
  \makeatletter
  \providecommand\color[2][]{%
    \GenericError{(gnuplot) \space\space\space\@spaces}{%
      Package color not loaded in conjunction with
      terminal option `colourtext'%
    }{See the gnuplot documentation for explanation.%
    }{Either use 'blacktext' in gnuplot or load the package
      color.sty in LaTeX.}%
    \renewcommand\color[2][]{}%
  }%
  \providecommand\includegraphics[2][]{%
    \GenericError{(gnuplot) \space\space\space\@spaces}{%
      Package graphicx or graphics not loaded%
    }{See the gnuplot documentation for explanation.%
    }{The gnuplot epslatex terminal needs graphicx.sty or graphics.sty.}%
    \renewcommand\includegraphics[2][]{}%
  }%
  \providecommand\rotatebox[2]{#2}%
  \@ifundefined{ifGPcolor}{%
    \newif\ifGPcolor
    \GPcolorfalse
  }{}%
  \@ifundefined{ifGPblacktext}{%
    \newif\ifGPblacktext
    \GPblacktexttrue
  }{}%
  \let\gplgaddtomacro\g@addto@macro
  \gdef\gplbacktext{}%
  \gdef\gplfronttext{}%
  \makeatother
  \ifGPblacktext
    \def\colorrgb#1{}%
    \def\colorgray#1{}%
  \else
    \ifGPcolor
      \def\colorrgb#1{\color[rgb]{#1}}%
      \def\colorgray#1{\color[gray]{#1}}%
      \expandafter\def\csname LTw\endcsname{\color{white}}%
      \expandafter\def\csname LTb\endcsname{\color{black}}%
      \expandafter\def\csname LTa\endcsname{\color{black}}%
      \expandafter\def\csname LT0\endcsname{\color[rgb]{1,0,0}}%
      \expandafter\def\csname LT1\endcsname{\color[rgb]{0,1,0}}%
      \expandafter\def\csname LT2\endcsname{\color[rgb]{0,0,1}}%
      \expandafter\def\csname LT3\endcsname{\color[rgb]{1,0,1}}%
      \expandafter\def\csname LT4\endcsname{\color[rgb]{0,1,1}}%
      \expandafter\def\csname LT5\endcsname{\color[rgb]{1,1,0}}%
      \expandafter\def\csname LT6\endcsname{\color[rgb]{0,0,0}}%
      \expandafter\def\csname LT7\endcsname{\color[rgb]{1,0.3,0}}%
      \expandafter\def\csname LT8\endcsname{\color[rgb]{0.5,0.5,0.5}}%
    \else
      \def\colorrgb#1{\color{black}}%
      \def\colorgray#1{\color[gray]{#1}}%
      \expandafter\def\csname LTw\endcsname{\color{white}}%
      \expandafter\def\csname LTb\endcsname{\color{black}}%
      \expandafter\def\csname LTa\endcsname{\color{black}}%
      \expandafter\def\csname LT0\endcsname{\color{black}}%
      \expandafter\def\csname LT1\endcsname{\color{black}}%
      \expandafter\def\csname LT2\endcsname{\color{black}}%
      \expandafter\def\csname LT3\endcsname{\color{black}}%
      \expandafter\def\csname LT4\endcsname{\color{black}}%
      \expandafter\def\csname LT5\endcsname{\color{black}}%
      \expandafter\def\csname LT6\endcsname{\color{black}}%
      \expandafter\def\csname LT7\endcsname{\color{black}}%
      \expandafter\def\csname LT8\endcsname{\color{black}}%
    \fi
  \fi
  \setlength{\unitlength}{0.0500bp}%
  \begin{picture}(6236.00,3968.00)%
    \gplgaddtomacro\gplbacktext{%
      \csname LTb\endcsname%
      \put(1220,640){\makebox(0,0)[r]{\strut{}$10^{-14}$}}%
      \put(1220,983){\makebox(0,0)[r]{\strut{}$10^{-12}$}}%
      \put(1220,1326){\makebox(0,0)[r]{\strut{}$10^{-10}$}}%
      \put(1220,1669){\makebox(0,0)[r]{\strut{}$10^{-8}$}}%
      \put(1220,2012){\makebox(0,0)[r]{\strut{}$10^{-6}$}}%
      \put(1220,2355){\makebox(0,0)[r]{\strut{}$10^{-4}$}}%
      \put(1220,2698){\makebox(0,0)[r]{\strut{}$10^{-2}$}}%
      \put(1220,3041){\makebox(0,0)[r]{\strut{}$1$}}%
      \put(1220,3384){\makebox(0,0)[r]{\strut{}$10^{2}$}}%
      \put(1220,3727){\makebox(0,0)[r]{\strut{}$10^{4}$}}%
      \put(1340,440){\makebox(0,0){\strut{} 0}}%
      \put(2096,440){\makebox(0,0){\strut{} 50}}%
      \put(2852,440){\makebox(0,0){\strut{} 100}}%
      \put(3608,440){\makebox(0,0){\strut{} 150}}%
      \put(4363,440){\makebox(0,0){\strut{} 200}}%
      \put(5119,440){\makebox(0,0){\strut{} 250}}%
      \put(5875,440){\makebox(0,0){\strut{} 300}}%
      \put(160,2183){\rotatebox{-270}{\makebox(0,0){\strut{}residual}}}%
      \put(3607,140){\makebox(0,0){\strut{}SUMR iterations}}%
    }%
    \gplgaddtomacro\gplfronttext{%
      \csname LTb\endcsname%
      \put(4972,3564){\makebox(0,0)[r]{\strut{}relSUMR}}%
      \csname LTb\endcsname%
      \put(4972,3364){\makebox(0,0)[r]{\strut{}deflated relSUMR}}%
    }%
    \gplbacktext
    \put(0,0){\includegraphics{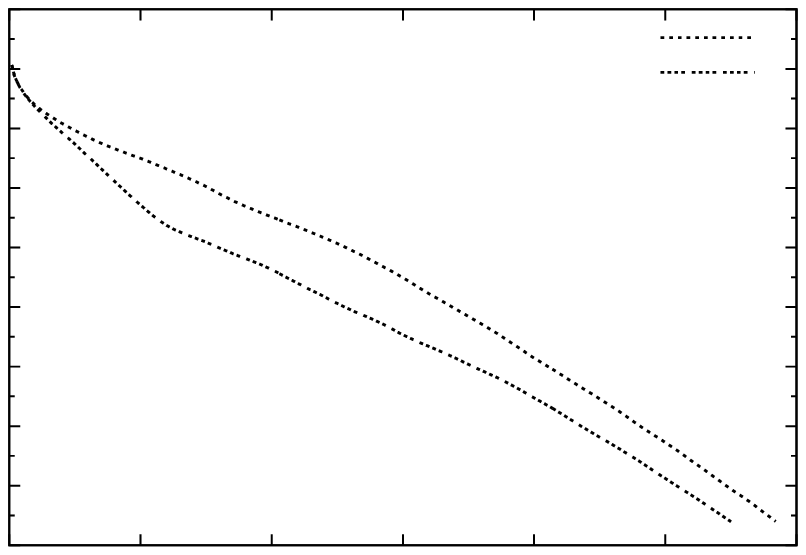}}%
    \gplfronttext
  \end{picture}%
\endgroup

%% file: figs/s8t32m0.03_c55_n30cgvsumr.tex
\begingroup
  \makeatletter
  \providecommand\color[2][]{%
    \GenericError{(gnuplot) \space\space\space\@spaces}{%
      Package color not loaded in conjunction with
      terminal option `colourtext'%
    }{See the gnuplot documentation for explanation.%
    }{Either use 'blacktext' in gnuplot or load the package
      color.sty in LaTeX.}%
    \renewcommand\color[2][]{}%
  }%
  \providecommand\includegraphics[2][]{%
    \GenericError{(gnuplot) \space\space\space\@spaces}{%
      Package graphicx or graphics not loaded%
    }{See the gnuplot documentation for explanation.%
    }{The gnuplot epslatex terminal needs graphicx.sty or graphics.sty.}%
    \renewcommand\includegraphics[2][]{}%
  }%
  \providecommand\rotatebox[2]{#2}%
  \@ifundefined{ifGPcolor}{%
    \newif\ifGPcolor
    \GPcolorfalse
  }{}%
  \@ifundefined{ifGPblacktext}{%
    \newif\ifGPblacktext
    \GPblacktexttrue
  }{}%
  \let\gplgaddtomacro\g@addto@macro
  \gdef\gplbacktext{}%
  \gdef\gplfronttext{}%
  \makeatother
  \ifGPblacktext
    \def\colorrgb#1{}%
    \def\colorgray#1{}%
  \else
    \ifGPcolor
      \def\colorrgb#1{\color[rgb]{#1}}%
      \def\colorgray#1{\color[gray]{#1}}%
      \expandafter\def\csname LTw\endcsname{\color{white}}%
      \expandafter\def\csname LTb\endcsname{\color{black}}%
      \expandafter\def\csname LTa\endcsname{\color{black}}%
      \expandafter\def\csname LT0\endcsname{\color[rgb]{1,0,0}}%
      \expandafter\def\csname LT1\endcsname{\color[rgb]{0,1,0}}%
      \expandafter\def\csname LT2\endcsname{\color[rgb]{0,0,1}}%
      \expandafter\def\csname LT3\endcsname{\color[rgb]{1,0,1}}%
      \expandafter\def\csname LT4\endcsname{\color[rgb]{0,1,1}}%
      \expandafter\def\csname LT5\endcsname{\color[rgb]{1,1,0}}%
      \expandafter\def\csname LT6\endcsname{\color[rgb]{0,0,0}}%
      \expandafter\def\csname LT7\endcsname{\color[rgb]{1,0.3,0}}%
      \expandafter\def\csname LT8\endcsname{\color[rgb]{0.5,0.5,0.5}}%
    \else
      \def\colorrgb#1{\color{black}}%
      \def\colorgray#1{\color[gray]{#1}}%
      \expandafter\def\csname LTw\endcsname{\color{white}}%
      \expandafter\def\csname LTb\endcsname{\color{black}}%
      \expandafter\def\csname LTa\endcsname{\color{black}}%
      \expandafter\def\csname LT0\endcsname{\color{black}}%
      \expandafter\def\csname LT1\endcsname{\color{black}}%
      \expandafter\def\csname LT2\endcsname{\color{black}}%
      \expandafter\def\csname LT3\endcsname{\color{black}}%
      \expandafter\def\csname LT4\endcsname{\color{black}}%
      \expandafter\def\csname LT5\endcsname{\color{black}}%
      \expandafter\def\csname LT6\endcsname{\color{black}}%
      \expandafter\def\csname LT7\endcsname{\color{black}}%
      \expandafter\def\csname LT8\endcsname{\color{black}}%
    \fi
  \fi
  \setlength{\unitlength}{0.0500bp}%
  \begin{picture}(6236.00,3968.00)%
    \gplgaddtomacro\gplbacktext{%
      \csname LTb\endcsname%
      \put(1220,640){\makebox(0,0)[r]{\strut{}$10^{-14}$}}%
      \put(1220,983){\makebox(0,0)[r]{\strut{}$10^{-12}$}}%
      \put(1220,1326){\makebox(0,0)[r]{\strut{}$10^{-10}$}}%
      \put(1220,1669){\makebox(0,0)[r]{\strut{}$10^{-8}$}}%
      \put(1220,2012){\makebox(0,0)[r]{\strut{}$10^{-6}$}}%
      \put(1220,2355){\makebox(0,0)[r]{\strut{}$10^{-4}$}}%
      \put(1220,2698){\makebox(0,0)[r]{\strut{}$10^{-2}$}}%
      \put(1220,3041){\makebox(0,0)[r]{\strut{}$1$}}%
      \put(1220,3384){\makebox(0,0)[r]{\strut{}$10^{2}$}}%
      \put(1220,3727){\makebox(0,0)[r]{\strut{}$10^{4}$}}%
      \put(1339,440){\makebox(0,0){\strut{} 0}}%
      \put(2246,440){\makebox(0,0){\strut{} 40000}}%
      \put(3153,440){\makebox(0,0){\strut{} 80000}}%
      \put(4061,440){\makebox(0,0){\strut{} 120000}}%
      \put(4968,440){\makebox(0,0){\strut{} 160000}}%
      \put(5875,440){\makebox(0,0){\strut{} 200000}}%
      \put(160,2183){\rotatebox{-270}{\makebox(0,0){\strut{}residual}}}%
      \put(3607,140){\makebox(0,0){\strut{}Calls to Kernel operator}}%
    }%
    \gplgaddtomacro\gplfronttext{%
      \csname LTb\endcsname%
      \put(4972,3564){\makebox(0,0)[r]{\strut{}GMRES(relCG)}}%
      \csname LTb\endcsname%
      \put(4972,3364){\makebox(0,0)[r]{\strut{}relCG}}%
      \csname LTb\endcsname%
      \put(4972,3164){\makebox(0,0)[r]{\strut{}GMRES(relSUMR)}}%
      \csname LTb\endcsname%
      \put(4972,2964){\makebox(0,0)[r]{\strut{}relSUMR}}%
    }%
    \gplbacktext
    \put(0,0){\includegraphics{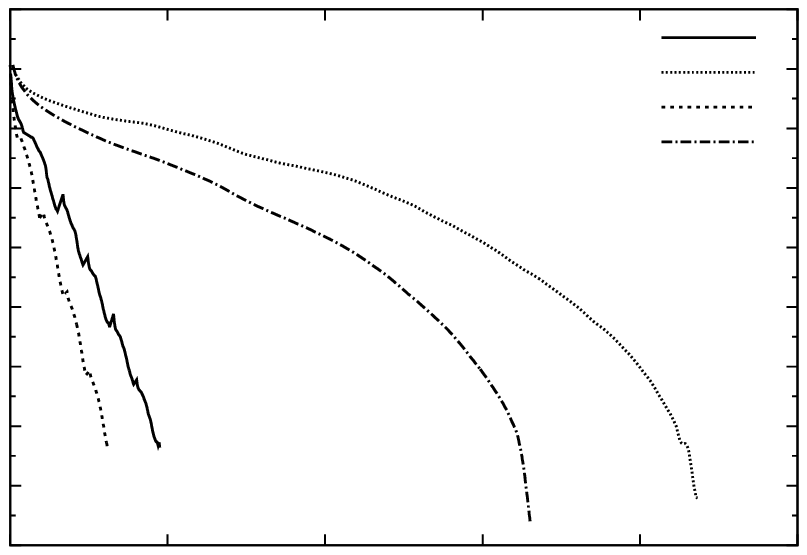}}%
    \gplfronttext
  \end{picture}%
\endgroup

%% file: figs/s8t32m0.03_c55_n30deflate.tex
\begingroup
  \makeatletter
  \providecommand\color[2][]{%
    \GenericError{(gnuplot) \space\space\space\@spaces}{%
      Package color not loaded in conjunction with
      terminal option `colourtext'%
    }{See the gnuplot documentation for explanation.%
    }{Either use 'blacktext' in gnuplot or load the package
      color.sty in LaTeX.}%
    \renewcommand\color[2][]{}%
  }%
  \providecommand\includegraphics[2][]{%
    \GenericError{(gnuplot) \space\space\space\@spaces}{%
      Package graphicx or graphics not loaded%
    }{See the gnuplot documentation for explanation.%
    }{The gnuplot epslatex terminal needs graphicx.sty or graphics.sty.}%
    \renewcommand\includegraphics[2][]{}%
  }%
  \providecommand\rotatebox[2]{#2}%
  \@ifundefined{ifGPcolor}{%
    \newif\ifGPcolor
    \GPcolorfalse
  }{}%
  \@ifundefined{ifGPblacktext}{%
    \newif\ifGPblacktext
    \GPblacktexttrue
  }{}%
  \let\gplgaddtomacro\g@addto@macro
  \gdef\gplbacktext{}%
  \gdef\gplfronttext{}%
  \makeatother
  \ifGPblacktext
    \def\colorrgb#1{}%
    \def\colorgray#1{}%
  \else
    \ifGPcolor
      \def\colorrgb#1{\color[rgb]{#1}}%
      \def\colorgray#1{\color[gray]{#1}}%
      \expandafter\def\csname LTw\endcsname{\color{white}}%
      \expandafter\def\csname LTb\endcsname{\color{black}}%
      \expandafter\def\csname LTa\endcsname{\color{black}}%
      \expandafter\def\csname LT0\endcsname{\color[rgb]{1,0,0}}%
      \expandafter\def\csname LT1\endcsname{\color[rgb]{0,1,0}}%
      \expandafter\def\csname LT2\endcsname{\color[rgb]{0,0,1}}%
      \expandafter\def\csname LT3\endcsname{\color[rgb]{1,0,1}}%
      \expandafter\def\csname LT4\endcsname{\color[rgb]{0,1,1}}%
      \expandafter\def\csname LT5\endcsname{\color[rgb]{1,1,0}}%
      \expandafter\def\csname LT6\endcsname{\color[rgb]{0,0,0}}%
      \expandafter\def\csname LT7\endcsname{\color[rgb]{1,0.3,0}}%
      \expandafter\def\csname LT8\endcsname{\color[rgb]{0.5,0.5,0.5}}%
    \else
      \def\colorrgb#1{\color{black}}%
      \def\colorgray#1{\color[gray]{#1}}%
      \expandafter\def\csname LTw\endcsname{\color{white}}%
      \expandafter\def\csname LTb\endcsname{\color{black}}%
      \expandafter\def\csname LTa\endcsname{\color{black}}%
      \expandafter\def\csname LT0\endcsname{\color{black}}%
      \expandafter\def\csname LT1\endcsname{\color{black}}%
      \expandafter\def\csname LT2\endcsname{\color{black}}%
      \expandafter\def\csname LT3\endcsname{\color{black}}%
      \expandafter\def\csname LT4\endcsname{\color{black}}%
      \expandafter\def\csname LT5\endcsname{\color{black}}%
      \expandafter\def\csname LT6\endcsname{\color{black}}%
      \expandafter\def\csname LT7\endcsname{\color{black}}%
      \expandafter\def\csname LT8\endcsname{\color{black}}%
    \fi
  \fi
  \setlength{\unitlength}{0.0500bp}%
  \begin{picture}(6236.00,3968.00)%
    \gplgaddtomacro\gplbacktext{%
      \csname LTb\endcsname%
      \put(1220,640){\makebox(0,0)[r]{\strut{}$10^{-12}$}}%
      \put(1220,1026){\makebox(0,0)[r]{\strut{}$10^{-10}$}}%
      \put(1220,1412){\makebox(0,0)[r]{\strut{}$10^{-8}$}}%
      \put(1220,1798){\makebox(0,0)[r]{\strut{}$10^{-6}$}}%
      \put(1220,2184){\makebox(0,0)[r]{\strut{}$10^{-4}$}}%
      \put(1220,2569){\makebox(0,0)[r]{\strut{}$10^{-2}$}}%
      \put(1220,2955){\makebox(0,0)[r]{\strut{}$1$}}%
      \put(1220,3341){\makebox(0,0)[r]{\strut{}$10^{2}$}}%
      \put(1220,3727){\makebox(0,0)[r]{\strut{}$10^{4}$}}%
      \put(1335,440){\makebox(0,0){\strut{} 0}}%
      \put(2470,440){\makebox(0,0){\strut{} 10000}}%
      \put(3605,440){\makebox(0,0){\strut{} 20000}}%
      \put(4740,440){\makebox(0,0){\strut{} 30000}}%
      \put(5875,440){\makebox(0,0){\strut{} 40000}}%
      \put(160,2183){\rotatebox{-270}{\makebox(0,0){\strut{}residual}}}%
      \put(3607,140){\makebox(0,0){\strut{}Calls to Kernel operator}}%
    }%
    \gplgaddtomacro\gplfronttext{%
      \csname LTb\endcsname%
      \put(4972,3564){\makebox(0,0)[r]{\strut{}GMRES(relCG)}}%
      \csname LTb\endcsname%
      \put(4972,3364){\makebox(0,0)[r]{\strut{}GMRES(deflated relCG)}}%
      \csname LTb\endcsname%
      \put(4972,3164){\makebox(0,0)[r]{\strut{}GMRES(relSUMR)}}%
      \csname LTb\endcsname%
      \put(4972,2964){\makebox(0,0)[r]{\strut{}GMRES(deflated relSUMR)}}%
    }%
    \gplbacktext
    \put(0,0){\includegraphics{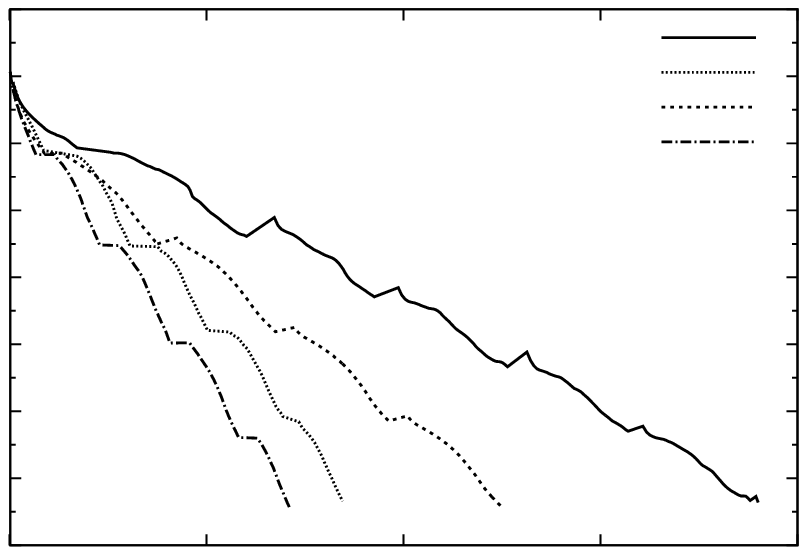}}%
    \gplfronttext
  \end{picture}%
\endgroup

%% file: figs/s8t32m0.03_c55_n30GMRESSUMRWilsonCalls.tex
\begingroup
  \makeatletter
  \providecommand\color[2][]{%
    \GenericError{(gnuplot) \space\space\space\@spaces}{%
      Package color not loaded in conjunction with
      terminal option `colourtext'%
    }{See the gnuplot documentation for explanation.%
    }{Either use 'blacktext' in gnuplot or load the package
      color.sty in LaTeX.}%
    \renewcommand\color[2][]{}%
  }%
  \providecommand\includegraphics[2][]{%
    \GenericError{(gnuplot) \space\space\space\@spaces}{%
      Package graphicx or graphics not loaded%
    }{See the gnuplot documentation for explanation.%
    }{The gnuplot epslatex terminal needs graphicx.sty or graphics.sty.}%
    \renewcommand\includegraphics[2][]{}%
  }%
  \providecommand\rotatebox[2]{#2}%
  \@ifundefined{ifGPcolor}{%
    \newif\ifGPcolor
    \GPcolorfalse
  }{}%
  \@ifundefined{ifGPblacktext}{%
    \newif\ifGPblacktext
    \GPblacktexttrue
  }{}%
  \let\gplgaddtomacro\g@addto@macro
  \gdef\gplbacktext{}%
  \gdef\gplfronttext{}%
  \makeatother
  \ifGPblacktext
    \def\colorrgb#1{}%
    \def\colorgray#1{}%
  \else
    \ifGPcolor
      \def\colorrgb#1{\color[rgb]{#1}}%
      \def\colorgray#1{\color[gray]{#1}}%
      \expandafter\def\csname LTw\endcsname{\color{white}}%
      \expandafter\def\csname LTb\endcsname{\color{black}}%
      \expandafter\def\csname LTa\endcsname{\color{black}}%
      \expandafter\def\csname LT0\endcsname{\color[rgb]{1,0,0}}%
      \expandafter\def\csname LT1\endcsname{\color[rgb]{0,1,0}}%
      \expandafter\def\csname LT2\endcsname{\color[rgb]{0,0,1}}%
      \expandafter\def\csname LT3\endcsname{\color[rgb]{1,0,1}}%
      \expandafter\def\csname LT4\endcsname{\color[rgb]{0,1,1}}%
      \expandafter\def\csname LT5\endcsname{\color[rgb]{1,1,0}}%
      \expandafter\def\csname LT6\endcsname{\color[rgb]{0,0,0}}%
      \expandafter\def\csname LT7\endcsname{\color[rgb]{1,0.3,0}}%
      \expandafter\def\csname LT8\endcsname{\color[rgb]{0.5,0.5,0.5}}%
    \else
      \def\colorrgb#1{\color{black}}%
      \def\colorgray#1{\color[gray]{#1}}%
      \expandafter\def\csname LTw\endcsname{\color{white}}%
      \expandafter\def\csname LTb\endcsname{\color{black}}%
      \expandafter\def\csname LTa\endcsname{\color{black}}%
      \expandafter\def\csname LT0\endcsname{\color{black}}%
      \expandafter\def\csname LT1\endcsname{\color{black}}%
      \expandafter\def\csname LT2\endcsname{\color{black}}%
      \expandafter\def\csname LT3\endcsname{\color{black}}%
      \expandafter\def\csname LT4\endcsname{\color{black}}%
      \expandafter\def\csname LT5\endcsname{\color{black}}%
      \expandafter\def\csname LT6\endcsname{\color{black}}%
      \expandafter\def\csname LT7\endcsname{\color{black}}%
      \expandafter\def\csname LT8\endcsname{\color{black}}%
    \fi
  \fi
  \setlength{\unitlength}{0.0500bp}%
  \begin{picture}(6236.00,3968.00)%
    \gplgaddtomacro\gplbacktext{%
      \csname LTb\endcsname%
      \put(1220,640){\makebox(0,0)[r]{\strut{}$10^{-12}$}}%
      \put(1220,983){\makebox(0,0)[r]{\strut{}$10^{-10}$}}%
      \put(1220,1326){\makebox(0,0)[r]{\strut{}$10^{-8}$}}%
      \put(1220,1669){\makebox(0,0)[r]{\strut{}$10^{-6}$}}%
      \put(1220,2012){\makebox(0,0)[r]{\strut{}$10^{-4}$}}%
      \put(1220,2355){\makebox(0,0)[r]{\strut{}$10^{-2}$}}%
      \put(1220,2698){\makebox(0,0)[r]{\strut{}$1$}}%
      \put(1220,3041){\makebox(0,0)[r]{\strut{}$10^{2}$}}%
      \put(1220,3384){\makebox(0,0)[r]{\strut{}$10^{4}$}}%
      \put(1220,3727){\makebox(0,0)[r]{\strut{}$10^{6}$}}%
      \put(1339,440){\makebox(0,0){\strut{} 0}}%
      \put(2095,440){\makebox(0,0){\strut{} 40000}}%
      \put(2851,440){\makebox(0,0){\strut{} 80000}}%
      \put(3607,440){\makebox(0,0){\strut{} 120000}}%
      \put(4363,440){\makebox(0,0){\strut{} 160000}}%
      \put(5119,440){\makebox(0,0){\strut{} 200000}}%
      \put(5875,440){\makebox(0,0){\strut{} 240000}}%
      \put(160,2183){\rotatebox{-270}{\makebox(0,0){\strut{}residual}}}%
      \put(3607,140){\makebox(0,0){\strut{}Calls to Kernel operator}}%
    }%
    \gplgaddtomacro\gplfronttext{%
      \csname LTb\endcsname%
      \put(4972,3564){\makebox(0,0)[r]{\strut{}GMRES(relSUMR)}}%
      \csname LTb\endcsname%
      \put(4972,3364){\makebox(0,0)[r]{\strut{}GMRES(calculating relSUMR 1)}}%
      \csname LTb\endcsname%
      \put(4972,3164){\makebox(0,0)[r]{\strut{}GMRES(calculating relSUMR 3)}}%
      \csname LTb\endcsname%
      \put(4972,2964){\makebox(0,0)[r]{\strut{}GMRES(calculating relSUMR 5)}}%
      \csname LTb\endcsname%
      \put(4972,2764){\makebox(0,0)[r]{\strut{}GMRES(deflated relSUMR)}}%
    }%
    \gplbacktext
    \put(0,0){\includegraphics{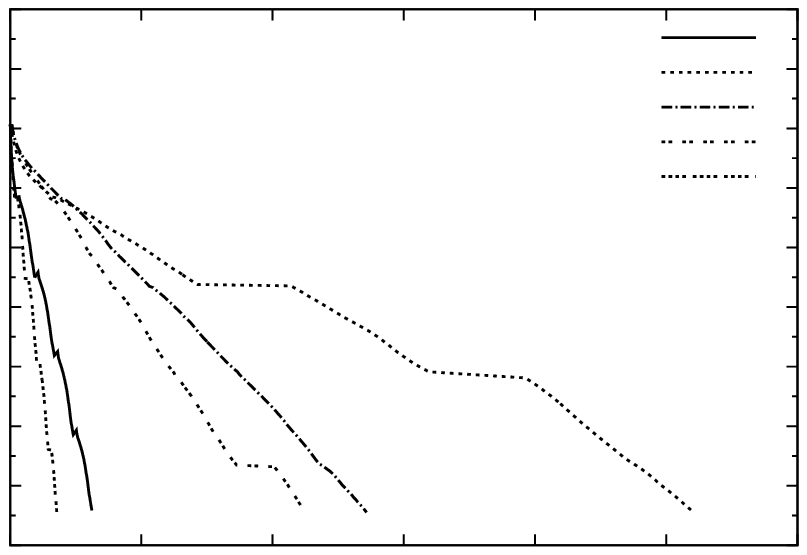}}%
    \gplfronttext
  \end{picture}%
\endgroup

%% file: figs/s8t32m0.03_c55_n30GMRESSUMRiterations.tex
\begingroup
  \makeatletter
  \providecommand\color[2][]{%
    \GenericError{(gnuplot) \space\space\space\@spaces}{%
      Package color not loaded in conjunction with
      terminal option `colourtext'%
    }{See the gnuplot documentation for explanation.%
    }{Either use 'blacktext' in gnuplot or load the package
      color.sty in LaTeX.}%
    \renewcommand\color[2][]{}%
  }%
  \providecommand\includegraphics[2][]{%
    \GenericError{(gnuplot) \space\space\space\@spaces}{%
      Package graphicx or graphics not loaded%
    }{See the gnuplot documentation for explanation.%
    }{The gnuplot epslatex terminal needs graphicx.sty or graphics.sty.}%
    \renewcommand\includegraphics[2][]{}%
  }%
  \providecommand\rotatebox[2]{#2}%
  \@ifundefined{ifGPcolor}{%
    \newif\ifGPcolor
    \GPcolorfalse
  }{}%
  \@ifundefined{ifGPblacktext}{%
    \newif\ifGPblacktext
    \GPblacktexttrue
  }{}%
  \let\gplgaddtomacro\g@addto@macro
  \gdef\gplbacktext{}%
  \gdef\gplfronttext{}%
  \makeatother
  \ifGPblacktext
    \def\colorrgb#1{}%
    \def\colorgray#1{}%
  \else
    \ifGPcolor
      \def\colorrgb#1{\color[rgb]{#1}}%
      \def\colorgray#1{\color[gray]{#1}}%
      \expandafter\def\csname LTw\endcsname{\color{white}}%
      \expandafter\def\csname LTb\endcsname{\color{black}}%
      \expandafter\def\csname LTa\endcsname{\color{black}}%
      \expandafter\def\csname LT0\endcsname{\color[rgb]{1,0,0}}%
      \expandafter\def\csname LT1\endcsname{\color[rgb]{0,1,0}}%
      \expandafter\def\csname LT2\endcsname{\color[rgb]{0,0,1}}%
      \expandafter\def\csname LT3\endcsname{\color[rgb]{1,0,1}}%
      \expandafter\def\csname LT4\endcsname{\color[rgb]{0,1,1}}%
      \expandafter\def\csname LT5\endcsname{\color[rgb]{1,1,0}}%
      \expandafter\def\csname LT6\endcsname{\color[rgb]{0,0,0}}%
      \expandafter\def\csname LT7\endcsname{\color[rgb]{1,0.3,0}}%
      \expandafter\def\csname LT8\endcsname{\color[rgb]{0.5,0.5,0.5}}%
    \else
      \def\colorrgb#1{\color{black}}%
      \def\colorgray#1{\color[gray]{#1}}%
      \expandafter\def\csname LTw\endcsname{\color{white}}%
      \expandafter\def\csname LTb\endcsname{\color{black}}%
      \expandafter\def\csname LTa\endcsname{\color{black}}%
      \expandafter\def\csname LT0\endcsname{\color{black}}%
      \expandafter\def\csname LT1\endcsname{\color{black}}%
      \expandafter\def\csname LT2\endcsname{\color{black}}%
      \expandafter\def\csname LT3\endcsname{\color{black}}%
      \expandafter\def\csname LT4\endcsname{\color{black}}%
      \expandafter\def\csname LT5\endcsname{\color{black}}%
      \expandafter\def\csname LT6\endcsname{\color{black}}%
      \expandafter\def\csname LT7\endcsname{\color{black}}%
      \expandafter\def\csname LT8\endcsname{\color{black}}%
    \fi
  \fi
  \setlength{\unitlength}{0.0500bp}%
  \begin{picture}(6236.00,3968.00)%
    \gplgaddtomacro\gplbacktext{%
      \csname LTb\endcsname%
      \put(1220,640){\makebox(0,0)[r]{\strut{}$10^{-12}$}}%
      \put(1220,1026){\makebox(0,0)[r]{\strut{}$10^{-10}$}}%
      \put(1220,1412){\makebox(0,0)[r]{\strut{}$10^{-8}$}}%
      \put(1220,1798){\makebox(0,0)[r]{\strut{}$10^{-6}$}}%
      \put(1220,2184){\makebox(0,0)[r]{\strut{}$10^{-4}$}}%
      \put(1220,2569){\makebox(0,0)[r]{\strut{}$10^{-2}$}}%
      \put(1220,2955){\makebox(0,0)[r]{\strut{}$1$}}%
      \put(1220,3341){\makebox(0,0)[r]{\strut{}$10^{2}$}}%
      \put(1220,3727){\makebox(0,0)[r]{\strut{}$10^{4}$}}%
      \put(1340,440){\makebox(0,0){\strut{} 0}}%
      \put(1988,440){\makebox(0,0){\strut{} 50}}%
      \put(2636,440){\makebox(0,0){\strut{} 100}}%
      \put(3284,440){\makebox(0,0){\strut{} 150}}%
      \put(3931,440){\makebox(0,0){\strut{} 200}}%
      \put(4579,440){\makebox(0,0){\strut{} 250}}%
      \put(5227,440){\makebox(0,0){\strut{} 300}}%
      \put(5875,440){\makebox(0,0){\strut{} 350}}%
      \put(160,2183){\rotatebox{-270}{\makebox(0,0){\strut{}residual}}}%
      \put(3607,140){\makebox(0,0){\strut{}SUMR iterations}}%
    }%
    \gplgaddtomacro\gplfronttext{%
      \csname LTb\endcsname%
      \put(4972,3564){\makebox(0,0)[r]{\strut{}GMRES(relSUMR)}}%
      \csname LTb\endcsname%
      \put(4972,3364){\makebox(0,0)[r]{\strut{}GMRES(calculating relSUMR 1)}}%
      \csname LTb\endcsname%
      \put(4972,3164){\makebox(0,0)[r]{\strut{}GMRES(calculating relSUMR 3)}}%
      \csname LTb\endcsname%
      \put(4972,2964){\makebox(0,0)[r]{\strut{}GMRES(calculating relSUMR 5)}}%
      \csname LTb\endcsname%
      \put(4972,2764){\makebox(0,0)[r]{\strut{}GMRES(deflated relSUMR)}}%
    }%
    \gplbacktext
    \put(0,0){\includegraphics{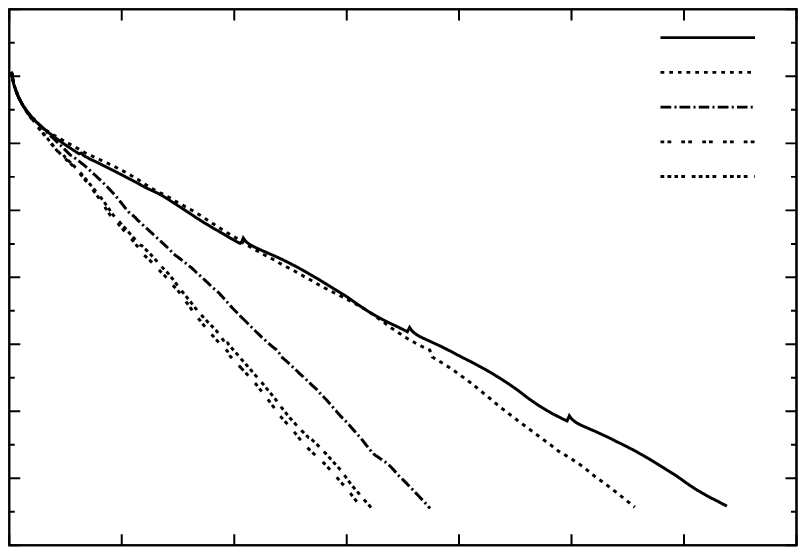}}%
    \gplfronttext
  \end{picture}%
\endgroup

%% file: figs/s8t32m0.03_c55_n30GMRESCGWilsonCalls.tex
\begingroup
  \makeatletter
  \providecommand\color[2][]{%
    \GenericError{(gnuplot) \space\space\space\@spaces}{%
      Package color not loaded in conjunction with
      terminal option `colourtext'%
    }{See the gnuplot documentation for explanation.%
    }{Either use 'blacktext' in gnuplot or load the package
      color.sty in LaTeX.}%
    \renewcommand\color[2][]{}%
  }%
  \providecommand\includegraphics[2][]{%
    \GenericError{(gnuplot) \space\space\space\@spaces}{%
      Package graphicx or graphics not loaded%
    }{See the gnuplot documentation for explanation.%
    }{The gnuplot epslatex terminal needs graphicx.sty or graphics.sty.}%
    \renewcommand\includegraphics[2][]{}%
  }%
  \providecommand\rotatebox[2]{#2}%
  \@ifundefined{ifGPcolor}{%
    \newif\ifGPcolor
    \GPcolorfalse
  }{}%
  \@ifundefined{ifGPblacktext}{%
    \newif\ifGPblacktext
    \GPblacktexttrue
  }{}%
  \let\gplgaddtomacro\g@addto@macro
  \gdef\gplbacktext{}%
  \gdef\gplfronttext{}%
  \makeatother
  \ifGPblacktext
    \def\colorrgb#1{}%
    \def\colorgray#1{}%
  \else
    \ifGPcolor
      \def\colorrgb#1{\color[rgb]{#1}}%
      \def\colorgray#1{\color[gray]{#1}}%
      \expandafter\def\csname LTw\endcsname{\color{white}}%
      \expandafter\def\csname LTb\endcsname{\color{black}}%
      \expandafter\def\csname LTa\endcsname{\color{black}}%
      \expandafter\def\csname LT0\endcsname{\color[rgb]{1,0,0}}%
      \expandafter\def\csname LT1\endcsname{\color[rgb]{0,1,0}}%
      \expandafter\def\csname LT2\endcsname{\color[rgb]{0,0,1}}%
      \expandafter\def\csname LT3\endcsname{\color[rgb]{1,0,1}}%
      \expandafter\def\csname LT4\endcsname{\color[rgb]{0,1,1}}%
      \expandafter\def\csname LT5\endcsname{\color[rgb]{1,1,0}}%
      \expandafter\def\csname LT6\endcsname{\color[rgb]{0,0,0}}%
      \expandafter\def\csname LT7\endcsname{\color[rgb]{1,0.3,0}}%
      \expandafter\def\csname LT8\endcsname{\color[rgb]{0.5,0.5,0.5}}%
    \else
      \def\colorrgb#1{\color{black}}%
      \def\colorgray#1{\color[gray]{#1}}%
      \expandafter\def\csname LTw\endcsname{\color{white}}%
      \expandafter\def\csname LTb\endcsname{\color{black}}%
      \expandafter\def\csname LTa\endcsname{\color{black}}%
      \expandafter\def\csname LT0\endcsname{\color{black}}%
      \expandafter\def\csname LT1\endcsname{\color{black}}%
      \expandafter\def\csname LT2\endcsname{\color{black}}%
      \expandafter\def\csname LT3\endcsname{\color{black}}%
      \expandafter\def\csname LT4\endcsname{\color{black}}%
      \expandafter\def\csname LT5\endcsname{\color{black}}%
      \expandafter\def\csname LT6\endcsname{\color{black}}%
      \expandafter\def\csname LT7\endcsname{\color{black}}%
      \expandafter\def\csname LT8\endcsname{\color{black}}%
    \fi
  \fi
  \setlength{\unitlength}{0.0500bp}%
  \begin{picture}(6236.00,3968.00)%
    \gplgaddtomacro\gplbacktext{%
      \csname LTb\endcsname%
      \put(1220,640){\makebox(0,0)[r]{\strut{}$10^{-12}$}}%
      \put(1220,1026){\makebox(0,0)[r]{\strut{}$10^{-10}$}}%
      \put(1220,1412){\makebox(0,0)[r]{\strut{}$10^{-8}$}}%
      \put(1220,1798){\makebox(0,0)[r]{\strut{}$10^{-6}$}}%
      \put(1220,2184){\makebox(0,0)[r]{\strut{}$10^{-4}$}}%
      \put(1220,2569){\makebox(0,0)[r]{\strut{}$10^{-2}$}}%
      \put(1220,2955){\makebox(0,0)[r]{\strut{}$1$}}%
      \put(1220,3341){\makebox(0,0)[r]{\strut{}$10^{2}$}}%
      \put(1220,3727){\makebox(0,0)[r]{\strut{}$10^{4}$}}%
      \put(1336,440){\makebox(0,0){\strut{} 0}}%
      \put(2471,440){\makebox(0,0){\strut{} 50000}}%
      \put(3605,440){\makebox(0,0){\strut{} 100000}}%
      \put(4740,440){\makebox(0,0){\strut{} 150000}}%
      \put(5875,440){\makebox(0,0){\strut{} 200000}}%
      \put(160,2183){\rotatebox{-270}{\makebox(0,0){\strut{}residual}}}%
      \put(3607,140){\makebox(0,0){\strut{}Calls to Kernel operator}}%
    }%
    \gplgaddtomacro\gplfronttext{%
      \csname LTb\endcsname%
      \put(4972,3564){\makebox(0,0)[r]{\strut{}GMRES(relCG)}}%
      \csname LTb\endcsname%
      \put(4972,3364){\makebox(0,0)[r]{\strut{}GMRES(calculating relCG 1)}}%
      \csname LTb\endcsname%
      \put(4972,3164){\makebox(0,0)[r]{\strut{}GMRES(calculating relCG 2)}}%
      \csname LTb\endcsname%
      \put(4972,2964){\makebox(0,0)[r]{\strut{}GMRES(calculating relCG 3)}}%
      \csname LTb\endcsname%
      \put(4972,2764){\makebox(0,0)[r]{\strut{}GMRES(deflated relCG)}}%
    }%
    \gplbacktext
    \put(0,0){\includegraphics{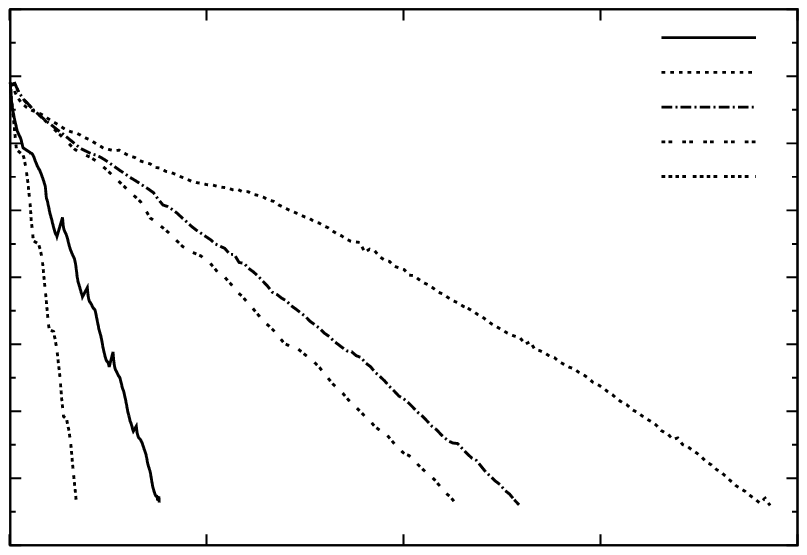}}%
    \gplfronttext
  \end{picture}%
\endgroup

%% file: figs/s8t32m0.03_c55_n30GMRESCGiterations.tex
\begingroup
  \makeatletter
  \providecommand\color[2][]{%
    \GenericError{(gnuplot) \space\space\space\@spaces}{%
      Package color not loaded in conjunction with
      terminal option `colourtext'%
    }{See the gnuplot documentation for explanation.%
    }{Either use 'blacktext' in gnuplot or load the package
      color.sty in LaTeX.}%
    \renewcommand\color[2][]{}%
  }%
  \providecommand\includegraphics[2][]{%
    \GenericError{(gnuplot) \space\space\space\@spaces}{%
      Package graphicx or graphics not loaded%
    }{See the gnuplot documentation for explanation.%
    }{The gnuplot epslatex terminal needs graphicx.sty or graphics.sty.}%
    \renewcommand\includegraphics[2][]{}%
  }%
  \providecommand\rotatebox[2]{#2}%
  \@ifundefined{ifGPcolor}{%
    \newif\ifGPcolor
    \GPcolorfalse
  }{}%
  \@ifundefined{ifGPblacktext}{%
    \newif\ifGPblacktext
    \GPblacktexttrue
  }{}%
  \let\gplgaddtomacro\g@addto@macro
  \gdef\gplbacktext{}%
  \gdef\gplfronttext{}%
  \makeatother
  \ifGPblacktext
    \def\colorrgb#1{}%
    \def\colorgray#1{}%
  \else
    \ifGPcolor
      \def\colorrgb#1{\color[rgb]{#1}}%
      \def\colorgray#1{\color[gray]{#1}}%
      \expandafter\def\csname LTw\endcsname{\color{white}}%
      \expandafter\def\csname LTb\endcsname{\color{black}}%
      \expandafter\def\csname LTa\endcsname{\color{black}}%
      \expandafter\def\csname LT0\endcsname{\color[rgb]{1,0,0}}%
      \expandafter\def\csname LT1\endcsname{\color[rgb]{0,1,0}}%
      \expandafter\def\csname LT2\endcsname{\color[rgb]{0,0,1}}%
      \expandafter\def\csname LT3\endcsname{\color[rgb]{1,0,1}}%
      \expandafter\def\csname LT4\endcsname{\color[rgb]{0,1,1}}%
      \expandafter\def\csname LT5\endcsname{\color[rgb]{1,1,0}}%
      \expandafter\def\csname LT6\endcsname{\color[rgb]{0,0,0}}%
      \expandafter\def\csname LT7\endcsname{\color[rgb]{1,0.3,0}}%
      \expandafter\def\csname LT8\endcsname{\color[rgb]{0.5,0.5,0.5}}%
    \else
      \def\colorrgb#1{\color{black}}%
      \def\colorgray#1{\color[gray]{#1}}%
      \expandafter\def\csname LTw\endcsname{\color{white}}%
      \expandafter\def\csname LTb\endcsname{\color{black}}%
      \expandafter\def\csname LTa\endcsname{\color{black}}%
      \expandafter\def\csname LT0\endcsname{\color{black}}%
      \expandafter\def\csname LT1\endcsname{\color{black}}%
      \expandafter\def\csname LT2\endcsname{\color{black}}%
      \expandafter\def\csname LT3\endcsname{\color{black}}%
      \expandafter\def\csname LT4\endcsname{\color{black}}%
      \expandafter\def\csname LT5\endcsname{\color{black}}%
      \expandafter\def\csname LT6\endcsname{\color{black}}%
      \expandafter\def\csname LT7\endcsname{\color{black}}%
      \expandafter\def\csname LT8\endcsname{\color{black}}%
    \fi
  \fi
  \setlength{\unitlength}{0.0500bp}%
  \begin{picture}(6236.00,3968.00)%
    \gplgaddtomacro\gplbacktext{%
      \csname LTb\endcsname%
      \put(1220,640){\makebox(0,0)[r]{\strut{}$10^{-12}$}}%
      \put(1220,1026){\makebox(0,0)[r]{\strut{}$10^{-10}$}}%
      \put(1220,1412){\makebox(0,0)[r]{\strut{}$10^{-8}$}}%
      \put(1220,1798){\makebox(0,0)[r]{\strut{}$10^{-6}$}}%
      \put(1220,2184){\makebox(0,0)[r]{\strut{}$10^{-4}$}}%
      \put(1220,2569){\makebox(0,0)[r]{\strut{}$10^{-2}$}}%
      \put(1220,2955){\makebox(0,0)[r]{\strut{}$1$}}%
      \put(1220,3341){\makebox(0,0)[r]{\strut{}$10^{2}$}}%
      \put(1220,3727){\makebox(0,0)[r]{\strut{}$10^{4}$}}%
      \put(1340,440){\makebox(0,0){\strut{} 0}}%
      \put(2096,440){\makebox(0,0){\strut{} 50}}%
      \put(2852,440){\makebox(0,0){\strut{} 100}}%
      \put(3608,440){\makebox(0,0){\strut{} 150}}%
      \put(4363,440){\makebox(0,0){\strut{} 200}}%
      \put(5119,440){\makebox(0,0){\strut{} 250}}%
      \put(5875,440){\makebox(0,0){\strut{} 300}}%
      \put(160,2183){\rotatebox{-270}{\makebox(0,0){\strut{}residual}}}%
      \put(3607,140){\makebox(0,0){\strut{}CG iterations}}%
    }%
    \gplgaddtomacro\gplfronttext{%
      \csname LTb\endcsname%
      \put(4972,3564){\makebox(0,0)[r]{\strut{}GMRES(relCG)}}%
      \csname LTb\endcsname%
      \put(4972,3364){\makebox(0,0)[r]{\strut{}GMRES(calculating relCG 1)}}%
      \csname LTb\endcsname%
      \put(4972,3164){\makebox(0,0)[r]{\strut{}GMRES(calculating relCG 2)}}%
      \csname LTb\endcsname%
      \put(4972,2964){\makebox(0,0)[r]{\strut{}GMRES(calculating relCG 3)}}%
      \csname LTb\endcsname%
      \put(4972,2764){\makebox(0,0)[r]{\strut{}GMRES(deflated relCG)}}%
    }%
    \gplbacktext
    \put(0,0){\includegraphics{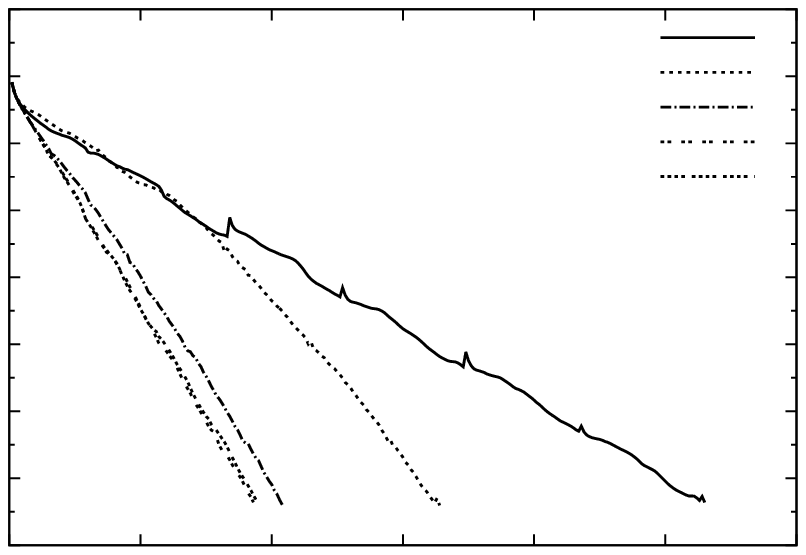}}%
    \gplfronttext
  \end{picture}%
\endgroup

%% file: figs/s8t32m0.03_c55_n30UnitaryResidual_0_29.tex
\begingroup
  \makeatletter
  \providecommand\color[2][]{%
    \GenericError{(gnuplot) \space\space\space\@spaces}{%
      Package color not loaded in conjunction with
      terminal option `colourtext'%
    }{See the gnuplot documentation for explanation.%
    }{Either use 'blacktext' in gnuplot or load the package
      color.sty in LaTeX.}%
    \renewcommand\color[2][]{}%
  }%
  \providecommand\includegraphics[2][]{%
    \GenericError{(gnuplot) \space\space\space\@spaces}{%
      Package graphicx or graphics not loaded%
    }{See the gnuplot documentation for explanation.%
    }{The gnuplot epslatex terminal needs graphicx.sty or graphics.sty.}%
    \renewcommand\includegraphics[2][]{}%
  }%
  \providecommand\rotatebox[2]{#2}%
  \@ifundefined{ifGPcolor}{%
    \newif\ifGPcolor
    \GPcolorfalse
  }{}%
  \@ifundefined{ifGPblacktext}{%
    \newif\ifGPblacktext
    \GPblacktexttrue
  }{}%
  \let\gplgaddtomacro\g@addto@macro
  \gdef\gplbacktext{}%
  \gdef\gplfronttext{}%
  \makeatother
  \ifGPblacktext
    \def\colorrgb#1{}%
    \def\colorgray#1{}%
  \else
    \ifGPcolor
      \def\colorrgb#1{\color[rgb]{#1}}%
      \def\colorgray#1{\color[gray]{#1}}%
      \expandafter\def\csname LTw\endcsname{\color{white}}%
      \expandafter\def\csname LTb\endcsname{\color{black}}%
      \expandafter\def\csname LTa\endcsname{\color{black}}%
      \expandafter\def\csname LT0\endcsname{\color[rgb]{1,0,0}}%
      \expandafter\def\csname LT1\endcsname{\color[rgb]{0,1,0}}%
      \expandafter\def\csname LT2\endcsname{\color[rgb]{0,0,1}}%
      \expandafter\def\csname LT3\endcsname{\color[rgb]{1,0,1}}%
      \expandafter\def\csname LT4\endcsname{\color[rgb]{0,1,1}}%
      \expandafter\def\csname LT5\endcsname{\color[rgb]{1,1,0}}%
      \expandafter\def\csname LT6\endcsname{\color[rgb]{0,0,0}}%
      \expandafter\def\csname LT7\endcsname{\color[rgb]{1,0.3,0}}%
      \expandafter\def\csname LT8\endcsname{\color[rgb]{0.5,0.5,0.5}}%
    \else
      \def\colorrgb#1{\color{black}}%
      \def\colorgray#1{\color[gray]{#1}}%
      \expandafter\def\csname LTw\endcsname{\color{white}}%
      \expandafter\def\csname LTb\endcsname{\color{black}}%
      \expandafter\def\csname LTa\endcsname{\color{black}}%
      \expandafter\def\csname LT0\endcsname{\color{black}}%
      \expandafter\def\csname LT1\endcsname{\color{black}}%
      \expandafter\def\csname LT2\endcsname{\color{black}}%
      \expandafter\def\csname LT3\endcsname{\color{black}}%
      \expandafter\def\csname LT4\endcsname{\color{black}}%
      \expandafter\def\csname LT5\endcsname{\color{black}}%
      \expandafter\def\csname LT6\endcsname{\color{black}}%
      \expandafter\def\csname LT7\endcsname{\color{black}}%
      \expandafter\def\csname LT8\endcsname{\color{black}}%
    \fi
  \fi
  \setlength{\unitlength}{0.0500bp}%
  \begin{picture}(6236.00,3968.00)%
    \gplgaddtomacro\gplbacktext{%
      \csname LTb\endcsname%
      \put(1220,640){\makebox(0,0)[r]{\strut{}$10^{-5}$}}%
      \put(1220,1101){\makebox(0,0)[r]{\strut{}$10^{-4}$}}%
      \put(1220,1562){\makebox(0,0)[r]{\strut{}$10^{-3}$}}%
      \put(1220,2022){\makebox(0,0)[r]{\strut{}$10^{-2}$}}%
      \put(1220,2483){\makebox(0,0)[r]{\strut{}$10^{-1}$}}%
      \put(1220,2944){\makebox(0,0)[r]{\strut{}$1$}}%
      \put(1220,3405){\makebox(0,0)[r]{\strut{}$10$}}%
      \put(2087,440){\makebox(0,0){\strut{} 3$\times 10^{6}$}}%
      \put(2845,440){\makebox(0,0){\strut{} 6$\times 10^{6}$}}%
      \put(3602,440){\makebox(0,0){\strut{} 9$\times 10^{6}$}}%
      \put(4360,440){\makebox(0,0){\strut{} 1.2$\times 10^{7}$}}%
      \put(5117,440){\makebox(0,0){\strut{} 1.5$\times 10^{7}$}}%
      \put(5875,440){\makebox(0,0){\strut{} 1.8$\times 10^{7}$}}%
      \put(160,2183){\rotatebox{-270}{\makebox(0,0){\strut{}residual}}}%
      \put(3607,140){\makebox(0,0){\strut{}Calls to Kernel operator}}%
    }%
    \gplgaddtomacro\gplfronttext{%
      \csname LTb\endcsname%
      \put(4972,3564){\makebox(0,0)[r]{\strut{}Eigenvalue 1 residual (relaxed)}}%
      \csname LTb\endcsname%
      \put(4972,3364){\makebox(0,0)[r]{\strut{}Eigenvalue 1 residual (full accuracy)}}%
      \csname LTb\endcsname%
      \put(4972,3164){\makebox(0,0)[r]{\strut{}Eigenvalue 30 residual (relaxed)}}%
      \csname LTb\endcsname%
      \put(4972,2964){\makebox(0,0)[r]{\strut{}Eigenvalue 30 residual (full accuracy)}}%
    }%
    \gplbacktext
    \put(0,0){\includegraphics{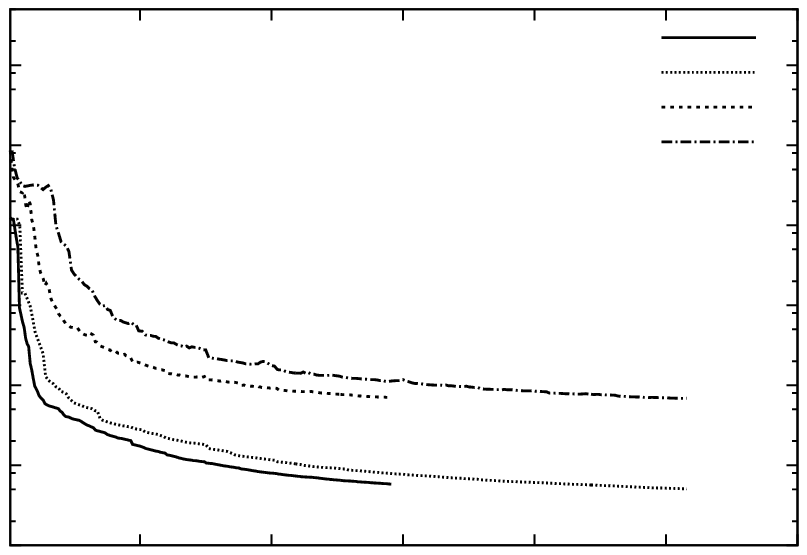}}%
    \gplfronttext
  \end{picture}%
\endgroup

%% file: figs/s8t32m0.03_c55_n30UnitaryResidualIter_0_29.tex
\begingroup
  \makeatletter
  \providecommand\color[2][]{%
    \GenericError{(gnuplot) \space\space\space\@spaces}{%
      Package color not loaded in conjunction with
      terminal option `colourtext'%
    }{See the gnuplot documentation for explanation.%
    }{Either use 'blacktext' in gnuplot or load the package
      color.sty in LaTeX.}%
    \renewcommand\color[2][]{}%
  }%
  \providecommand\includegraphics[2][]{%
    \GenericError{(gnuplot) \space\space\space\@spaces}{%
      Package graphicx or graphics not loaded%
    }{See the gnuplot documentation for explanation.%
    }{The gnuplot epslatex terminal needs graphicx.sty or graphics.sty.}%
    \renewcommand\includegraphics[2][]{}%
  }%
  \providecommand\rotatebox[2]{#2}%
  \@ifundefined{ifGPcolor}{%
    \newif\ifGPcolor
    \GPcolorfalse
  }{}%
  \@ifundefined{ifGPblacktext}{%
    \newif\ifGPblacktext
    \GPblacktexttrue
  }{}%
  \let\gplgaddtomacro\g@addto@macro
  \gdef\gplbacktext{}%
  \gdef\gplfronttext{}%
  \makeatother
  \ifGPblacktext
    \def\colorrgb#1{}%
    \def\colorgray#1{}%
  \else
    \ifGPcolor
      \def\colorrgb#1{\color[rgb]{#1}}%
      \def\colorgray#1{\color[gray]{#1}}%
      \expandafter\def\csname LTw\endcsname{\color{white}}%
      \expandafter\def\csname LTb\endcsname{\color{black}}%
      \expandafter\def\csname LTa\endcsname{\color{black}}%
      \expandafter\def\csname LT0\endcsname{\color[rgb]{1,0,0}}%
      \expandafter\def\csname LT1\endcsname{\color[rgb]{0,1,0}}%
      \expandafter\def\csname LT2\endcsname{\color[rgb]{0,0,1}}%
      \expandafter\def\csname LT3\endcsname{\color[rgb]{1,0,1}}%
      \expandafter\def\csname LT4\endcsname{\color[rgb]{0,1,1}}%
      \expandafter\def\csname LT5\endcsname{\color[rgb]{1,1,0}}%
      \expandafter\def\csname LT6\endcsname{\color[rgb]{0,0,0}}%
      \expandafter\def\csname LT7\endcsname{\color[rgb]{1,0.3,0}}%
      \expandafter\def\csname LT8\endcsname{\color[rgb]{0.5,0.5,0.5}}%
    \else
      \def\colorrgb#1{\color{black}}%
      \def\colorgray#1{\color[gray]{#1}}%
      \expandafter\def\csname LTw\endcsname{\color{white}}%
      \expandafter\def\csname LTb\endcsname{\color{black}}%
      \expandafter\def\csname LTa\endcsname{\color{black}}%
      \expandafter\def\csname LT0\endcsname{\color{black}}%
      \expandafter\def\csname LT1\endcsname{\color{black}}%
      \expandafter\def\csname LT2\endcsname{\color{black}}%
      \expandafter\def\csname LT3\endcsname{\color{black}}%
      \expandafter\def\csname LT4\endcsname{\color{black}}%
      \expandafter\def\csname LT5\endcsname{\color{black}}%
      \expandafter\def\csname LT6\endcsname{\color{black}}%
      \expandafter\def\csname LT7\endcsname{\color{black}}%
      \expandafter\def\csname LT8\endcsname{\color{black}}%
    \fi
  \fi
  \setlength{\unitlength}{0.0500bp}%
  \begin{picture}(6236.00,3968.00)%
    \gplgaddtomacro\gplbacktext{%
      \csname LTb\endcsname%
      \put(1220,640){\makebox(0,0)[r]{\strut{}$10^{-5}$}}%
      \put(1220,1101){\makebox(0,0)[r]{\strut{}$10^{-4}$}}%
      \put(1220,1562){\makebox(0,0)[r]{\strut{}$10^{-3}$}}%
      \put(1220,2022){\makebox(0,0)[r]{\strut{}$10^{-2}$}}%
      \put(1220,2483){\makebox(0,0)[r]{\strut{}$10^{-1}$}}%
      \put(1220,2944){\makebox(0,0)[r]{\strut{}$1$}}%
      \put(1220,3405){\makebox(0,0)[r]{\strut{}$10$}}%
      \put(1340,440){\makebox(0,0){\strut{} 0}}%
      \put(2247,440){\makebox(0,0){\strut{} 50}}%
      \put(3154,440){\makebox(0,0){\strut{} 100}}%
      \put(4061,440){\makebox(0,0){\strut{} 150}}%
      \put(4968,440){\makebox(0,0){\strut{} 200}}%
      \put(5875,440){\makebox(0,0){\strut{} 250}}%
      \put(160,2183){\rotatebox{-270}{\makebox(0,0){\strut{}residual}}}%
      \put(3607,140){\makebox(0,0){\strut{}Iterations}}%
    }%
    \gplgaddtomacro\gplfronttext{%
      \csname LTb\endcsname%
      \put(4972,3564){\makebox(0,0)[r]{\strut{}Eigenvalue 1 residual (relaxed)}}%
      \csname LTb\endcsname%
      \put(4972,3364){\makebox(0,0)[r]{\strut{}Eigenvalue 1 residual (full accuracy)}}%
      \csname LTb\endcsname%
      \put(4972,3164){\makebox(0,0)[r]{\strut{}Eigenvalue 30 residual (relaxed)}}%
      \csname LTb\endcsname%
      \put(4972,2964){\makebox(0,0)[r]{\strut{}Eigenvalue 30 residual (full accuracy)}}%
    }%
    \gplbacktext
    \put(0,0){\includegraphics{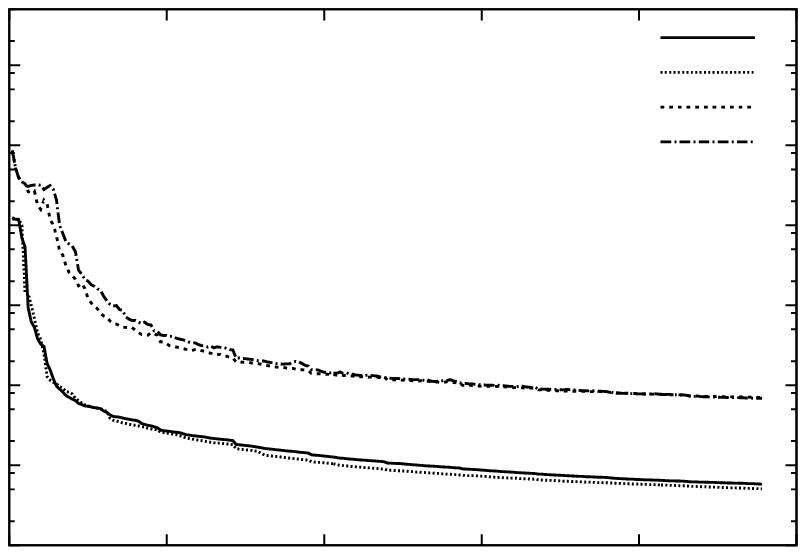}}%
    \gplfronttext
  \end{picture}%
\endgroup

%% file: figs/s8t32m0.03_c55_n30UnitaryAccuracy_29.tex
\begingroup
  \makeatletter
  \providecommand\color[2][]{%
    \GenericError{(gnuplot) \space\space\space\@spaces}{%
      Package color not loaded in conjunction with
      terminal option `colourtext'%
    }{See the gnuplot documentation for explanation.%
    }{Either use 'blacktext' in gnuplot or load the package
      color.sty in LaTeX.}%
    \renewcommand\color[2][]{}%
  }%
  \providecommand\includegraphics[2][]{%
    \GenericError{(gnuplot) \space\space\space\@spaces}{%
      Package graphicx or graphics not loaded%
    }{See the gnuplot documentation for explanation.%
    }{The gnuplot epslatex terminal needs graphicx.sty or graphics.sty.}%
    \renewcommand\includegraphics[2][]{}%
  }%
  \providecommand\rotatebox[2]{#2}%
  \@ifundefined{ifGPcolor}{%
    \newif\ifGPcolor
    \GPcolorfalse
  }{}%
  \@ifundefined{ifGPblacktext}{%
    \newif\ifGPblacktext
    \GPblacktexttrue
  }{}%
  \let\gplgaddtomacro\g@addto@macro
  \gdef\gplbacktext{}%
  \gdef\gplfronttext{}%
  \makeatother
  \ifGPblacktext
    \def\colorrgb#1{}%
    \def\colorgray#1{}%
  \else
    \ifGPcolor
      \def\colorrgb#1{\color[rgb]{#1}}%
      \def\colorgray#1{\color[gray]{#1}}%
      \expandafter\def\csname LTw\endcsname{\color{white}}%
      \expandafter\def\csname LTb\endcsname{\color{black}}%
      \expandafter\def\csname LTa\endcsname{\color{black}}%
      \expandafter\def\csname LT0\endcsname{\color[rgb]{1,0,0}}%
      \expandafter\def\csname LT1\endcsname{\color[rgb]{0,1,0}}%
      \expandafter\def\csname LT2\endcsname{\color[rgb]{0,0,1}}%
      \expandafter\def\csname LT3\endcsname{\color[rgb]{1,0,1}}%
      \expandafter\def\csname LT4\endcsname{\color[rgb]{0,1,1}}%
      \expandafter\def\csname LT5\endcsname{\color[rgb]{1,1,0}}%
      \expandafter\def\csname LT6\endcsname{\color[rgb]{0,0,0}}%
      \expandafter\def\csname LT7\endcsname{\color[rgb]{1,0.3,0}}%
      \expandafter\def\csname LT8\endcsname{\color[rgb]{0.5,0.5,0.5}}%
    \else
      \def\colorrgb#1{\color{black}}%
      \def\colorgray#1{\color[gray]{#1}}%
      \expandafter\def\csname LTw\endcsname{\color{white}}%
      \expandafter\def\csname LTb\endcsname{\color{black}}%
      \expandafter\def\csname LTa\endcsname{\color{black}}%
      \expandafter\def\csname LT0\endcsname{\color{black}}%
      \expandafter\def\csname LT1\endcsname{\color{black}}%
      \expandafter\def\csname LT2\endcsname{\color{black}}%
      \expandafter\def\csname LT3\endcsname{\color{black}}%
      \expandafter\def\csname LT4\endcsname{\color{black}}%
      \expandafter\def\csname LT5\endcsname{\color{black}}%
      \expandafter\def\csname LT6\endcsname{\color{black}}%
      \expandafter\def\csname LT7\endcsname{\color{black}}%
      \expandafter\def\csname LT8\endcsname{\color{black}}%
    \fi
  \fi
  \setlength{\unitlength}{0.0500bp}%
  \begin{picture}(6236.00,3968.00)%
    \gplgaddtomacro\gplbacktext{%
      \csname LTb\endcsname%
      \put(1220,640){\makebox(0,0)[r]{\strut{}$10^{-10}$}}%
      \put(1220,1155){\makebox(0,0)[r]{\strut{}$10^{-9}$}}%
      \put(1220,1669){\makebox(0,0)[r]{\strut{}$10^{-8}$}}%
      \put(1220,2184){\makebox(0,0)[r]{\strut{}$10^{-7}$}}%
      \put(1220,2698){\makebox(0,0)[r]{\strut{}$10^{-6}$}}%
      \put(1220,3213){\makebox(0,0)[r]{\strut{}$10^{-5}$}}%
      \put(1220,3727){\makebox(0,0)[r]{\strut{}$10^{-4}$}}%
      \put(1340,440){\makebox(0,0){\strut{} 0}}%
      \put(2852,440){\makebox(0,0){\strut{} 3$\times 10^{6}$}}%
      \put(4363,440){\makebox(0,0){\strut{} 6$\times 10^{6}$}}%
      \put(5875,440){\makebox(0,0){\strut{} 9$\times 10^{6}$}}%
      \put(160,2183){\rotatebox{-270}{\makebox(0,0){\strut{}residual}}}%
      \put(3607,140){\makebox(0,0){\strut{}Calls to Kernel operator}}%
    }%
    \gplgaddtomacro\gplfronttext{%
      \csname LTb\endcsname%
      \put(4972,3564){\makebox(0,0)[r]{\strut{}Accuracy of $v^D$ }}%
      \csname LTb\endcsname%
      \put(4972,3364){\makebox(0,0)[r]{\strut{}Target accuracy}}%
    }%
    \gplbacktext
    \put(0,0){\includegraphics{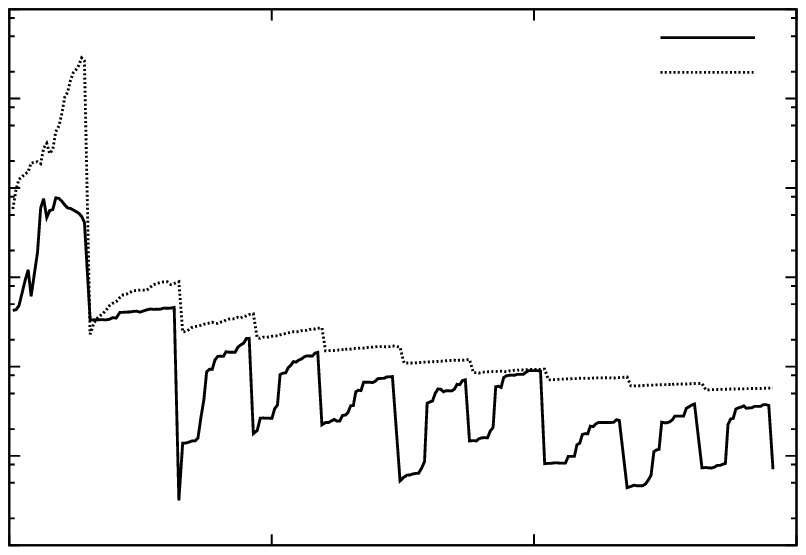}}%
    \gplfronttext
  \end{picture}%
\endgroup

%% file: figs/s8t32m0.03_c55_n30GMRESCGComparison1.tex
\begingroup
  \makeatletter
  \providecommand\color[2][]{%
    \GenericError{(gnuplot) \space\space\space\@spaces}{%
      Package color not loaded in conjunction with
      terminal option `colourtext'%
    }{See the gnuplot documentation for explanation.%
    }{Either use 'blacktext' in gnuplot or load the package
      color.sty in LaTeX.}%
    \renewcommand\color[2][]{}%
  }%
  \providecommand\includegraphics[2][]{%
    \GenericError{(gnuplot) \space\space\space\@spaces}{%
      Package graphicx or graphics not loaded%
    }{See the gnuplot documentation for explanation.%
    }{The gnuplot epslatex terminal needs graphicx.sty or graphics.sty.}%
    \renewcommand\includegraphics[2][]{}%
  }%
  \providecommand\rotatebox[2]{#2}%
  \@ifundefined{ifGPcolor}{%
    \newif\ifGPcolor
    \GPcolorfalse
  }{}%
  \@ifundefined{ifGPblacktext}{%
    \newif\ifGPblacktext
    \GPblacktexttrue
  }{}%
  \let\gplgaddtomacro\g@addto@macro
  \gdef\gplbacktext{}%
  \gdef\gplfronttext{}%
  \makeatother
  \ifGPblacktext
    \def\colorrgb#1{}%
    \def\colorgray#1{}%
  \else
    \ifGPcolor
      \def\colorrgb#1{\color[rgb]{#1}}%
      \def\colorgray#1{\color[gray]{#1}}%
      \expandafter\def\csname LTw\endcsname{\color{white}}%
      \expandafter\def\csname LTb\endcsname{\color{black}}%
      \expandafter\def\csname LTa\endcsname{\color{black}}%
      \expandafter\def\csname LT0\endcsname{\color[rgb]{1,0,0}}%
      \expandafter\def\csname LT1\endcsname{\color[rgb]{0,1,0}}%
      \expandafter\def\csname LT2\endcsname{\color[rgb]{0,0,1}}%
      \expandafter\def\csname LT3\endcsname{\color[rgb]{1,0,1}}%
      \expandafter\def\csname LT4\endcsname{\color[rgb]{0,1,1}}%
      \expandafter\def\csname LT5\endcsname{\color[rgb]{1,1,0}}%
      \expandafter\def\csname LT6\endcsname{\color[rgb]{0,0,0}}%
      \expandafter\def\csname LT7\endcsname{\color[rgb]{1,0.3,0}}%
      \expandafter\def\csname LT8\endcsname{\color[rgb]{0.5,0.5,0.5}}%
    \else
      \def\colorrgb#1{\color{black}}%
      \def\colorgray#1{\color[gray]{#1}}%
      \expandafter\def\csname LTw\endcsname{\color{white}}%
      \expandafter\def\csname LTb\endcsname{\color{black}}%
      \expandafter\def\csname LTa\endcsname{\color{black}}%
      \expandafter\def\csname LT0\endcsname{\color{black}}%
      \expandafter\def\csname LT1\endcsname{\color{black}}%
      \expandafter\def\csname LT2\endcsname{\color{black}}%
      \expandafter\def\csname LT3\endcsname{\color{black}}%
      \expandafter\def\csname LT4\endcsname{\color{black}}%
      \expandafter\def\csname LT5\endcsname{\color{black}}%
      \expandafter\def\csname LT6\endcsname{\color{black}}%
      \expandafter\def\csname LT7\endcsname{\color{black}}%
      \expandafter\def\csname LT8\endcsname{\color{black}}%
    \fi
  \fi
  \setlength{\unitlength}{0.0500bp}%
  \begin{picture}(6236.00,3968.00)%
    \gplgaddtomacro\gplbacktext{%
      \csname LTb\endcsname%
      \put(1220,640){\makebox(0,0)[r]{\strut{}$10^{-12}$}}%
      \put(1220,1081){\makebox(0,0)[r]{\strut{}$10^{-10}$}}%
      \put(1220,1522){\makebox(0,0)[r]{\strut{}$10^{-8}$}}%
      \put(1220,1963){\makebox(0,0)[r]{\strut{}$10^{-6}$}}%
      \put(1220,2404){\makebox(0,0)[r]{\strut{}$10^{-4}$}}%
      \put(1220,2845){\makebox(0,0)[r]{\strut{}$10^{-2}$}}%
      \put(1220,3286){\makebox(0,0)[r]{\strut{}$1$}}%
      \put(1220,3727){\makebox(0,0)[r]{\strut{}$10^{2}$}}%
      \put(1340,440){\makebox(0,0){\strut{} 0}}%
      \put(2474,440){\makebox(0,0){\strut{} 10000}}%
      \put(3608,440){\makebox(0,0){\strut{} 20000}}%
      \put(4741,440){\makebox(0,0){\strut{} 30000}}%
      \put(5875,440){\makebox(0,0){\strut{} 40000}}%
      \put(160,2183){\rotatebox{-270}{\makebox(0,0){\strut{}residual}}}%
      \put(3607,140){\makebox(0,0){\strut{}Calls to Kernel operator}}%
    }%
    \gplgaddtomacro\gplfronttext{%
      \csname LTb\endcsname%
      \put(4972,3564){\makebox(0,0)[r]{\strut{}GMRES(relCG)}}%
      \csname LTb\endcsname%
      \put(4972,3364){\makebox(0,0)[r]{\strut{}GMRES(deflated(15) relCG)}}%
      \csname LTb\endcsname%
      \put(4972,3164){\makebox(0,0)[r]{\strut{}GMRES(deflated(30) relCG)}}%
      \csname LTb\endcsname%
      \put(4972,2964){\makebox(0,0)[r]{\strut{}GMRES(deflated(45) relCG)}}%
    }%
    \gplbacktext
    \put(0,0){\includegraphics{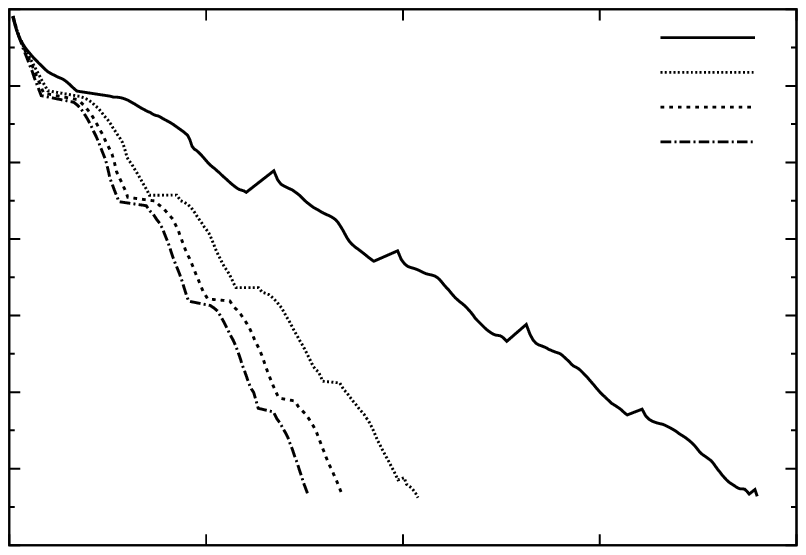}}%
    \gplfronttext
  \end{picture}%
\endgroup

%% file: figs/s8t32m0.03_c55_n30GMRESCGiterationsComparison1.tex
\begingroup
  \makeatletter
  \providecommand\color[2][]{%
    \GenericError{(gnuplot) \space\space\space\@spaces}{%
      Package color not loaded in conjunction with
      terminal option `colourtext'%
    }{See the gnuplot documentation for explanation.%
    }{Either use 'blacktext' in gnuplot or load the package
      color.sty in LaTeX.}%
    \renewcommand\color[2][]{}%
  }%
  \providecommand\includegraphics[2][]{%
    \GenericError{(gnuplot) \space\space\space\@spaces}{%
      Package graphicx or graphics not loaded%
    }{See the gnuplot documentation for explanation.%
    }{The gnuplot epslatex terminal needs graphicx.sty or graphics.sty.}%
    \renewcommand\includegraphics[2][]{}%
  }%
  \providecommand\rotatebox[2]{#2}%
  \@ifundefined{ifGPcolor}{%
    \newif\ifGPcolor
    \GPcolorfalse
  }{}%
  \@ifundefined{ifGPblacktext}{%
    \newif\ifGPblacktext
    \GPblacktexttrue
  }{}%
  \let\gplgaddtomacro\g@addto@macro
  \gdef\gplbacktext{}%
  \gdef\gplfronttext{}%
  \makeatother
  \ifGPblacktext
    \def\colorrgb#1{}%
    \def\colorgray#1{}%
  \else
    \ifGPcolor
      \def\colorrgb#1{\color[rgb]{#1}}%
      \def\colorgray#1{\color[gray]{#1}}%
      \expandafter\def\csname LTw\endcsname{\color{white}}%
      \expandafter\def\csname LTb\endcsname{\color{black}}%
      \expandafter\def\csname LTa\endcsname{\color{black}}%
      \expandafter\def\csname LT0\endcsname{\color[rgb]{1,0,0}}%
      \expandafter\def\csname LT1\endcsname{\color[rgb]{0,1,0}}%
      \expandafter\def\csname LT2\endcsname{\color[rgb]{0,0,1}}%
      \expandafter\def\csname LT3\endcsname{\color[rgb]{1,0,1}}%
      \expandafter\def\csname LT4\endcsname{\color[rgb]{0,1,1}}%
      \expandafter\def\csname LT5\endcsname{\color[rgb]{1,1,0}}%
      \expandafter\def\csname LT6\endcsname{\color[rgb]{0,0,0}}%
      \expandafter\def\csname LT7\endcsname{\color[rgb]{1,0.3,0}}%
      \expandafter\def\csname LT8\endcsname{\color[rgb]{0.5,0.5,0.5}}%
    \else
      \def\colorrgb#1{\color{black}}%
      \def\colorgray#1{\color[gray]{#1}}%
      \expandafter\def\csname LTw\endcsname{\color{white}}%
      \expandafter\def\csname LTb\endcsname{\color{black}}%
      \expandafter\def\csname LTa\endcsname{\color{black}}%
      \expandafter\def\csname LT0\endcsname{\color{black}}%
      \expandafter\def\csname LT1\endcsname{\color{black}}%
      \expandafter\def\csname LT2\endcsname{\color{black}}%
      \expandafter\def\csname LT3\endcsname{\color{black}}%
      \expandafter\def\csname LT4\endcsname{\color{black}}%
      \expandafter\def\csname LT5\endcsname{\color{black}}%
      \expandafter\def\csname LT6\endcsname{\color{black}}%
      \expandafter\def\csname LT7\endcsname{\color{black}}%
      \expandafter\def\csname LT8\endcsname{\color{black}}%
    \fi
  \fi
  \setlength{\unitlength}{0.0500bp}%
  \begin{picture}(6236.00,3968.00)%
    \gplgaddtomacro\gplbacktext{%
      \csname LTb\endcsname%
      \put(1220,640){\makebox(0,0)[r]{\strut{}$10^{-12}$}}%
      \put(1220,1081){\makebox(0,0)[r]{\strut{}$10^{-10}$}}%
      \put(1220,1522){\makebox(0,0)[r]{\strut{}$10^{-8}$}}%
      \put(1220,1963){\makebox(0,0)[r]{\strut{}$10^{-6}$}}%
      \put(1220,2404){\makebox(0,0)[r]{\strut{}$10^{-4}$}}%
      \put(1220,2845){\makebox(0,0)[r]{\strut{}$10^{-2}$}}%
      \put(1220,3286){\makebox(0,0)[r]{\strut{}$1$}}%
      \put(1220,3727){\makebox(0,0)[r]{\strut{}$10^{2}$}}%
      \put(1340,440){\makebox(0,0){\strut{} 0}}%
      \put(2096,440){\makebox(0,0){\strut{} 50}}%
      \put(2852,440){\makebox(0,0){\strut{} 100}}%
      \put(3608,440){\makebox(0,0){\strut{} 150}}%
      \put(4363,440){\makebox(0,0){\strut{} 200}}%
      \put(5119,440){\makebox(0,0){\strut{} 250}}%
      \put(5875,440){\makebox(0,0){\strut{} 300}}%
      \put(160,2183){\rotatebox{-270}{\makebox(0,0){\strut{}residual}}}%
      \put(3607,140){\makebox(0,0){\strut{}CG iterations}}%
    }%
    \gplgaddtomacro\gplfronttext{%
      \csname LTb\endcsname%
      \put(4972,3564){\makebox(0,0)[r]{\strut{}GMRES(relCG)}}%
      \csname LTb\endcsname%
      \put(4972,3364){\makebox(0,0)[r]{\strut{}GMRES(deflated(15) relCG)}}%
      \csname LTb\endcsname%
      \put(4972,3164){\makebox(0,0)[r]{\strut{}GMRES(deflated(30) relCG)}}%
      \csname LTb\endcsname%
      \put(4972,2964){\makebox(0,0)[r]{\strut{}GMRES(deflated(45) relCG)}}%
    }%
    \gplbacktext
    \put(0,0){\includegraphics{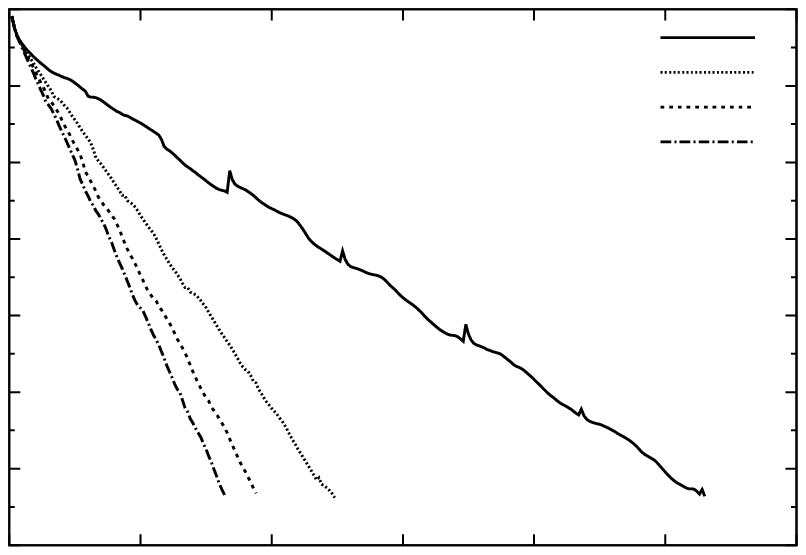}}%
    \gplfronttext
  \end{picture}%
\endgroup

%% file: figs/s8t32m0.03_c55_n30GMRESSUMRComparison1.tex
\begingroup
  \makeatletter
  \providecommand\color[2][]{%
    \GenericError{(gnuplot) \space\space\space\@spaces}{%
      Package color not loaded in conjunction with
      terminal option `colourtext'%
    }{See the gnuplot documentation for explanation.%
    }{Either use 'blacktext' in gnuplot or load the package
      color.sty in LaTeX.}%
    \renewcommand\color[2][]{}%
  }%
  \providecommand\includegraphics[2][]{%
    \GenericError{(gnuplot) \space\space\space\@spaces}{%
      Package graphicx or graphics not loaded%
    }{See the gnuplot documentation for explanation.%
    }{The gnuplot epslatex terminal needs graphicx.sty or graphics.sty.}%
    \renewcommand\includegraphics[2][]{}%
  }%
  \providecommand\rotatebox[2]{#2}%
  \@ifundefined{ifGPcolor}{%
    \newif\ifGPcolor
    \GPcolorfalse
  }{}%
  \@ifundefined{ifGPblacktext}{%
    \newif\ifGPblacktext
    \GPblacktexttrue
  }{}%
  \let\gplgaddtomacro\g@addto@macro
  \gdef\gplbacktext{}%
  \gdef\gplfronttext{}%
  \makeatother
  \ifGPblacktext
    \def\colorrgb#1{}%
    \def\colorgray#1{}%
  \else
    \ifGPcolor
      \def\colorrgb#1{\color[rgb]{#1}}%
      \def\colorgray#1{\color[gray]{#1}}%
      \expandafter\def\csname LTw\endcsname{\color{white}}%
      \expandafter\def\csname LTb\endcsname{\color{black}}%
      \expandafter\def\csname LTa\endcsname{\color{black}}%
      \expandafter\def\csname LT0\endcsname{\color[rgb]{1,0,0}}%
      \expandafter\def\csname LT1\endcsname{\color[rgb]{0,1,0}}%
      \expandafter\def\csname LT2\endcsname{\color[rgb]{0,0,1}}%
      \expandafter\def\csname LT3\endcsname{\color[rgb]{1,0,1}}%
      \expandafter\def\csname LT4\endcsname{\color[rgb]{0,1,1}}%
      \expandafter\def\csname LT5\endcsname{\color[rgb]{1,1,0}}%
      \expandafter\def\csname LT6\endcsname{\color[rgb]{0,0,0}}%
      \expandafter\def\csname LT7\endcsname{\color[rgb]{1,0.3,0}}%
      \expandafter\def\csname LT8\endcsname{\color[rgb]{0.5,0.5,0.5}}%
    \else
      \def\colorrgb#1{\color{black}}%
      \def\colorgray#1{\color[gray]{#1}}%
      \expandafter\def\csname LTw\endcsname{\color{white}}%
      \expandafter\def\csname LTb\endcsname{\color{black}}%
      \expandafter\def\csname LTa\endcsname{\color{black}}%
      \expandafter\def\csname LT0\endcsname{\color{black}}%
      \expandafter\def\csname LT1\endcsname{\color{black}}%
      \expandafter\def\csname LT2\endcsname{\color{black}}%
      \expandafter\def\csname LT3\endcsname{\color{black}}%
      \expandafter\def\csname LT4\endcsname{\color{black}}%
      \expandafter\def\csname LT5\endcsname{\color{black}}%
      \expandafter\def\csname LT6\endcsname{\color{black}}%
      \expandafter\def\csname LT7\endcsname{\color{black}}%
      \expandafter\def\csname LT8\endcsname{\color{black}}%
    \fi
  \fi
  \setlength{\unitlength}{0.0500bp}%
  \begin{picture}(6236.00,3968.00)%
    \gplgaddtomacro\gplbacktext{%
      \csname LTb\endcsname%
      \put(1220,640){\makebox(0,0)[r]{\strut{}$10^{-12}$}}%
      \put(1220,1026){\makebox(0,0)[r]{\strut{}$10^{-10}$}}%
      \put(1220,1412){\makebox(0,0)[r]{\strut{}$10^{-8}$}}%
      \put(1220,1798){\makebox(0,0)[r]{\strut{}$10^{-6}$}}%
      \put(1220,2184){\makebox(0,0)[r]{\strut{}$10^{-4}$}}%
      \put(1220,2569){\makebox(0,0)[r]{\strut{}$10^{-2}$}}%
      \put(1220,2955){\makebox(0,0)[r]{\strut{}$1$}}%
      \put(1220,3341){\makebox(0,0)[r]{\strut{}$10^{2}$}}%
      \put(1220,3727){\makebox(0,0)[r]{\strut{}$10^{4}$}}%
      \put(1340,440){\makebox(0,0){\strut{} 0}}%
      \put(5875,440){\makebox(0,0){\strut{} 30000}}%
      \put(160,2183){\rotatebox{-270}{\makebox(0,0){\strut{}residual}}}%
      \put(3607,140){\makebox(0,0){\strut{}Calls to Kernel operator}}%
    }%
    \gplgaddtomacro\gplfronttext{%
      \csname LTb\endcsname%
      \put(4972,3564){\makebox(0,0)[r]{\strut{}GMRES(relSUMR)}}%
      \csname LTb\endcsname%
      \put(4972,3364){\makebox(0,0)[r]{\strut{}GMRES(deflated(15) relSUMR)}}%
      \csname LTb\endcsname%
      \put(4972,3164){\makebox(0,0)[r]{\strut{}GMRES(deflated(30) relSUMR)}}%
      \csname LTb\endcsname%
      \put(4972,2964){\makebox(0,0)[r]{\strut{}GMRES(deflated(45) relSUMR)}}%
    }%
    \gplbacktext
    \put(0,0){\includegraphics{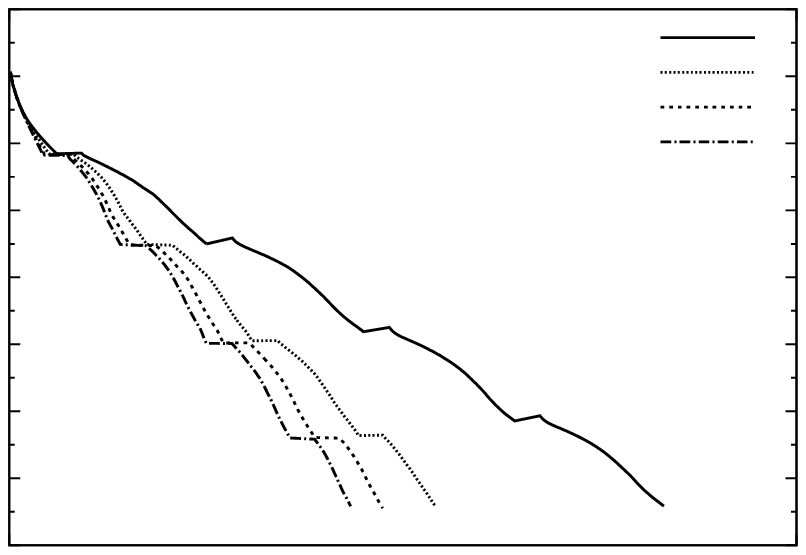}}%
    \gplfronttext
  \end{picture}%
\endgroup

%% file: figs/s8t32m0.03_c55_n30GMRESSUMRiterationsComparison1.tex
\begingroup
  \makeatletter
  \providecommand\color[2][]{%
    \GenericError{(gnuplot) \space\space\space\@spaces}{%
      Package color not loaded in conjunction with
      terminal option `colourtext'%
    }{See the gnuplot documentation for explanation.%
    }{Either use 'blacktext' in gnuplot or load the package
      color.sty in LaTeX.}%
    \renewcommand\color[2][]{}%
  }%
  \providecommand\includegraphics[2][]{%
    \GenericError{(gnuplot) \space\space\space\@spaces}{%
      Package graphicx or graphics not loaded%
    }{See the gnuplot documentation for explanation.%
    }{The gnuplot epslatex terminal needs graphicx.sty or graphics.sty.}%
    \renewcommand\includegraphics[2][]{}%
  }%
  \providecommand\rotatebox[2]{#2}%
  \@ifundefined{ifGPcolor}{%
    \newif\ifGPcolor
    \GPcolorfalse
  }{}%
  \@ifundefined{ifGPblacktext}{%
    \newif\ifGPblacktext
    \GPblacktexttrue
  }{}%
  \let\gplgaddtomacro\g@addto@macro
  \gdef\gplbacktext{}%
  \gdef\gplfronttext{}%
  \makeatother
  \ifGPblacktext
    \def\colorrgb#1{}%
    \def\colorgray#1{}%
  \else
    \ifGPcolor
      \def\colorrgb#1{\color[rgb]{#1}}%
      \def\colorgray#1{\color[gray]{#1}}%
      \expandafter\def\csname LTw\endcsname{\color{white}}%
      \expandafter\def\csname LTb\endcsname{\color{black}}%
      \expandafter\def\csname LTa\endcsname{\color{black}}%
      \expandafter\def\csname LT0\endcsname{\color[rgb]{1,0,0}}%
      \expandafter\def\csname LT1\endcsname{\color[rgb]{0,1,0}}%
      \expandafter\def\csname LT2\endcsname{\color[rgb]{0,0,1}}%
      \expandafter\def\csname LT3\endcsname{\color[rgb]{1,0,1}}%
      \expandafter\def\csname LT4\endcsname{\color[rgb]{0,1,1}}%
      \expandafter\def\csname LT5\endcsname{\color[rgb]{1,1,0}}%
      \expandafter\def\csname LT6\endcsname{\color[rgb]{0,0,0}}%
      \expandafter\def\csname LT7\endcsname{\color[rgb]{1,0.3,0}}%
      \expandafter\def\csname LT8\endcsname{\color[rgb]{0.5,0.5,0.5}}%
    \else
      \def\colorrgb#1{\color{black}}%
      \def\colorgray#1{\color[gray]{#1}}%
      \expandafter\def\csname LTw\endcsname{\color{white}}%
      \expandafter\def\csname LTb\endcsname{\color{black}}%
      \expandafter\def\csname LTa\endcsname{\color{black}}%
      \expandafter\def\csname LT0\endcsname{\color{black}}%
      \expandafter\def\csname LT1\endcsname{\color{black}}%
      \expandafter\def\csname LT2\endcsname{\color{black}}%
      \expandafter\def\csname LT3\endcsname{\color{black}}%
      \expandafter\def\csname LT4\endcsname{\color{black}}%
      \expandafter\def\csname LT5\endcsname{\color{black}}%
      \expandafter\def\csname LT6\endcsname{\color{black}}%
      \expandafter\def\csname LT7\endcsname{\color{black}}%
      \expandafter\def\csname LT8\endcsname{\color{black}}%
    \fi
  \fi
  \setlength{\unitlength}{0.0500bp}%
  \begin{picture}(6236.00,3968.00)%
    \gplgaddtomacro\gplbacktext{%
      \csname LTb\endcsname%
      \put(1220,640){\makebox(0,0)[r]{\strut{}$10^{-12}$}}%
      \put(1220,1026){\makebox(0,0)[r]{\strut{}$10^{-10}$}}%
      \put(1220,1412){\makebox(0,0)[r]{\strut{}$10^{-8}$}}%
      \put(1220,1798){\makebox(0,0)[r]{\strut{}$10^{-6}$}}%
      \put(1220,2184){\makebox(0,0)[r]{\strut{}$10^{-4}$}}%
      \put(1220,2569){\makebox(0,0)[r]{\strut{}$10^{-2}$}}%
      \put(1220,2955){\makebox(0,0)[r]{\strut{}$1$}}%
      \put(1220,3341){\makebox(0,0)[r]{\strut{}$10^{2}$}}%
      \put(1220,3727){\makebox(0,0)[r]{\strut{}$10^{4}$}}%
      \put(1340,440){\makebox(0,0){\strut{} 0}}%
      \put(1988,440){\makebox(0,0){\strut{} 50}}%
      \put(2636,440){\makebox(0,0){\strut{} 100}}%
      \put(3284,440){\makebox(0,0){\strut{} 150}}%
      \put(3931,440){\makebox(0,0){\strut{} 200}}%
      \put(4579,440){\makebox(0,0){\strut{} 250}}%
      \put(5227,440){\makebox(0,0){\strut{} 300}}%
      \put(5875,440){\makebox(0,0){\strut{} 350}}%
      \put(160,2183){\rotatebox{-270}{\makebox(0,0){\strut{}residual}}}%
      \put(3607,140){\makebox(0,0){\strut{}SUMR iterations}}%
    }%
    \gplgaddtomacro\gplfronttext{%
      \csname LTb\endcsname%
      \put(4972,3564){\makebox(0,0)[r]{\strut{}GMRES(relSUMR)}}%
      \csname LTb\endcsname%
      \put(4972,3364){\makebox(0,0)[r]{\strut{}GMRES(deflated(15) relSUMR)}}%
      \csname LTb\endcsname%
      \put(4972,3164){\makebox(0,0)[r]{\strut{}GMRES(deflated(30) relSUMR)}}%
      \csname LTb\endcsname%
      \put(4972,2964){\makebox(0,0)[r]{\strut{}GMRES(deflated(45) relSUMR)}}%
    }%
    \gplbacktext
    \put(0,0){\includegraphics{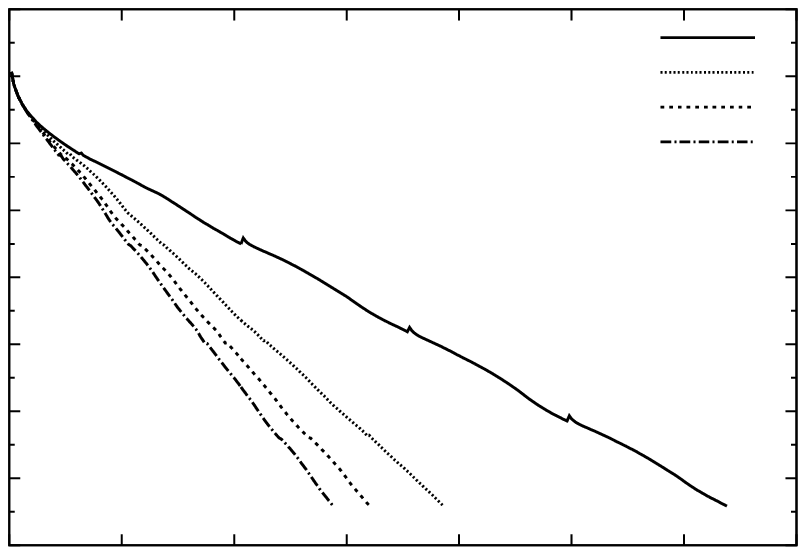}}%
    \gplfronttext
  \end{picture}%
\endgroup

%% file: figs/s8t32m0.03_c55_n30GMRESCGComparison2.tex
\begingroup
  \makeatletter
  \providecommand\color[2][]{%
    \GenericError{(gnuplot) \space\space\space\@spaces}{%
      Package color not loaded in conjunction with
      terminal option `colourtext'%
    }{See the gnuplot documentation for explanation.%
    }{Either use 'blacktext' in gnuplot or load the package
      color.sty in LaTeX.}%
    \renewcommand\color[2][]{}%
  }%
  \providecommand\includegraphics[2][]{%
    \GenericError{(gnuplot) \space\space\space\@spaces}{%
      Package graphicx or graphics not loaded%
    }{See the gnuplot documentation for explanation.%
    }{The gnuplot epslatex terminal needs graphicx.sty or graphics.sty.}%
    \renewcommand\includegraphics[2][]{}%
  }%
  \providecommand\rotatebox[2]{#2}%
  \@ifundefined{ifGPcolor}{%
    \newif\ifGPcolor
    \GPcolorfalse
  }{}%
  \@ifundefined{ifGPblacktext}{%
    \newif\ifGPblacktext
    \GPblacktexttrue
  }{}%
  \let\gplgaddtomacro\g@addto@macro
  \gdef\gplbacktext{}%
  \gdef\gplfronttext{}%
  \makeatother
  \ifGPblacktext
    \def\colorrgb#1{}%
    \def\colorgray#1{}%
  \else
    \ifGPcolor
      \def\colorrgb#1{\color[rgb]{#1}}%
      \def\colorgray#1{\color[gray]{#1}}%
      \expandafter\def\csname LTw\endcsname{\color{white}}%
      \expandafter\def\csname LTb\endcsname{\color{black}}%
      \expandafter\def\csname LTa\endcsname{\color{black}}%
      \expandafter\def\csname LT0\endcsname{\color[rgb]{1,0,0}}%
      \expandafter\def\csname LT1\endcsname{\color[rgb]{0,1,0}}%
      \expandafter\def\csname LT2\endcsname{\color[rgb]{0,0,1}}%
      \expandafter\def\csname LT3\endcsname{\color[rgb]{1,0,1}}%
      \expandafter\def\csname LT4\endcsname{\color[rgb]{0,1,1}}%
      \expandafter\def\csname LT5\endcsname{\color[rgb]{1,1,0}}%
      \expandafter\def\csname LT6\endcsname{\color[rgb]{0,0,0}}%
      \expandafter\def\csname LT7\endcsname{\color[rgb]{1,0.3,0}}%
      \expandafter\def\csname LT8\endcsname{\color[rgb]{0.5,0.5,0.5}}%
    \else
      \def\colorrgb#1{\color{black}}%
      \def\colorgray#1{\color[gray]{#1}}%
      \expandafter\def\csname LTw\endcsname{\color{white}}%
      \expandafter\def\csname LTb\endcsname{\color{black}}%
      \expandafter\def\csname LTa\endcsname{\color{black}}%
      \expandafter\def\csname LT0\endcsname{\color{black}}%
      \expandafter\def\csname LT1\endcsname{\color{black}}%
      \expandafter\def\csname LT2\endcsname{\color{black}}%
      \expandafter\def\csname LT3\endcsname{\color{black}}%
      \expandafter\def\csname LT4\endcsname{\color{black}}%
      \expandafter\def\csname LT5\endcsname{\color{black}}%
      \expandafter\def\csname LT6\endcsname{\color{black}}%
      \expandafter\def\csname LT7\endcsname{\color{black}}%
      \expandafter\def\csname LT8\endcsname{\color{black}}%
    \fi
  \fi
  \setlength{\unitlength}{0.0500bp}%
  \begin{picture}(6236.00,3968.00)%
    \gplgaddtomacro\gplbacktext{%
      \csname LTb\endcsname%
      \put(1220,640){\makebox(0,0)[r]{\strut{}$10^{-12}$}}%
      \put(1220,983){\makebox(0,0)[r]{\strut{}$10^{-10}$}}%
      \put(1220,1326){\makebox(0,0)[r]{\strut{}$10^{-8}$}}%
      \put(1220,1669){\makebox(0,0)[r]{\strut{}$10^{-6}$}}%
      \put(1220,2012){\makebox(0,0)[r]{\strut{}$10^{-4}$}}%
      \put(1220,2355){\makebox(0,0)[r]{\strut{}$10^{-2}$}}%
      \put(1220,2698){\makebox(0,0)[r]{\strut{}$1$}}%
      \put(1220,3041){\makebox(0,0)[r]{\strut{}$10^{2}$}}%
      \put(1220,3384){\makebox(0,0)[r]{\strut{}$10^{4}$}}%
      \put(1220,3727){\makebox(0,0)[r]{\strut{}$10^{6}$}}%
      \put(1340,440){\makebox(0,0){\strut{} 0}}%
      \put(2247,440){\makebox(0,0){\strut{} 30000}}%
      \put(3154,440){\makebox(0,0){\strut{} 60000}}%
      \put(4061,440){\makebox(0,0){\strut{} 90000}}%
      \put(4968,440){\makebox(0,0){\strut{} 120000}}%
      \put(5875,440){\makebox(0,0){\strut{} 150000}}%
      \put(160,2183){\rotatebox{-270}{\makebox(0,0){\strut{}residual}}}%
      \put(3607,140){\makebox(0,0){\strut{}Calls to Kernel operator}}%
    }%
    \gplgaddtomacro\gplfronttext{%
      \csname LTb\endcsname%
      \put(4972,3564){\makebox(0,0)[r]{\strut{}GMRES(relCG)}}%
      \csname LTb\endcsname%
      \put(4972,3364){\makebox(0,0)[r]{\strut{}GMRES(calculating(15) relCG 3)}}%
      \csname LTb\endcsname%
      \put(4972,3164){\makebox(0,0)[r]{\strut{}GMRES(calculating(30) relCG 3)}}%
      \csname LTb\endcsname%
      \put(4972,2964){\makebox(0,0)[r]{\strut{}GMRES(calculating(45) relCG 3)}}%
    }%
    \gplbacktext
    \put(0,0){\includegraphics{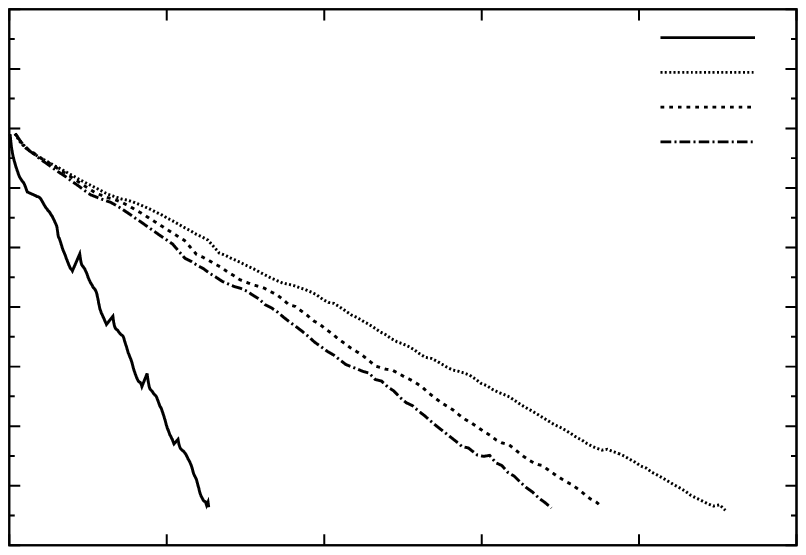}}%
    \gplfronttext
  \end{picture}%
\endgroup

%% file: figs/s8t32m0.03_c55_n30GMRESCGiterationsComparison2.tex
\begingroup
  \makeatletter
  \providecommand\color[2][]{%
    \GenericError{(gnuplot) \space\space\space\@spaces}{%
      Package color not loaded in conjunction with
      terminal option `colourtext'%
    }{See the gnuplot documentation for explanation.%
    }{Either use 'blacktext' in gnuplot or load the package
      color.sty in LaTeX.}%
    \renewcommand\color[2][]{}%
  }%
  \providecommand\includegraphics[2][]{%
    \GenericError{(gnuplot) \space\space\space\@spaces}{%
      Package graphicx or graphics not loaded%
    }{See the gnuplot documentation for explanation.%
    }{The gnuplot epslatex terminal needs graphicx.sty or graphics.sty.}%
    \renewcommand\includegraphics[2][]{}%
  }%
  \providecommand\rotatebox[2]{#2}%
  \@ifundefined{ifGPcolor}{%
    \newif\ifGPcolor
    \GPcolorfalse
  }{}%
  \@ifundefined{ifGPblacktext}{%
    \newif\ifGPblacktext
    \GPblacktexttrue
  }{}%
  \let\gplgaddtomacro\g@addto@macro
  \gdef\gplbacktext{}%
  \gdef\gplfronttext{}%
  \makeatother
  \ifGPblacktext
    \def\colorrgb#1{}%
    \def\colorgray#1{}%
  \else
    \ifGPcolor
      \def\colorrgb#1{\color[rgb]{#1}}%
      \def\colorgray#1{\color[gray]{#1}}%
      \expandafter\def\csname LTw\endcsname{\color{white}}%
      \expandafter\def\csname LTb\endcsname{\color{black}}%
      \expandafter\def\csname LTa\endcsname{\color{black}}%
      \expandafter\def\csname LT0\endcsname{\color[rgb]{1,0,0}}%
      \expandafter\def\csname LT1\endcsname{\color[rgb]{0,1,0}}%
      \expandafter\def\csname LT2\endcsname{\color[rgb]{0,0,1}}%
      \expandafter\def\csname LT3\endcsname{\color[rgb]{1,0,1}}%
      \expandafter\def\csname LT4\endcsname{\color[rgb]{0,1,1}}%
      \expandafter\def\csname LT5\endcsname{\color[rgb]{1,1,0}}%
      \expandafter\def\csname LT6\endcsname{\color[rgb]{0,0,0}}%
      \expandafter\def\csname LT7\endcsname{\color[rgb]{1,0.3,0}}%
      \expandafter\def\csname LT8\endcsname{\color[rgb]{0.5,0.5,0.5}}%
    \else
      \def\colorrgb#1{\color{black}}%
      \def\colorgray#1{\color[gray]{#1}}%
      \expandafter\def\csname LTw\endcsname{\color{white}}%
      \expandafter\def\csname LTb\endcsname{\color{black}}%
      \expandafter\def\csname LTa\endcsname{\color{black}}%
      \expandafter\def\csname LT0\endcsname{\color{black}}%
      \expandafter\def\csname LT1\endcsname{\color{black}}%
      \expandafter\def\csname LT2\endcsname{\color{black}}%
      \expandafter\def\csname LT3\endcsname{\color{black}}%
      \expandafter\def\csname LT4\endcsname{\color{black}}%
      \expandafter\def\csname LT5\endcsname{\color{black}}%
      \expandafter\def\csname LT6\endcsname{\color{black}}%
      \expandafter\def\csname LT7\endcsname{\color{black}}%
      \expandafter\def\csname LT8\endcsname{\color{black}}%
    \fi
  \fi
  \setlength{\unitlength}{0.0500bp}%
  \begin{picture}(6236.00,3968.00)%
    \gplgaddtomacro\gplbacktext{%
      \csname LTb\endcsname%
      \put(1220,640){\makebox(0,0)[r]{\strut{}$10^{-12}$}}%
      \put(1220,983){\makebox(0,0)[r]{\strut{}$10^{-10}$}}%
      \put(1220,1326){\makebox(0,0)[r]{\strut{}$10^{-8}$}}%
      \put(1220,1669){\makebox(0,0)[r]{\strut{}$10^{-6}$}}%
      \put(1220,2012){\makebox(0,0)[r]{\strut{}$10^{-4}$}}%
      \put(1220,2355){\makebox(0,0)[r]{\strut{}$10^{-2}$}}%
      \put(1220,2698){\makebox(0,0)[r]{\strut{}$1$}}%
      \put(1220,3041){\makebox(0,0)[r]{\strut{}$10^{2}$}}%
      \put(1220,3384){\makebox(0,0)[r]{\strut{}$10^{4}$}}%
      \put(1220,3727){\makebox(0,0)[r]{\strut{}$10^{6}$}}%
      \put(1340,440){\makebox(0,0){\strut{} 0}}%
      \put(2096,440){\makebox(0,0){\strut{} 50}}%
      \put(2852,440){\makebox(0,0){\strut{} 100}}%
      \put(3608,440){\makebox(0,0){\strut{} 150}}%
      \put(4363,440){\makebox(0,0){\strut{} 200}}%
      \put(5119,440){\makebox(0,0){\strut{} 250}}%
      \put(5875,440){\makebox(0,0){\strut{} 300}}%
      \put(160,2183){\rotatebox{-270}{\makebox(0,0){\strut{}residual}}}%
      \put(3607,140){\makebox(0,0){\strut{}CG iterations}}%
    }%
    \gplgaddtomacro\gplfronttext{%
      \csname LTb\endcsname%
      \put(4972,3564){\makebox(0,0)[r]{\strut{}GMRES(relCG)}}%
      \csname LTb\endcsname%
      \put(4972,3364){\makebox(0,0)[r]{\strut{}GMRES(calculating(15) relCG 3)}}%
      \csname LTb\endcsname%
      \put(4972,3164){\makebox(0,0)[r]{\strut{}GMRES(calculating(30) relCG 3)}}%
      \csname LTb\endcsname%
      \put(4972,2964){\makebox(0,0)[r]{\strut{}GMRES(calculating(45) relCG 3)}}%
    }%
    \gplbacktext
    \put(0,0){\includegraphics{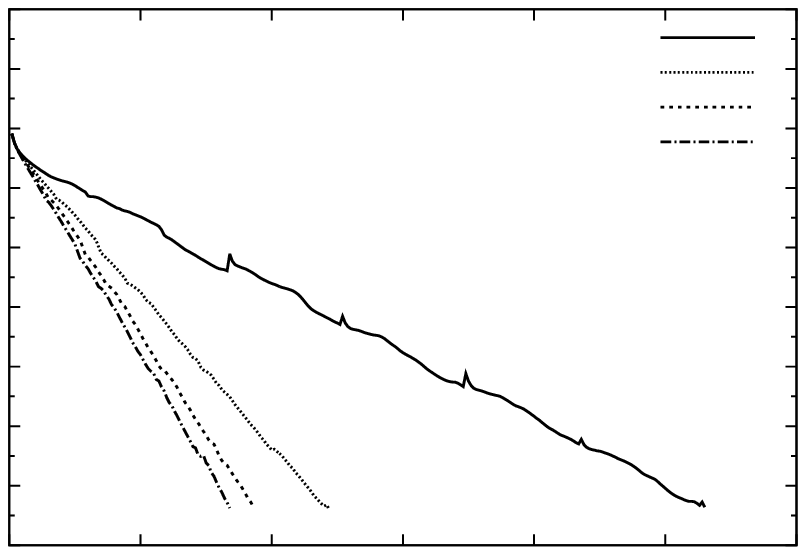}}%
    \gplfronttext
  \end{picture}%
\endgroup

%% file: figs/s8t32m0.03_c55_n30GMRESSUMRComparison2.tex
\begingroup
  \makeatletter
  \providecommand\color[2][]{%
    \GenericError{(gnuplot) \space\space\space\@spaces}{%
      Package color not loaded in conjunction with
      terminal option `colourtext'%
    }{See the gnuplot documentation for explanation.%
    }{Either use 'blacktext' in gnuplot or load the package
      color.sty in LaTeX.}%
    \renewcommand\color[2][]{}%
  }%
  \providecommand\includegraphics[2][]{%
    \GenericError{(gnuplot) \space\space\space\@spaces}{%
      Package graphicx or graphics not loaded%
    }{See the gnuplot documentation for explanation.%
    }{The gnuplot epslatex terminal needs graphicx.sty or graphics.sty.}%
    \renewcommand\includegraphics[2][]{}%
  }%
  \providecommand\rotatebox[2]{#2}%
  \@ifundefined{ifGPcolor}{%
    \newif\ifGPcolor
    \GPcolorfalse
  }{}%
  \@ifundefined{ifGPblacktext}{%
    \newif\ifGPblacktext
    \GPblacktexttrue
  }{}%
  \let\gplgaddtomacro\g@addto@macro
  \gdef\gplbacktext{}%
  \gdef\gplfronttext{}%
  \makeatother
  \ifGPblacktext
    \def\colorrgb#1{}%
    \def\colorgray#1{}%
  \else
    \ifGPcolor
      \def\colorrgb#1{\color[rgb]{#1}}%
      \def\colorgray#1{\color[gray]{#1}}%
      \expandafter\def\csname LTw\endcsname{\color{white}}%
      \expandafter\def\csname LTb\endcsname{\color{black}}%
      \expandafter\def\csname LTa\endcsname{\color{black}}%
      \expandafter\def\csname LT0\endcsname{\color[rgb]{1,0,0}}%
      \expandafter\def\csname LT1\endcsname{\color[rgb]{0,1,0}}%
      \expandafter\def\csname LT2\endcsname{\color[rgb]{0,0,1}}%
      \expandafter\def\csname LT3\endcsname{\color[rgb]{1,0,1}}%
      \expandafter\def\csname LT4\endcsname{\color[rgb]{0,1,1}}%
      \expandafter\def\csname LT5\endcsname{\color[rgb]{1,1,0}}%
      \expandafter\def\csname LT6\endcsname{\color[rgb]{0,0,0}}%
      \expandafter\def\csname LT7\endcsname{\color[rgb]{1,0.3,0}}%
      \expandafter\def\csname LT8\endcsname{\color[rgb]{0.5,0.5,0.5}}%
    \else
      \def\colorrgb#1{\color{black}}%
      \def\colorgray#1{\color[gray]{#1}}%
      \expandafter\def\csname LTw\endcsname{\color{white}}%
      \expandafter\def\csname LTb\endcsname{\color{black}}%
      \expandafter\def\csname LTa\endcsname{\color{black}}%
      \expandafter\def\csname LT0\endcsname{\color{black}}%
      \expandafter\def\csname LT1\endcsname{\color{black}}%
      \expandafter\def\csname LT2\endcsname{\color{black}}%
      \expandafter\def\csname LT3\endcsname{\color{black}}%
      \expandafter\def\csname LT4\endcsname{\color{black}}%
      \expandafter\def\csname LT5\endcsname{\color{black}}%
      \expandafter\def\csname LT6\endcsname{\color{black}}%
      \expandafter\def\csname LT7\endcsname{\color{black}}%
      \expandafter\def\csname LT8\endcsname{\color{black}}%
    \fi
  \fi
  \setlength{\unitlength}{0.0500bp}%
  \begin{picture}(6236.00,3968.00)%
    \gplgaddtomacro\gplbacktext{%
      \csname LTb\endcsname%
      \put(1220,640){\makebox(0,0)[r]{\strut{}$10^{-12}$}}%
      \put(1220,983){\makebox(0,0)[r]{\strut{}$10^{-10}$}}%
      \put(1220,1326){\makebox(0,0)[r]{\strut{}$10^{-8}$}}%
      \put(1220,1669){\makebox(0,0)[r]{\strut{}$10^{-6}$}}%
      \put(1220,2012){\makebox(0,0)[r]{\strut{}$10^{-4}$}}%
      \put(1220,2355){\makebox(0,0)[r]{\strut{}$10^{-2}$}}%
      \put(1220,2698){\makebox(0,0)[r]{\strut{}$1$}}%
      \put(1220,3041){\makebox(0,0)[r]{\strut{}$10^{2}$}}%
      \put(1220,3384){\makebox(0,0)[r]{\strut{}$10^{4}$}}%
      \put(1220,3727){\makebox(0,0)[r]{\strut{}$10^{6}$}}%
      \put(1340,440){\makebox(0,0){\strut{} 0}}%
      \put(2474,440){\makebox(0,0){\strut{} 30000}}%
      \put(3608,440){\makebox(0,0){\strut{} 60000}}%
      \put(4741,440){\makebox(0,0){\strut{} 90000}}%
      \put(5875,440){\makebox(0,0){\strut{} 120000}}%
      \put(160,2183){\rotatebox{-270}{\makebox(0,0){\strut{}residual}}}%
      \put(3607,140){\makebox(0,0){\strut{}Calls to Kernel operator}}%
    }%
    \gplgaddtomacro\gplfronttext{%
      \csname LTb\endcsname%
      \put(4972,3564){\makebox(0,0)[r]{\strut{}GMRES(relSUMR)}}%
      \csname LTb\endcsname%
      \put(4972,3364){\makebox(0,0)[r]{\strut{}GMRES(calculating(15) relSUMR 3)}}%
      \csname LTb\endcsname%
      \put(4972,3164){\makebox(0,0)[r]{\strut{}GMRES(calculating(30) relSUMR 3)}}%
      \csname LTb\endcsname%
      \put(4972,2964){\makebox(0,0)[r]{\strut{}GMRES(calculating(45) relSUMR 3)}}%
    }%
    \gplbacktext
    \put(0,0){\includegraphics{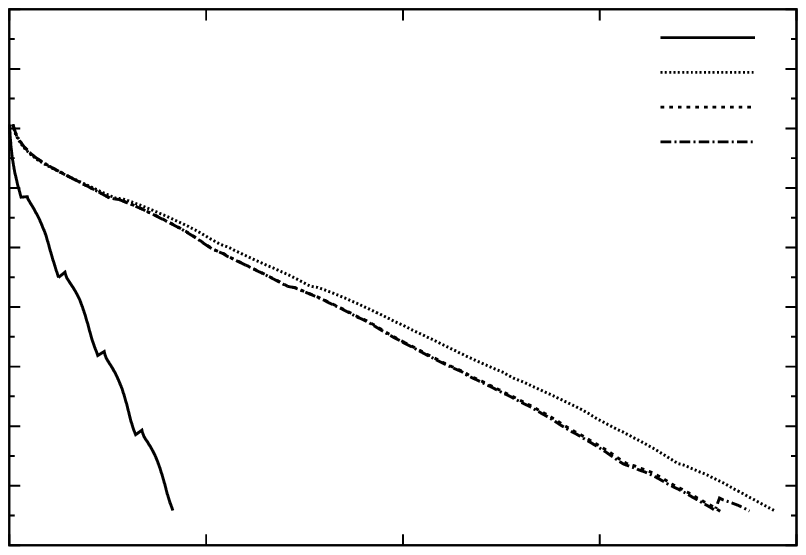}}%
    \gplfronttext
  \end{picture}%
\endgroup

%% file: figs/s8t32m0.03_c55_n30GMRESSUMRiterationsComparison2.tex
\begingroup
  \makeatletter
  \providecommand\color[2][]{%
    \GenericError{(gnuplot) \space\space\space\@spaces}{%
      Package color not loaded in conjunction with
      terminal option `colourtext'%
    }{See the gnuplot documentation for explanation.%
    }{Either use 'blacktext' in gnuplot or load the package
      color.sty in LaTeX.}%
    \renewcommand\color[2][]{}%
  }%
  \providecommand\includegraphics[2][]{%
    \GenericError{(gnuplot) \space\space\space\@spaces}{%
      Package graphicx or graphics not loaded%
    }{See the gnuplot documentation for explanation.%
    }{The gnuplot epslatex terminal needs graphicx.sty or graphics.sty.}%
    \renewcommand\includegraphics[2][]{}%
  }%
  \providecommand\rotatebox[2]{#2}%
  \@ifundefined{ifGPcolor}{%
    \newif\ifGPcolor
    \GPcolorfalse
  }{}%
  \@ifundefined{ifGPblacktext}{%
    \newif\ifGPblacktext
    \GPblacktexttrue
  }{}%
  \let\gplgaddtomacro\g@addto@macro
  \gdef\gplbacktext{}%
  \gdef\gplfronttext{}%
  \makeatother
  \ifGPblacktext
    \def\colorrgb#1{}%
    \def\colorgray#1{}%
  \else
    \ifGPcolor
      \def\colorrgb#1{\color[rgb]{#1}}%
      \def\colorgray#1{\color[gray]{#1}}%
      \expandafter\def\csname LTw\endcsname{\color{white}}%
      \expandafter\def\csname LTb\endcsname{\color{black}}%
      \expandafter\def\csname LTa\endcsname{\color{black}}%
      \expandafter\def\csname LT0\endcsname{\color[rgb]{1,0,0}}%
      \expandafter\def\csname LT1\endcsname{\color[rgb]{0,1,0}}%
      \expandafter\def\csname LT2\endcsname{\color[rgb]{0,0,1}}%
      \expandafter\def\csname LT3\endcsname{\color[rgb]{1,0,1}}%
      \expandafter\def\csname LT4\endcsname{\color[rgb]{0,1,1}}%
      \expandafter\def\csname LT5\endcsname{\color[rgb]{1,1,0}}%
      \expandafter\def\csname LT6\endcsname{\color[rgb]{0,0,0}}%
      \expandafter\def\csname LT7\endcsname{\color[rgb]{1,0.3,0}}%
      \expandafter\def\csname LT8\endcsname{\color[rgb]{0.5,0.5,0.5}}%
    \else
      \def\colorrgb#1{\color{black}}%
      \def\colorgray#1{\color[gray]{#1}}%
      \expandafter\def\csname LTw\endcsname{\color{white}}%
      \expandafter\def\csname LTb\endcsname{\color{black}}%
      \expandafter\def\csname LTa\endcsname{\color{black}}%
      \expandafter\def\csname LT0\endcsname{\color{black}}%
      \expandafter\def\csname LT1\endcsname{\color{black}}%
      \expandafter\def\csname LT2\endcsname{\color{black}}%
      \expandafter\def\csname LT3\endcsname{\color{black}}%
      \expandafter\def\csname LT4\endcsname{\color{black}}%
      \expandafter\def\csname LT5\endcsname{\color{black}}%
      \expandafter\def\csname LT6\endcsname{\color{black}}%
      \expandafter\def\csname LT7\endcsname{\color{black}}%
      \expandafter\def\csname LT8\endcsname{\color{black}}%
    \fi
  \fi
  \setlength{\unitlength}{0.0500bp}%
  \begin{picture}(6236.00,3968.00)%
    \gplgaddtomacro\gplbacktext{%
      \csname LTb\endcsname%
      \put(1220,640){\makebox(0,0)[r]{\strut{}$10^{-12}$}}%
      \put(1220,983){\makebox(0,0)[r]{\strut{}$10^{-10}$}}%
      \put(1220,1326){\makebox(0,0)[r]{\strut{}$10^{-8}$}}%
      \put(1220,1669){\makebox(0,0)[r]{\strut{}$10^{-6}$}}%
      \put(1220,2012){\makebox(0,0)[r]{\strut{}$10^{-4}$}}%
      \put(1220,2355){\makebox(0,0)[r]{\strut{}$10^{-2}$}}%
      \put(1220,2698){\makebox(0,0)[r]{\strut{}$1$}}%
      \put(1220,3041){\makebox(0,0)[r]{\strut{}$10^{2}$}}%
      \put(1220,3384){\makebox(0,0)[r]{\strut{}$10^{4}$}}%
      \put(1220,3727){\makebox(0,0)[r]{\strut{}$10^{6}$}}%
      \put(1340,440){\makebox(0,0){\strut{} 0}}%
      \put(1988,440){\makebox(0,0){\strut{} 50}}%
      \put(2636,440){\makebox(0,0){\strut{} 100}}%
      \put(3284,440){\makebox(0,0){\strut{} 150}}%
      \put(3931,440){\makebox(0,0){\strut{} 200}}%
      \put(4579,440){\makebox(0,0){\strut{} 250}}%
      \put(5227,440){\makebox(0,0){\strut{} 300}}%
      \put(5875,440){\makebox(0,0){\strut{} 350}}%
      \put(160,2183){\rotatebox{-270}{\makebox(0,0){\strut{}residual}}}%
      \put(3607,140){\makebox(0,0){\strut{}SUMR iterations}}%
    }%
    \gplgaddtomacro\gplfronttext{%
      \csname LTb\endcsname%
      \put(4972,3564){\makebox(0,0)[r]{\strut{}GMRES(relSUMR)}}%
      \csname LTb\endcsname%
      \put(4972,3364){\makebox(0,0)[r]{\strut{}GMRES(calculating(15) relSUMR 3)}}%
      \csname LTb\endcsname%
      \put(4972,3164){\makebox(0,0)[r]{\strut{}GMRES(calculating(30) relSUMR 3)}}%
      \csname LTb\endcsname%
      \put(4972,2964){\makebox(0,0)[r]{\strut{}GMRES(calculating(45) relSUMR 3)}}%
    }%
    \gplbacktext
    \put(0,0){\includegraphics{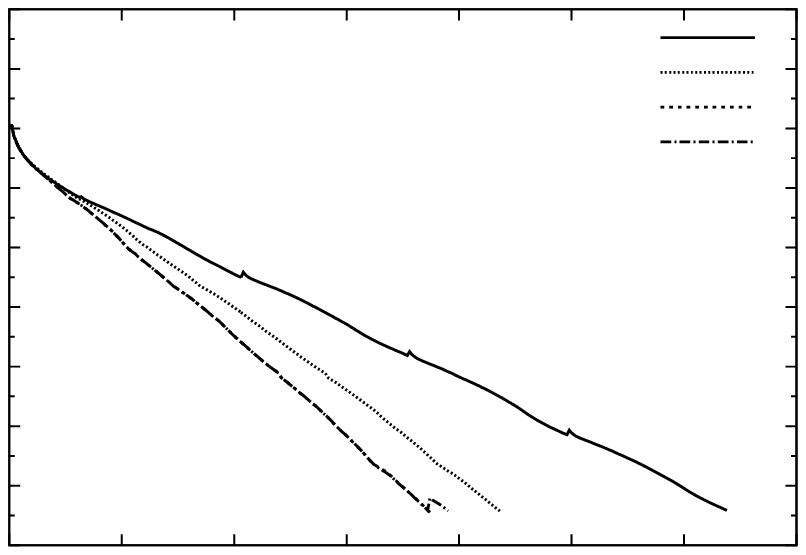}}%
    \gplfronttext
  \end{picture}%
\endgroup

%% file: figs/s8t32m0.03_c55_n30GMRESCGComparison3.tex
\begingroup
  \makeatletter
  \providecommand\color[2][]{%
    \GenericError{(gnuplot) \space\space\space\@spaces}{%
      Package color not loaded in conjunction with
      terminal option `colourtext'%
    }{See the gnuplot documentation for explanation.%
    }{Either use 'blacktext' in gnuplot or load the package
      color.sty in LaTeX.}%
    \renewcommand\color[2][]{}%
  }%
  \providecommand\includegraphics[2][]{%
    \GenericError{(gnuplot) \space\space\space\@spaces}{%
      Package graphicx or graphics not loaded%
    }{See the gnuplot documentation for explanation.%
    }{The gnuplot epslatex terminal needs graphicx.sty or graphics.sty.}%
    \renewcommand\includegraphics[2][]{}%
  }%
  \providecommand\rotatebox[2]{#2}%
  \@ifundefined{ifGPcolor}{%
    \newif\ifGPcolor
    \GPcolorfalse
  }{}%
  \@ifundefined{ifGPblacktext}{%
    \newif\ifGPblacktext
    \GPblacktexttrue
  }{}%
  \let\gplgaddtomacro\g@addto@macro
  \gdef\gplbacktext{}%
  \gdef\gplfronttext{}%
  \makeatother
  \ifGPblacktext
    \def\colorrgb#1{}%
    \def\colorgray#1{}%
  \else
    \ifGPcolor
      \def\colorrgb#1{\color[rgb]{#1}}%
      \def\colorgray#1{\color[gray]{#1}}%
      \expandafter\def\csname LTw\endcsname{\color{white}}%
      \expandafter\def\csname LTb\endcsname{\color{black}}%
      \expandafter\def\csname LTa\endcsname{\color{black}}%
      \expandafter\def\csname LT0\endcsname{\color[rgb]{1,0,0}}%
      \expandafter\def\csname LT1\endcsname{\color[rgb]{0,1,0}}%
      \expandafter\def\csname LT2\endcsname{\color[rgb]{0,0,1}}%
      \expandafter\def\csname LT3\endcsname{\color[rgb]{1,0,1}}%
      \expandafter\def\csname LT4\endcsname{\color[rgb]{0,1,1}}%
      \expandafter\def\csname LT5\endcsname{\color[rgb]{1,1,0}}%
      \expandafter\def\csname LT6\endcsname{\color[rgb]{0,0,0}}%
      \expandafter\def\csname LT7\endcsname{\color[rgb]{1,0.3,0}}%
      \expandafter\def\csname LT8\endcsname{\color[rgb]{0.5,0.5,0.5}}%
    \else
      \def\colorrgb#1{\color{black}}%
      \def\colorgray#1{\color[gray]{#1}}%
      \expandafter\def\csname LTw\endcsname{\color{white}}%
      \expandafter\def\csname LTb\endcsname{\color{black}}%
      \expandafter\def\csname LTa\endcsname{\color{black}}%
      \expandafter\def\csname LT0\endcsname{\color{black}}%
      \expandafter\def\csname LT1\endcsname{\color{black}}%
      \expandafter\def\csname LT2\endcsname{\color{black}}%
      \expandafter\def\csname LT3\endcsname{\color{black}}%
      \expandafter\def\csname LT4\endcsname{\color{black}}%
      \expandafter\def\csname LT5\endcsname{\color{black}}%
      \expandafter\def\csname LT6\endcsname{\color{black}}%
      \expandafter\def\csname LT7\endcsname{\color{black}}%
      \expandafter\def\csname LT8\endcsname{\color{black}}%
    \fi
  \fi
  \setlength{\unitlength}{0.0500bp}%
  \begin{picture}(6236.00,3968.00)%
    \gplgaddtomacro\gplbacktext{%
      \csname LTb\endcsname%
      \put(1220,640){\makebox(0,0)[r]{\strut{}$10^{-12}$}}%
      \put(1220,1026){\makebox(0,0)[r]{\strut{}$10^{-10}$}}%
      \put(1220,1412){\makebox(0,0)[r]{\strut{}$10^{-8}$}}%
      \put(1220,1798){\makebox(0,0)[r]{\strut{}$10^{-6}$}}%
      \put(1220,2184){\makebox(0,0)[r]{\strut{}$10^{-4}$}}%
      \put(1220,2569){\makebox(0,0)[r]{\strut{}$10^{-2}$}}%
      \put(1220,2955){\makebox(0,0)[r]{\strut{}$1$}}%
      \put(1220,3341){\makebox(0,0)[r]{\strut{}$10^{2}$}}%
      \put(1220,3727){\makebox(0,0)[r]{\strut{}$10^{4}$}}%
      \put(1340,440){\makebox(0,0){\strut{} 0}}%
      \put(2474,440){\makebox(0,0){\strut{} 50000}}%
      \put(3608,440){\makebox(0,0){\strut{} 100000}}%
      \put(4741,440){\makebox(0,0){\strut{} 150000}}%
      \put(5875,440){\makebox(0,0){\strut{} 200000}}%
      \put(160,2183){\rotatebox{-270}{\makebox(0,0){\strut{}residual}}}%
      \put(3607,140){\makebox(0,0){\strut{}Calls to Kernel operator}}%
    }%
    \gplgaddtomacro\gplfronttext{%
      \csname LTb\endcsname%
      \put(4972,3564){\makebox(0,0)[r]{\strut{}GMRES(relCG)}}%
      \csname LTb\endcsname%
      \put(4972,3364){\makebox(0,0)[r]{\strut{}GMRES(calculating(15) relCG 1)}}%
      \csname LTb\endcsname%
      \put(4972,3164){\makebox(0,0)[r]{\strut{}GMRES(calculating(30) relCG 1)}}%
      \csname LTb\endcsname%
      \put(4972,2964){\makebox(0,0)[r]{\strut{}GMRES(calculating(45) relCG 1)}}%
    }%
    \gplbacktext
    \put(0,0){\includegraphics{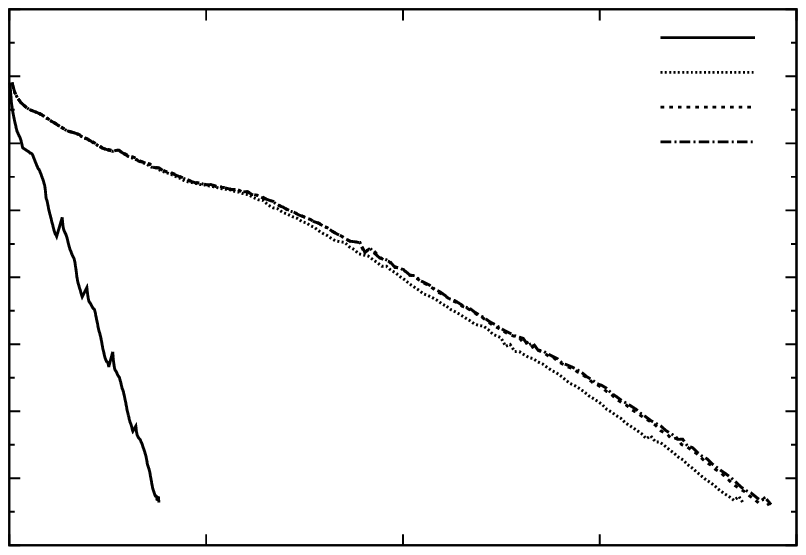}}%
    \gplfronttext
  \end{picture}%
\endgroup

%% file: figs/s8t32m0.03_c55_n30GMRESCGiterationsComparison3.tex
\begingroup
  \makeatletter
  \providecommand\color[2][]{%
    \GenericError{(gnuplot) \space\space\space\@spaces}{%
      Package color not loaded in conjunction with
      terminal option `colourtext'%
    }{See the gnuplot documentation for explanation.%
    }{Either use 'blacktext' in gnuplot or load the package
      color.sty in LaTeX.}%
    \renewcommand\color[2][]{}%
  }%
  \providecommand\includegraphics[2][]{%
    \GenericError{(gnuplot) \space\space\space\@spaces}{%
      Package graphicx or graphics not loaded%
    }{See the gnuplot documentation for explanation.%
    }{The gnuplot epslatex terminal needs graphicx.sty or graphics.sty.}%
    \renewcommand\includegraphics[2][]{}%
  }%
  \providecommand\rotatebox[2]{#2}%
  \@ifundefined{ifGPcolor}{%
    \newif\ifGPcolor
    \GPcolorfalse
  }{}%
  \@ifundefined{ifGPblacktext}{%
    \newif\ifGPblacktext
    \GPblacktexttrue
  }{}%
  \let\gplgaddtomacro\g@addto@macro
  \gdef\gplbacktext{}%
  \gdef\gplfronttext{}%
  \makeatother
  \ifGPblacktext
    \def\colorrgb#1{}%
    \def\colorgray#1{}%
  \else
    \ifGPcolor
      \def\colorrgb#1{\color[rgb]{#1}}%
      \def\colorgray#1{\color[gray]{#1}}%
      \expandafter\def\csname LTw\endcsname{\color{white}}%
      \expandafter\def\csname LTb\endcsname{\color{black}}%
      \expandafter\def\csname LTa\endcsname{\color{black}}%
      \expandafter\def\csname LT0\endcsname{\color[rgb]{1,0,0}}%
      \expandafter\def\csname LT1\endcsname{\color[rgb]{0,1,0}}%
      \expandafter\def\csname LT2\endcsname{\color[rgb]{0,0,1}}%
      \expandafter\def\csname LT3\endcsname{\color[rgb]{1,0,1}}%
      \expandafter\def\csname LT4\endcsname{\color[rgb]{0,1,1}}%
      \expandafter\def\csname LT5\endcsname{\color[rgb]{1,1,0}}%
      \expandafter\def\csname LT6\endcsname{\color[rgb]{0,0,0}}%
      \expandafter\def\csname LT7\endcsname{\color[rgb]{1,0.3,0}}%
      \expandafter\def\csname LT8\endcsname{\color[rgb]{0.5,0.5,0.5}}%
    \else
      \def\colorrgb#1{\color{black}}%
      \def\colorgray#1{\color[gray]{#1}}%
      \expandafter\def\csname LTw\endcsname{\color{white}}%
      \expandafter\def\csname LTb\endcsname{\color{black}}%
      \expandafter\def\csname LTa\endcsname{\color{black}}%
      \expandafter\def\csname LT0\endcsname{\color{black}}%
      \expandafter\def\csname LT1\endcsname{\color{black}}%
      \expandafter\def\csname LT2\endcsname{\color{black}}%
      \expandafter\def\csname LT3\endcsname{\color{black}}%
      \expandafter\def\csname LT4\endcsname{\color{black}}%
      \expandafter\def\csname LT5\endcsname{\color{black}}%
      \expandafter\def\csname LT6\endcsname{\color{black}}%
      \expandafter\def\csname LT7\endcsname{\color{black}}%
      \expandafter\def\csname LT8\endcsname{\color{black}}%
    \fi
  \fi
  \setlength{\unitlength}{0.0500bp}%
  \begin{picture}(6236.00,3968.00)%
    \gplgaddtomacro\gplbacktext{%
      \csname LTb\endcsname%
      \put(1220,640){\makebox(0,0)[r]{\strut{}$10^{-12}$}}%
      \put(1220,1026){\makebox(0,0)[r]{\strut{}$10^{-10}$}}%
      \put(1220,1412){\makebox(0,0)[r]{\strut{}$10^{-8}$}}%
      \put(1220,1798){\makebox(0,0)[r]{\strut{}$10^{-6}$}}%
      \put(1220,2184){\makebox(0,0)[r]{\strut{}$10^{-4}$}}%
      \put(1220,2569){\makebox(0,0)[r]{\strut{}$10^{-2}$}}%
      \put(1220,2955){\makebox(0,0)[r]{\strut{}$1$}}%
      \put(1220,3341){\makebox(0,0)[r]{\strut{}$10^{2}$}}%
      \put(1220,3727){\makebox(0,0)[r]{\strut{}$10^{4}$}}%
      \put(1340,440){\makebox(0,0){\strut{} 0}}%
      \put(2096,440){\makebox(0,0){\strut{} 50}}%
      \put(2852,440){\makebox(0,0){\strut{} 100}}%
      \put(3608,440){\makebox(0,0){\strut{} 150}}%
      \put(4363,440){\makebox(0,0){\strut{} 200}}%
      \put(5119,440){\makebox(0,0){\strut{} 250}}%
      \put(5875,440){\makebox(0,0){\strut{} 300}}%
      \put(160,2183){\rotatebox{-270}{\makebox(0,0){\strut{}residual}}}%
      \put(3607,140){\makebox(0,0){\strut{}CG iterations}}%
    }%
    \gplgaddtomacro\gplfronttext{%
      \csname LTb\endcsname%
      \put(4972,3564){\makebox(0,0)[r]{\strut{}GMRES(relCG)}}%
      \csname LTb\endcsname%
      \put(4972,3364){\makebox(0,0)[r]{\strut{}GMRES(calculating(15) relCG 1)}}%
      \csname LTb\endcsname%
      \put(4972,3164){\makebox(0,0)[r]{\strut{}GMRES(calculating(30) relCG 1)}}%
      \csname LTb\endcsname%
      \put(4972,2964){\makebox(0,0)[r]{\strut{}GMRES(calculating(45) relCG 1)}}%
    }%
    \gplbacktext
    \put(0,0){\includegraphics{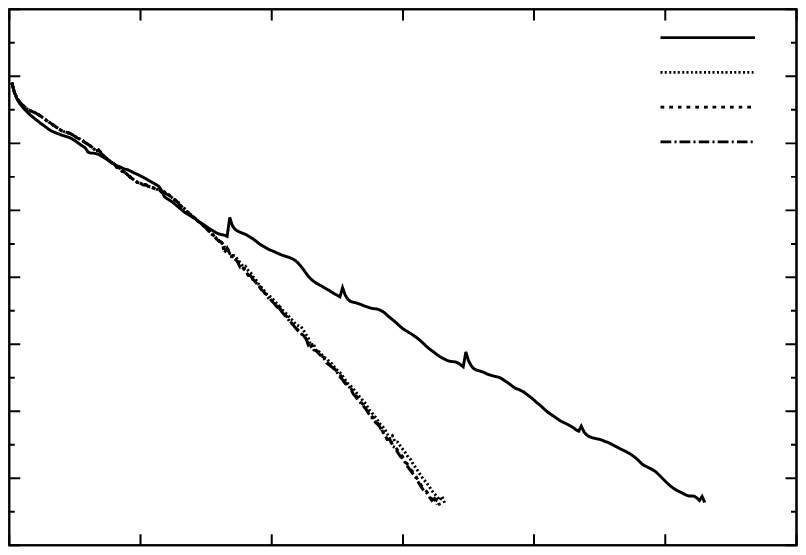}}%
    \gplfronttext
  \end{picture}%
\endgroup

%% file: figs/s8t32m0.03_c55_n30GMRESSUMRComparison3.tex
\begingroup
  \makeatletter
  \providecommand\color[2][]{%
    \GenericError{(gnuplot) \space\space\space\@spaces}{%
      Package color not loaded in conjunction with
      terminal option `colourtext'%
    }{See the gnuplot documentation for explanation.%
    }{Either use 'blacktext' in gnuplot or load the package
      color.sty in LaTeX.}%
    \renewcommand\color[2][]{}%
  }%
  \providecommand\includegraphics[2][]{%
    \GenericError{(gnuplot) \space\space\space\@spaces}{%
      Package graphicx or graphics not loaded%
    }{See the gnuplot documentation for explanation.%
    }{The gnuplot epslatex terminal needs graphicx.sty or graphics.sty.}%
    \renewcommand\includegraphics[2][]{}%
  }%
  \providecommand\rotatebox[2]{#2}%
  \@ifundefined{ifGPcolor}{%
    \newif\ifGPcolor
    \GPcolorfalse
  }{}%
  \@ifundefined{ifGPblacktext}{%
    \newif\ifGPblacktext
    \GPblacktexttrue
  }{}%
  \let\gplgaddtomacro\g@addto@macro
  \gdef\gplbacktext{}%
  \gdef\gplfronttext{}%
  \makeatother
  \ifGPblacktext
    \def\colorrgb#1{}%
    \def\colorgray#1{}%
  \else
    \ifGPcolor
      \def\colorrgb#1{\color[rgb]{#1}}%
      \def\colorgray#1{\color[gray]{#1}}%
      \expandafter\def\csname LTw\endcsname{\color{white}}%
      \expandafter\def\csname LTb\endcsname{\color{black}}%
      \expandafter\def\csname LTa\endcsname{\color{black}}%
      \expandafter\def\csname LT0\endcsname{\color[rgb]{1,0,0}}%
      \expandafter\def\csname LT1\endcsname{\color[rgb]{0,1,0}}%
      \expandafter\def\csname LT2\endcsname{\color[rgb]{0,0,1}}%
      \expandafter\def\csname LT3\endcsname{\color[rgb]{1,0,1}}%
      \expandafter\def\csname LT4\endcsname{\color[rgb]{0,1,1}}%
      \expandafter\def\csname LT5\endcsname{\color[rgb]{1,1,0}}%
      \expandafter\def\csname LT6\endcsname{\color[rgb]{0,0,0}}%
      \expandafter\def\csname LT7\endcsname{\color[rgb]{1,0.3,0}}%
      \expandafter\def\csname LT8\endcsname{\color[rgb]{0.5,0.5,0.5}}%
    \else
      \def\colorrgb#1{\color{black}}%
      \def\colorgray#1{\color[gray]{#1}}%
      \expandafter\def\csname LTw\endcsname{\color{white}}%
      \expandafter\def\csname LTb\endcsname{\color{black}}%
      \expandafter\def\csname LTa\endcsname{\color{black}}%
      \expandafter\def\csname LT0\endcsname{\color{black}}%
      \expandafter\def\csname LT1\endcsname{\color{black}}%
      \expandafter\def\csname LT2\endcsname{\color{black}}%
      \expandafter\def\csname LT3\endcsname{\color{black}}%
      \expandafter\def\csname LT4\endcsname{\color{black}}%
      \expandafter\def\csname LT5\endcsname{\color{black}}%
      \expandafter\def\csname LT6\endcsname{\color{black}}%
      \expandafter\def\csname LT7\endcsname{\color{black}}%
      \expandafter\def\csname LT8\endcsname{\color{black}}%
    \fi
  \fi
  \setlength{\unitlength}{0.0500bp}%
  \begin{picture}(6236.00,3968.00)%
    \gplgaddtomacro\gplbacktext{%
      \csname LTb\endcsname%
      \put(1220,640){\makebox(0,0)[r]{\strut{}$10^{-12}$}}%
      \put(1220,1003){\makebox(0,0)[r]{\strut{}$10^{-10}$}}%
      \put(1220,1366){\makebox(0,0)[r]{\strut{}$10^{-8}$}}%
      \put(1220,1730){\makebox(0,0)[r]{\strut{}$10^{-6}$}}%
      \put(1220,2093){\makebox(0,0)[r]{\strut{}$10^{-4}$}}%
      \put(1220,2456){\makebox(0,0)[r]{\strut{}$10^{-2}$}}%
      \put(1220,2819){\makebox(0,0)[r]{\strut{}$1$}}%
      \put(1220,3182){\makebox(0,0)[r]{\strut{}$10^{2}$}}%
      \put(1220,3545){\makebox(0,0)[r]{\strut{}$10^{4}$}}%
      \put(1340,440){\makebox(0,0){\strut{} 0}}%
      \put(1988,440){\makebox(0,0){\strut{} 30000}}%
      \put(2636,440){\makebox(0,0){\strut{} 60000}}%
      \put(3284,440){\makebox(0,0){\strut{} 90000}}%
      \put(3931,440){\makebox(0,0){\strut{} 120000}}%
      \put(4579,440){\makebox(0,0){\strut{} 150000}}%
      \put(5227,440){\makebox(0,0){\strut{} 180000}}%
      \put(5875,440){\makebox(0,0){\strut{} 210000}}%
      \put(160,2183){\rotatebox{-270}{\makebox(0,0){\strut{}residual}}}%
      \put(3607,140){\makebox(0,0){\strut{}Calls to Kernel operator}}%
    }%
    \gplgaddtomacro\gplfronttext{%
      \csname LTb\endcsname%
      \put(4972,3564){\makebox(0,0)[r]{\strut{}GMRES(relSUMR)}}%
      \csname LTb\endcsname%
      \put(4972,3364){\makebox(0,0)[r]{\strut{}GMRES(calculating(15) relSUMR 1)}}%
      \csname LTb\endcsname%
      \put(4972,3164){\makebox(0,0)[r]{\strut{}GMRES(calculating(30) relSUMR 1)}}%
      \csname LTb\endcsname%
      \put(4972,2964){\makebox(0,0)[r]{\strut{}GMRES(calculating(45) relSUMR 1)}}%
    }%
    \gplbacktext
    \put(0,0){\includegraphics{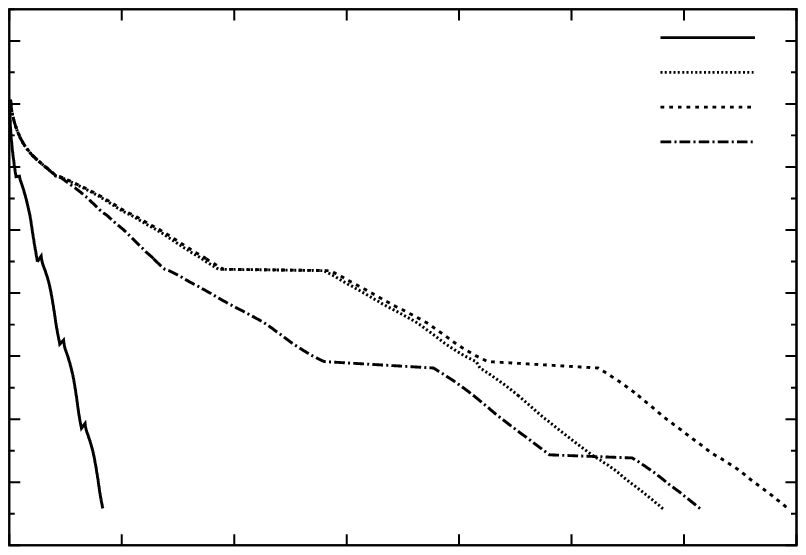}}%
    \gplfronttext
  \end{picture}%
\endgroup

%% file: figs/s8t32m0.03_c55_n30GMRESSUMRiterationsComparison3.tex
\begingroup
  \makeatletter
  \providecommand\color[2][]{%
    \GenericError{(gnuplot) \space\space\space\@spaces}{%
      Package color not loaded in conjunction with
      terminal option `colourtext'%
    }{See the gnuplot documentation for explanation.%
    }{Either use 'blacktext' in gnuplot or load the package
      color.sty in LaTeX.}%
    \renewcommand\color[2][]{}%
  }%
  \providecommand\includegraphics[2][]{%
    \GenericError{(gnuplot) \space\space\space\@spaces}{%
      Package graphicx or graphics not loaded%
    }{See the gnuplot documentation for explanation.%
    }{The gnuplot epslatex terminal needs graphicx.sty or graphics.sty.}%
    \renewcommand\includegraphics[2][]{}%
  }%
  \providecommand\rotatebox[2]{#2}%
  \@ifundefined{ifGPcolor}{%
    \newif\ifGPcolor
    \GPcolorfalse
  }{}%
  \@ifundefined{ifGPblacktext}{%
    \newif\ifGPblacktext
    \GPblacktexttrue
  }{}%
  \let\gplgaddtomacro\g@addto@macro
  \gdef\gplbacktext{}%
  \gdef\gplfronttext{}%
  \makeatother
  \ifGPblacktext
    \def\colorrgb#1{}%
    \def\colorgray#1{}%
  \else
    \ifGPcolor
      \def\colorrgb#1{\color[rgb]{#1}}%
      \def\colorgray#1{\color[gray]{#1}}%
      \expandafter\def\csname LTw\endcsname{\color{white}}%
      \expandafter\def\csname LTb\endcsname{\color{black}}%
      \expandafter\def\csname LTa\endcsname{\color{black}}%
      \expandafter\def\csname LT0\endcsname{\color[rgb]{1,0,0}}%
      \expandafter\def\csname LT1\endcsname{\color[rgb]{0,1,0}}%
      \expandafter\def\csname LT2\endcsname{\color[rgb]{0,0,1}}%
      \expandafter\def\csname LT3\endcsname{\color[rgb]{1,0,1}}%
      \expandafter\def\csname LT4\endcsname{\color[rgb]{0,1,1}}%
      \expandafter\def\csname LT5\endcsname{\color[rgb]{1,1,0}}%
      \expandafter\def\csname LT6\endcsname{\color[rgb]{0,0,0}}%
      \expandafter\def\csname LT7\endcsname{\color[rgb]{1,0.3,0}}%
      \expandafter\def\csname LT8\endcsname{\color[rgb]{0.5,0.5,0.5}}%
    \else
      \def\colorrgb#1{\color{black}}%
      \def\colorgray#1{\color[gray]{#1}}%
      \expandafter\def\csname LTw\endcsname{\color{white}}%
      \expandafter\def\csname LTb\endcsname{\color{black}}%
      \expandafter\def\csname LTa\endcsname{\color{black}}%
      \expandafter\def\csname LT0\endcsname{\color{black}}%
      \expandafter\def\csname LT1\endcsname{\color{black}}%
      \expandafter\def\csname LT2\endcsname{\color{black}}%
      \expandafter\def\csname LT3\endcsname{\color{black}}%
      \expandafter\def\csname LT4\endcsname{\color{black}}%
      \expandafter\def\csname LT5\endcsname{\color{black}}%
      \expandafter\def\csname LT6\endcsname{\color{black}}%
      \expandafter\def\csname LT7\endcsname{\color{black}}%
      \expandafter\def\csname LT8\endcsname{\color{black}}%
    \fi
  \fi
  \setlength{\unitlength}{0.0500bp}%
  \begin{picture}(6236.00,3968.00)%
    \gplgaddtomacro\gplbacktext{%
      \csname LTb\endcsname%
      \put(1220,640){\makebox(0,0)[r]{\strut{}$10^{-12}$}}%
      \put(1220,1003){\makebox(0,0)[r]{\strut{}$10^{-10}$}}%
      \put(1220,1366){\makebox(0,0)[r]{\strut{}$10^{-8}$}}%
      \put(1220,1730){\makebox(0,0)[r]{\strut{}$10^{-6}$}}%
      \put(1220,2093){\makebox(0,0)[r]{\strut{}$10^{-4}$}}%
      \put(1220,2456){\makebox(0,0)[r]{\strut{}$10^{-2}$}}%
      \put(1220,2819){\makebox(0,0)[r]{\strut{}$1$}}%
      \put(1220,3182){\makebox(0,0)[r]{\strut{}$10^{2}$}}%
      \put(1220,3545){\makebox(0,0)[r]{\strut{}$10^{4}$}}%
      \put(1340,440){\makebox(0,0){\strut{} 0}}%
      \put(1988,440){\makebox(0,0){\strut{} 50}}%
      \put(2636,440){\makebox(0,0){\strut{} 100}}%
      \put(3284,440){\makebox(0,0){\strut{} 150}}%
      \put(3931,440){\makebox(0,0){\strut{} 200}}%
      \put(4579,440){\makebox(0,0){\strut{} 250}}%
      \put(5227,440){\makebox(0,0){\strut{} 300}}%
      \put(5875,440){\makebox(0,0){\strut{} 350}}%
      \put(160,2183){\rotatebox{-270}{\makebox(0,0){\strut{}residual}}}%
      \put(3607,140){\makebox(0,0){\strut{}SUMR iterations}}%
    }%
    \gplgaddtomacro\gplfronttext{%
      \csname LTb\endcsname%
      \put(4972,3564){\makebox(0,0)[r]{\strut{}GMRES(relSUMR)}}%
      \csname LTb\endcsname%
      \put(4972,3364){\makebox(0,0)[r]{\strut{}GMRES(calculating(15) relSUMR 1)}}%
      \csname LTb\endcsname%
      \put(4972,3164){\makebox(0,0)[r]{\strut{}GMRES(calculating(30) relSUMR 1)}}%
      \csname LTb\endcsname%
      \put(4972,2964){\makebox(0,0)[r]{\strut{}GMRES(calculating(45) relSUMR 1)}}%
    }%
    \gplbacktext
    \put(0,0){\includegraphics{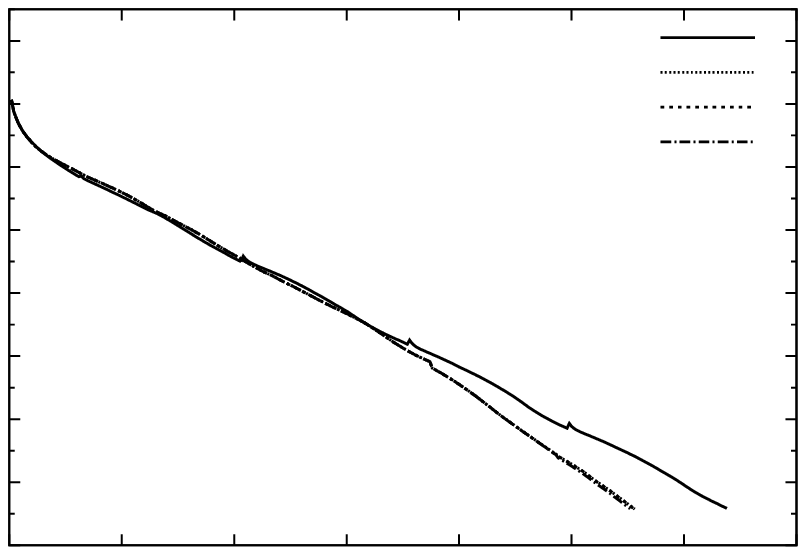}}%
    \gplfronttext
  \end{picture}%
\endgroup

%% file: figs/s8t32m0.03_c55_n30UnitaryResidual_10Comparison.tex
\begingroup
  \makeatletter
  \providecommand\color[2][]{%
    \GenericError{(gnuplot) \space\space\space\@spaces}{%
      Package color not loaded in conjunction with
      terminal option `colourtext'%
    }{See the gnuplot documentation for explanation.%
    }{Either use 'blacktext' in gnuplot or load the package
      color.sty in LaTeX.}%
    \renewcommand\color[2][]{}%
  }%
  \providecommand\includegraphics[2][]{%
    \GenericError{(gnuplot) \space\space\space\@spaces}{%
      Package graphicx or graphics not loaded%
    }{See the gnuplot documentation for explanation.%
    }{The gnuplot epslatex terminal needs graphicx.sty or graphics.sty.}%
    \renewcommand\includegraphics[2][]{}%
  }%
  \providecommand\rotatebox[2]{#2}%
  \@ifundefined{ifGPcolor}{%
    \newif\ifGPcolor
    \GPcolorfalse
  }{}%
  \@ifundefined{ifGPblacktext}{%
    \newif\ifGPblacktext
    \GPblacktexttrue
  }{}%
  \let\gplgaddtomacro\g@addto@macro
  \gdef\gplbacktext{}%
  \gdef\gplfronttext{}%
  \makeatother
  \ifGPblacktext
    \def\colorrgb#1{}%
    \def\colorgray#1{}%
  \else
    \ifGPcolor
      \def\colorrgb#1{\color[rgb]{#1}}%
      \def\colorgray#1{\color[gray]{#1}}%
      \expandafter\def\csname LTw\endcsname{\color{white}}%
      \expandafter\def\csname LTb\endcsname{\color{black}}%
      \expandafter\def\csname LTa\endcsname{\color{black}}%
      \expandafter\def\csname LT0\endcsname{\color[rgb]{1,0,0}}%
      \expandafter\def\csname LT1\endcsname{\color[rgb]{0,1,0}}%
      \expandafter\def\csname LT2\endcsname{\color[rgb]{0,0,1}}%
      \expandafter\def\csname LT3\endcsname{\color[rgb]{1,0,1}}%
      \expandafter\def\csname LT4\endcsname{\color[rgb]{0,1,1}}%
      \expandafter\def\csname LT5\endcsname{\color[rgb]{1,1,0}}%
      \expandafter\def\csname LT6\endcsname{\color[rgb]{0,0,0}}%
      \expandafter\def\csname LT7\endcsname{\color[rgb]{1,0.3,0}}%
      \expandafter\def\csname LT8\endcsname{\color[rgb]{0.5,0.5,0.5}}%
    \else
      \def\colorrgb#1{\color{black}}%
      \def\colorgray#1{\color[gray]{#1}}%
      \expandafter\def\csname LTw\endcsname{\color{white}}%
      \expandafter\def\csname LTb\endcsname{\color{black}}%
      \expandafter\def\csname LTa\endcsname{\color{black}}%
      \expandafter\def\csname LT0\endcsname{\color{black}}%
      \expandafter\def\csname LT1\endcsname{\color{black}}%
      \expandafter\def\csname LT2\endcsname{\color{black}}%
      \expandafter\def\csname LT3\endcsname{\color{black}}%
      \expandafter\def\csname LT4\endcsname{\color{black}}%
      \expandafter\def\csname LT5\endcsname{\color{black}}%
      \expandafter\def\csname LT6\endcsname{\color{black}}%
      \expandafter\def\csname LT7\endcsname{\color{black}}%
      \expandafter\def\csname LT8\endcsname{\color{black}}%
    \fi
  \fi
  \setlength{\unitlength}{0.0500bp}%
  \begin{picture}(6236.00,3968.00)%
    \gplgaddtomacro\gplbacktext{%
      \csname LTb\endcsname%
      \put(1220,640){\makebox(0,0)[r]{\strut{}$10^{-4}$}}%
      \put(1220,1412){\makebox(0,0)[r]{\strut{}$10^{-3}$}}%
      \put(1220,2184){\makebox(0,0)[r]{\strut{}$10^{-2}$}}%
      \put(1220,2955){\makebox(0,0)[r]{\strut{}$10^{-1}$}}%
      \put(1220,3727){\makebox(0,0)[r]{\strut{}$1$}}%
      \put(1340,440){\makebox(0,0){\strut{} 0}}%
      \put(2474,440){\makebox(0,0){\strut{} 1$\times 10^{6}$}}%
      \put(3608,440){\makebox(0,0){\strut{} 2$\times 10^{6}$}}%
      \put(4741,440){\makebox(0,0){\strut{} 3$\times 10^{6}$}}%
      \put(5875,440){\makebox(0,0){\strut{} 4$\times 10^{6}$}}%
      \put(160,2183){\rotatebox{-270}{\makebox(0,0){\strut{}residual}}}%
      \put(3607,140){\makebox(0,0){\strut{}Calls to Kernel operator}}%
    }%
    \gplgaddtomacro\gplfronttext{%
      \csname LTb\endcsname%
      \put(4972,3564){\makebox(0,0)[r]{\strut{}Eigenvalue 10 residual (15)}}%
      \csname LTb\endcsname%
      \put(4972,3364){\makebox(0,0)[r]{\strut{}Eigenvalue 10 residual (30)}}%
      \csname LTb\endcsname%
      \put(4972,3164){\makebox(0,0)[r]{\strut{}Eigenvalue 10 residual (45)}}%
    }%
    \gplbacktext
    \put(0,0){\includegraphics{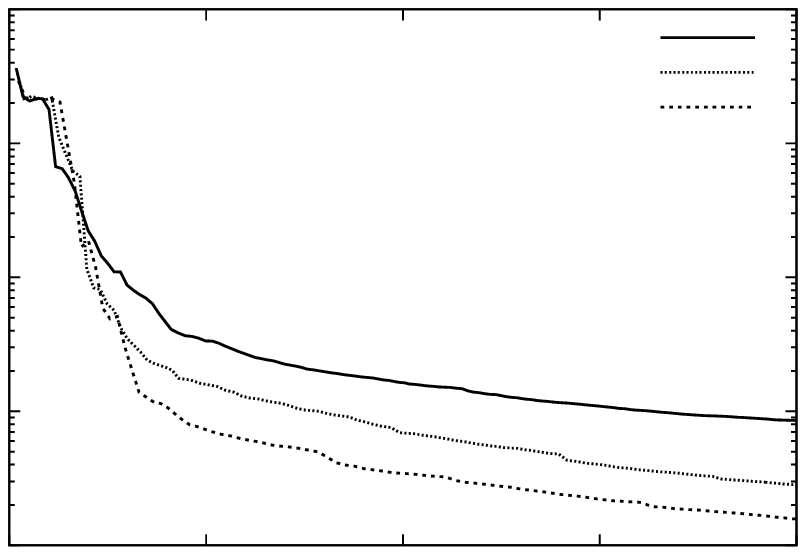}}%
    \gplfronttext
  \end{picture}%
\endgroup

%% file: figs/s8t32m0.03_c55_n30UnitaryResidual_10ComparisonIterations.tex
\begingroup
  \makeatletter
  \providecommand\color[2][]{%
    \GenericError{(gnuplot) \space\space\space\@spaces}{%
      Package color not loaded in conjunction with
      terminal option `colourtext'%
    }{See the gnuplot documentation for explanation.%
    }{Either use 'blacktext' in gnuplot or load the package
      color.sty in LaTeX.}%
    \renewcommand\color[2][]{}%
  }%
  \providecommand\includegraphics[2][]{%
    \GenericError{(gnuplot) \space\space\space\@spaces}{%
      Package graphicx or graphics not loaded%
    }{See the gnuplot documentation for explanation.%
    }{The gnuplot epslatex terminal needs graphicx.sty or graphics.sty.}%
    \renewcommand\includegraphics[2][]{}%
  }%
  \providecommand\rotatebox[2]{#2}%
  \@ifundefined{ifGPcolor}{%
    \newif\ifGPcolor
    \GPcolorfalse
  }{}%
  \@ifundefined{ifGPblacktext}{%
    \newif\ifGPblacktext
    \GPblacktexttrue
  }{}%
  \let\gplgaddtomacro\g@addto@macro
  \gdef\gplbacktext{}%
  \gdef\gplfronttext{}%
  \makeatother
  \ifGPblacktext
    \def\colorrgb#1{}%
    \def\colorgray#1{}%
  \else
    \ifGPcolor
      \def\colorrgb#1{\color[rgb]{#1}}%
      \def\colorgray#1{\color[gray]{#1}}%
      \expandafter\def\csname LTw\endcsname{\color{white}}%
      \expandafter\def\csname LTb\endcsname{\color{black}}%
      \expandafter\def\csname LTa\endcsname{\color{black}}%
      \expandafter\def\csname LT0\endcsname{\color[rgb]{1,0,0}}%
      \expandafter\def\csname LT1\endcsname{\color[rgb]{0,1,0}}%
      \expandafter\def\csname LT2\endcsname{\color[rgb]{0,0,1}}%
      \expandafter\def\csname LT3\endcsname{\color[rgb]{1,0,1}}%
      \expandafter\def\csname LT4\endcsname{\color[rgb]{0,1,1}}%
      \expandafter\def\csname LT5\endcsname{\color[rgb]{1,1,0}}%
      \expandafter\def\csname LT6\endcsname{\color[rgb]{0,0,0}}%
      \expandafter\def\csname LT7\endcsname{\color[rgb]{1,0.3,0}}%
      \expandafter\def\csname LT8\endcsname{\color[rgb]{0.5,0.5,0.5}}%
    \else
      \def\colorrgb#1{\color{black}}%
      \def\colorgray#1{\color[gray]{#1}}%
      \expandafter\def\csname LTw\endcsname{\color{white}}%
      \expandafter\def\csname LTb\endcsname{\color{black}}%
      \expandafter\def\csname LTa\endcsname{\color{black}}%
      \expandafter\def\csname LT0\endcsname{\color{black}}%
      \expandafter\def\csname LT1\endcsname{\color{black}}%
      \expandafter\def\csname LT2\endcsname{\color{black}}%
      \expandafter\def\csname LT3\endcsname{\color{black}}%
      \expandafter\def\csname LT4\endcsname{\color{black}}%
      \expandafter\def\csname LT5\endcsname{\color{black}}%
      \expandafter\def\csname LT6\endcsname{\color{black}}%
      \expandafter\def\csname LT7\endcsname{\color{black}}%
      \expandafter\def\csname LT8\endcsname{\color{black}}%
    \fi
  \fi
  \setlength{\unitlength}{0.0500bp}%
  \begin{picture}(6236.00,3968.00)%
    \gplgaddtomacro\gplbacktext{%
      \csname LTb\endcsname%
      \put(1220,640){\makebox(0,0)[r]{\strut{}$10^{-4}$}}%
      \put(1220,1412){\makebox(0,0)[r]{\strut{}$10^{-3}$}}%
      \put(1220,2184){\makebox(0,0)[r]{\strut{}$10^{-2}$}}%
      \put(1220,2955){\makebox(0,0)[r]{\strut{}$10^{-1}$}}%
      \put(1220,3727){\makebox(0,0)[r]{\strut{}$1$}}%
      \put(1340,440){\makebox(0,0){\strut{} 0}}%
      \put(1907,440){\makebox(0,0){\strut{} 10}}%
      \put(2474,440){\makebox(0,0){\strut{} 20}}%
      \put(3041,440){\makebox(0,0){\strut{} 30}}%
      \put(3608,440){\makebox(0,0){\strut{} 40}}%
      \put(4174,440){\makebox(0,0){\strut{} 50}}%
      \put(4741,440){\makebox(0,0){\strut{} 60}}%
      \put(5308,440){\makebox(0,0){\strut{} 70}}%
      \put(5875,440){\makebox(0,0){\strut{} 80}}%
      \put(160,2183){\rotatebox{-270}{\makebox(0,0){\strut{}residual}}}%
      \put(3607,140){\makebox(0,0){\strut{}Iterations}}%
    }%
    \gplgaddtomacro\gplfronttext{%
      \csname LTb\endcsname%
      \put(4972,3564){\makebox(0,0)[r]{\strut{}Eigenvalue 10 residual (15)}}%
      \csname LTb\endcsname%
      \put(4972,3364){\makebox(0,0)[r]{\strut{}Eigenvalue 10 residual (30)}}%
      \csname LTb\endcsname%
      \put(4972,3164){\makebox(0,0)[r]{\strut{}Eigenvalue 10 residual (45)}}%
    }%
    \gplbacktext
    \put(0,0){\includegraphics{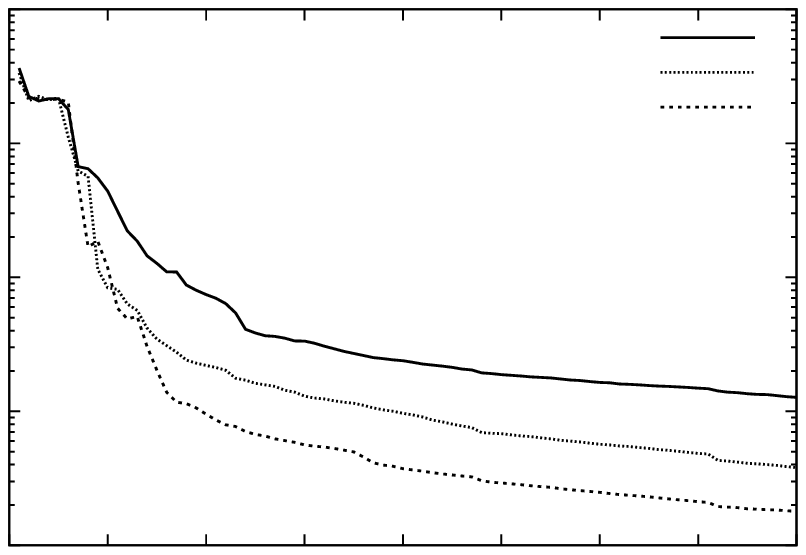}}%
    \gplfronttext
  \end{picture}%
\endgroup

%% file: figs/s8t32m0.03_c55_n30UnitaryResidual_LastComparison.tex
\begingroup
  \makeatletter
  \providecommand\color[2][]{%
    \GenericError{(gnuplot) \space\space\space\@spaces}{%
      Package color not loaded in conjunction with
      terminal option `colourtext'%
    }{See the gnuplot documentation for explanation.%
    }{Either use 'blacktext' in gnuplot or load the package
      color.sty in LaTeX.}%
    \renewcommand\color[2][]{}%
  }%
  \providecommand\includegraphics[2][]{%
    \GenericError{(gnuplot) \space\space\space\@spaces}{%
      Package graphicx or graphics not loaded%
    }{See the gnuplot documentation for explanation.%
    }{The gnuplot epslatex terminal needs graphicx.sty or graphics.sty.}%
    \renewcommand\includegraphics[2][]{}%
  }%
  \providecommand\rotatebox[2]{#2}%
  \@ifundefined{ifGPcolor}{%
    \newif\ifGPcolor
    \GPcolorfalse
  }{}%
  \@ifundefined{ifGPblacktext}{%
    \newif\ifGPblacktext
    \GPblacktexttrue
  }{}%
  \let\gplgaddtomacro\g@addto@macro
  \gdef\gplbacktext{}%
  \gdef\gplfronttext{}%
  \makeatother
  \ifGPblacktext
    \def\colorrgb#1{}%
    \def\colorgray#1{}%
  \else
    \ifGPcolor
      \def\colorrgb#1{\color[rgb]{#1}}%
      \def\colorgray#1{\color[gray]{#1}}%
      \expandafter\def\csname LTw\endcsname{\color{white}}%
      \expandafter\def\csname LTb\endcsname{\color{black}}%
      \expandafter\def\csname LTa\endcsname{\color{black}}%
      \expandafter\def\csname LT0\endcsname{\color[rgb]{1,0,0}}%
      \expandafter\def\csname LT1\endcsname{\color[rgb]{0,1,0}}%
      \expandafter\def\csname LT2\endcsname{\color[rgb]{0,0,1}}%
      \expandafter\def\csname LT3\endcsname{\color[rgb]{1,0,1}}%
      \expandafter\def\csname LT4\endcsname{\color[rgb]{0,1,1}}%
      \expandafter\def\csname LT5\endcsname{\color[rgb]{1,1,0}}%
      \expandafter\def\csname LT6\endcsname{\color[rgb]{0,0,0}}%
      \expandafter\def\csname LT7\endcsname{\color[rgb]{1,0.3,0}}%
      \expandafter\def\csname LT8\endcsname{\color[rgb]{0.5,0.5,0.5}}%
    \else
      \def\colorrgb#1{\color{black}}%
      \def\colorgray#1{\color[gray]{#1}}%
      \expandafter\def\csname LTw\endcsname{\color{white}}%
      \expandafter\def\csname LTb\endcsname{\color{black}}%
      \expandafter\def\csname LTa\endcsname{\color{black}}%
      \expandafter\def\csname LT0\endcsname{\color{black}}%
      \expandafter\def\csname LT1\endcsname{\color{black}}%
      \expandafter\def\csname LT2\endcsname{\color{black}}%
      \expandafter\def\csname LT3\endcsname{\color{black}}%
      \expandafter\def\csname LT4\endcsname{\color{black}}%
      \expandafter\def\csname LT5\endcsname{\color{black}}%
      \expandafter\def\csname LT6\endcsname{\color{black}}%
      \expandafter\def\csname LT7\endcsname{\color{black}}%
      \expandafter\def\csname LT8\endcsname{\color{black}}%
    \fi
  \fi
  \setlength{\unitlength}{0.0500bp}%
  \begin{picture}(6236.00,3968.00)%
    \gplgaddtomacro\gplbacktext{%
      \csname LTb\endcsname%
      \put(1220,640){\makebox(0,0)[r]{\strut{}$10^{-5}$}}%
      \put(1220,1155){\makebox(0,0)[r]{\strut{}$10^{-4}$}}%
      \put(1220,1669){\makebox(0,0)[r]{\strut{}$10^{-3}$}}%
      \put(1220,2184){\makebox(0,0)[r]{\strut{}$10^{-2}$}}%
      \put(1220,2698){\makebox(0,0)[r]{\strut{}$10^{-1}$}}%
      \put(1220,3213){\makebox(0,0)[r]{\strut{}$1$}}%
      \put(1220,3727){\makebox(0,0)[r]{\strut{}$10$}}%
      \put(1340,440){\makebox(0,0){\strut{} 0}}%
      \put(2474,440){\makebox(0,0){\strut{} 1$\times 10^{6}$}}%
      \put(3608,440){\makebox(0,0){\strut{} 2$\times 10^{6}$}}%
      \put(4741,440){\makebox(0,0){\strut{} 3$\times 10^{6}$}}%
      \put(5875,440){\makebox(0,0){\strut{} 4$\times 10^{6}$}}%
      \put(160,2183){\rotatebox{-270}{\makebox(0,0){\strut{}residual}}}%
      \put(3607,140){\makebox(0,0){\strut{}Calls to Kernel operator}}%
    }%
    \gplgaddtomacro\gplfronttext{%
      \csname LTb\endcsname%
      \put(4972,3564){\makebox(0,0)[r]{\strut{}Eigenvalue 15 residual (15)}}%
      \csname LTb\endcsname%
      \put(4972,3364){\makebox(0,0)[r]{\strut{}Eigenvalue 30 residual (30)}}%
      \csname LTb\endcsname%
      \put(4972,3164){\makebox(0,0)[r]{\strut{}Eigenvalue 45 residual (45)}}%
    }%
    \gplbacktext
    \put(0,0){\includegraphics{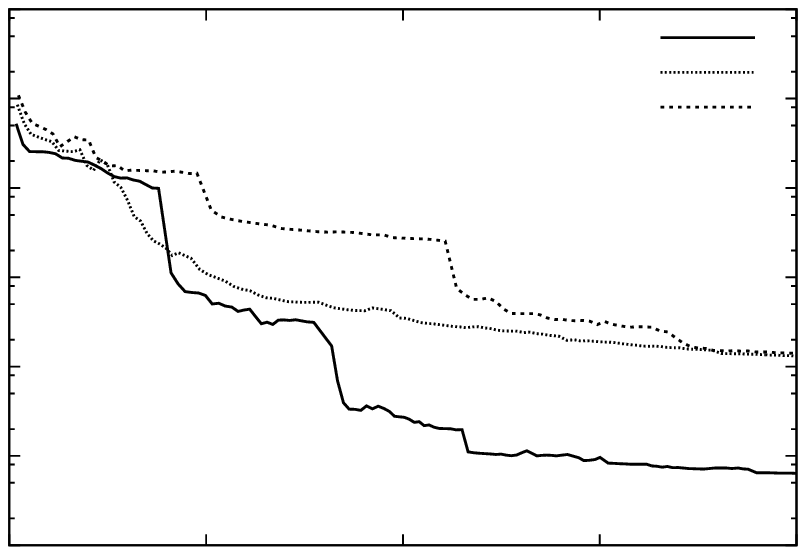}}%
    \gplfronttext
  \end{picture}%
\endgroup

%% file: figs/s8t32m0.03_c55_n30UnitaryResidual_LastComparisonIterations.tex
\begingroup
  \makeatletter
  \providecommand\color[2][]{%
    \GenericError{(gnuplot) \space\space\space\@spaces}{%
      Package color not loaded in conjunction with
      terminal option `colourtext'%
    }{See the gnuplot documentation for explanation.%
    }{Either use 'blacktext' in gnuplot or load the package
      color.sty in LaTeX.}%
    \renewcommand\color[2][]{}%
  }%
  \providecommand\includegraphics[2][]{%
    \GenericError{(gnuplot) \space\space\space\@spaces}{%
      Package graphicx or graphics not loaded%
    }{See the gnuplot documentation for explanation.%
    }{The gnuplot epslatex terminal needs graphicx.sty or graphics.sty.}%
    \renewcommand\includegraphics[2][]{}%
  }%
  \providecommand\rotatebox[2]{#2}%
  \@ifundefined{ifGPcolor}{%
    \newif\ifGPcolor
    \GPcolorfalse
  }{}%
  \@ifundefined{ifGPblacktext}{%
    \newif\ifGPblacktext
    \GPblacktexttrue
  }{}%
  \let\gplgaddtomacro\g@addto@macro
  \gdef\gplbacktext{}%
  \gdef\gplfronttext{}%
  \makeatother
  \ifGPblacktext
    \def\colorrgb#1{}%
    \def\colorgray#1{}%
  \else
    \ifGPcolor
      \def\colorrgb#1{\color[rgb]{#1}}%
      \def\colorgray#1{\color[gray]{#1}}%
      \expandafter\def\csname LTw\endcsname{\color{white}}%
      \expandafter\def\csname LTb\endcsname{\color{black}}%
      \expandafter\def\csname LTa\endcsname{\color{black}}%
      \expandafter\def\csname LT0\endcsname{\color[rgb]{1,0,0}}%
      \expandafter\def\csname LT1\endcsname{\color[rgb]{0,1,0}}%
      \expandafter\def\csname LT2\endcsname{\color[rgb]{0,0,1}}%
      \expandafter\def\csname LT3\endcsname{\color[rgb]{1,0,1}}%
      \expandafter\def\csname LT4\endcsname{\color[rgb]{0,1,1}}%
      \expandafter\def\csname LT5\endcsname{\color[rgb]{1,1,0}}%
      \expandafter\def\csname LT6\endcsname{\color[rgb]{0,0,0}}%
      \expandafter\def\csname LT7\endcsname{\color[rgb]{1,0.3,0}}%
      \expandafter\def\csname LT8\endcsname{\color[rgb]{0.5,0.5,0.5}}%
    \else
      \def\colorrgb#1{\color{black}}%
      \def\colorgray#1{\color[gray]{#1}}%
      \expandafter\def\csname LTw\endcsname{\color{white}}%
      \expandafter\def\csname LTb\endcsname{\color{black}}%
      \expandafter\def\csname LTa\endcsname{\color{black}}%
      \expandafter\def\csname LT0\endcsname{\color{black}}%
      \expandafter\def\csname LT1\endcsname{\color{black}}%
      \expandafter\def\csname LT2\endcsname{\color{black}}%
      \expandafter\def\csname LT3\endcsname{\color{black}}%
      \expandafter\def\csname LT4\endcsname{\color{black}}%
      \expandafter\def\csname LT5\endcsname{\color{black}}%
      \expandafter\def\csname LT6\endcsname{\color{black}}%
      \expandafter\def\csname LT7\endcsname{\color{black}}%
      \expandafter\def\csname LT8\endcsname{\color{black}}%
    \fi
  \fi
  \setlength{\unitlength}{0.0500bp}%
  \begin{picture}(6236.00,3968.00)%
    \gplgaddtomacro\gplbacktext{%
      \csname LTb\endcsname%
      \put(1220,640){\makebox(0,0)[r]{\strut{}$10^{-4}$}}%
      \put(1220,1257){\makebox(0,0)[r]{\strut{}$10^{-3}$}}%
      \put(1220,1875){\makebox(0,0)[r]{\strut{}$10^{-2}$}}%
      \put(1220,2492){\makebox(0,0)[r]{\strut{}$10^{-1}$}}%
      \put(1220,3110){\makebox(0,0)[r]{\strut{}$1$}}%
      \put(1220,3727){\makebox(0,0)[r]{\strut{}$10$}}%
      \put(1340,440){\makebox(0,0){\strut{} 0}}%
      \put(1907,440){\makebox(0,0){\strut{} 10}}%
      \put(2474,440){\makebox(0,0){\strut{} 20}}%
      \put(3041,440){\makebox(0,0){\strut{} 30}}%
      \put(3608,440){\makebox(0,0){\strut{} 40}}%
      \put(4174,440){\makebox(0,0){\strut{} 50}}%
      \put(4741,440){\makebox(0,0){\strut{} 60}}%
      \put(5308,440){\makebox(0,0){\strut{} 70}}%
      \put(5875,440){\makebox(0,0){\strut{} 80}}%
      \put(160,2183){\rotatebox{-270}{\makebox(0,0){\strut{}residual}}}%
      \put(3607,140){\makebox(0,0){\strut{}Iterations}}%
    }%
    \gplgaddtomacro\gplfronttext{%
      \csname LTb\endcsname%
      \put(4972,3564){\makebox(0,0)[r]{\strut{}Eigenvalue 15 residual (15)}}%
      \csname LTb\endcsname%
      \put(4972,3364){\makebox(0,0)[r]{\strut{}Eigenvalue 30 residual (30)}}%
      \csname LTb\endcsname%
      \put(4972,3164){\makebox(0,0)[r]{\strut{}Eigenvalue 45 residual (45)}}%
    }%
    \gplbacktext
    \put(0,0){\includegraphics{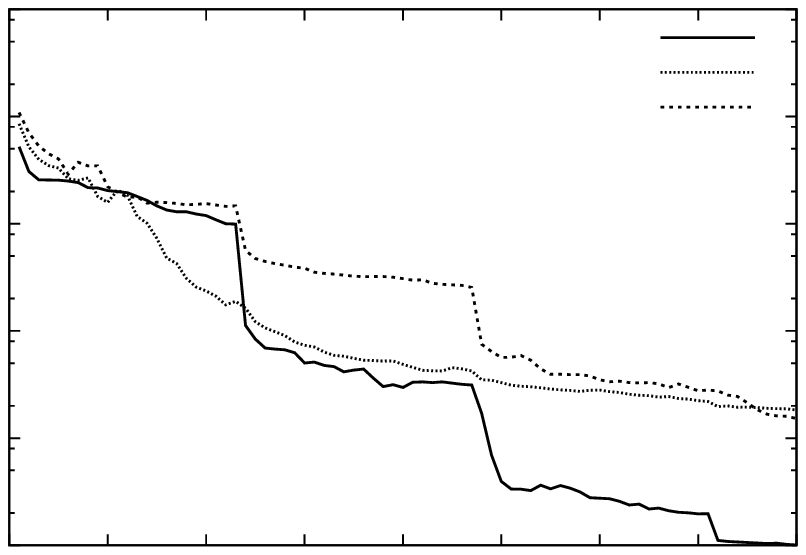}}%
    \gplfronttext
  \end{picture}%
\endgroup

%% file: Acknowledgements
Numerical calculations used servers at Seoul National University
funded by the BK21 program of the NRF (MEST), Republic of
Korea. NC was supported in part through the BK21 program of the NRF (MEST), Republic of Korea. This research was supported by Basic Science Research Program through the National Research Foundation of Korea(NRF) funded by the Ministry of Education(2013057640). W. Lee is supported by the Creative Research Initiatives
Program (2013-003454) of the NRF grant, and acknowledges the support
from the KISTI supercomputing center through the strategic support
program for the supercomputing application research
(KSC-2012-G2-01).

%% file: eigsumr.bbl
\begin{thebibliography}{10}
\expandafter\ifx\csname url\endcsname\relax
  \def\url#1{\texttt{#1}}\fi
\expandafter\ifx\csname urlprefix\endcsname\relax\def\urlprefix{URL }\fi
\expandafter\ifx\csname href\endcsname\relax
  \def\href#1#2{#2} \def\path#1{#1}\fi

\bibitem{Wilson:1974}
K.~G. Wilson, Phys. Rev. D10 (1974) 2445.

\bibitem{Ginsparg:1982bj}
P.~H. Ginsparg, K.~G. Wilson, {A remnant of chiral symetry on the lattice},
  Phys. Rev. D25 (1982) 2649.

\bibitem{Hasenfratz:1998ri}
P.~Hasenfratz, V.~Laliena, F.~Niedermayer, {The Index theorem in QCD with a
  finite cutoff}, Phys. Lett. B427 (1998) 125--131.
\newblock \href {http://arxiv.org/abs/hep-lat/9801021}
  {\path{arXiv:hep-lat/9801021}}, \href
  {http://dx.doi.org/10.1016/S0370-2693(98)00315-3}
  {\path{doi:10.1016/S0370-2693(98)00315-3}}.

\bibitem{Luscher:1998pqa}
M.~L{\"u}scher, {Exact chiral symmetry on the lattice and the Ginsparg-{W}ilson
  relation}, Phys. Lett. B428 (1998) 342--345.
\newblock \href {http://arxiv.org/abs/hep-lat/9802011}
  {\path{arXiv:hep-lat/9802011}}.

\bibitem{Nishy-Ninny}
H.~B. Nielsen, M.~Ninomiya, {Absence of Neutrinos on a Lattice. 1. Proof by
  Homotopy Theory }, Nucl. Phys B185 (1981) 20.

\bibitem{Narayanan:1993ss}
R.~Narayanan, H.~Neuberger, {Chiral fermions on the lattice}, Phys. Rev. Lett.
  71 (1993) 3251--3254.
\newblock \href {http://arxiv.org/abs/hep-lat/9308011}
  {\path{arXiv:hep-lat/9308011}}.

\bibitem{Neuberger:1998fp}
H.~Neuberger, {Exactly massless quarks on the lattice}, Phys. Lett. B417 (1998)
  141--144.
\newblock \href {http://arxiv.org/abs/hep-lat/9707022}
  {\path{arXiv:hep-lat/9707022}}.

\bibitem{Neuberger:1997bg}
H.~Neuberger, {Vector like gauge theories with almost massless fermions on the
  lattice}, Phys. Rev. D57 (1998) 5417--5433.
\newblock \href {http://arxiv.org/abs/hep-lat/9710089}
  {\path{arXiv:hep-lat/9710089}}, \href
  {http://dx.doi.org/10.1103/PhysRevD.57.5417}
  {\path{doi:10.1103/PhysRevD.57.5417}}.

\bibitem{Neuberger:1998my}
H.~Neuberger, {A practical implementation of the overlap-Dirac operator}, Phys.
  Rev. Lett. 81 (1998) 4060--4062.
\newblock \href {http://arxiv.org/abs/hep-lat/9806025}
  {\path{arXiv:hep-lat/9806025}}.

\bibitem{Hernandez:1999cu}
P.~Hernandez, K.~Jansen, L.~Lellouch, {Finite size scaling of the quark
  condensate in quenched lattice QCD}, Phys.Lett. B469 (1999) 198--204.
\newblock \href {http://arxiv.org/abs/hep-lat/9907022}
  {\path{arXiv:hep-lat/9907022}}, \href
  {http://dx.doi.org/10.1016/S0370-2693(99)01244-7}
  {\path{doi:10.1016/S0370-2693(99)01244-7}}.

\bibitem{vandenEshof:2002ms}
J.~van~den Eshof, A.~Frommer, T.~Lippert, K.~Schilling, H.~A. van~der Vorst,
  {Numerical methods for the {QCD} overlap operator. {I}: Sign- function and
  error bounds}, Comput. Phys. Commun. 146 (2002) 203--224.
\newblock \href {http://arxiv.org/abs/hep-lat/0202025}
  {\path{arXiv:hep-lat/0202025}}.

\bibitem{Borici:2007bp}
A.~Borici, {A Schur Complement Approach to Chiral Fermions}, PoS LAT2007 (2007)
  065.
\newblock \href {http://arxiv.org/abs/0711.0508} {\path{arXiv:0711.0508}}.

\bibitem{Narayanan:2000qx}
R.~Narayanan, H.~Neuberger, {An Alternative to domain wall fermions}, Phys.Rev.
  D62 (2000) 074504.
\newblock \href {http://arxiv.org/abs/hep-lat/0005004}
  {\path{arXiv:hep-lat/0005004}}, \href
  {http://dx.doi.org/10.1103/PhysRevD.62.074504}
  {\path{doi:10.1103/PhysRevD.62.074504}}.

\bibitem{Borici:2001ua}
A.~Borici, A.~Kennedy, B.~Pendleton, U.~Wenger, {The Overlap operator as a
  continued fraction}, Nucl.Phys.Proc.Suppl. 106 (2002) 757--759.
\newblock \href {http://arxiv.org/abs/hep-lat/0110070}
  {\path{arXiv:hep-lat/0110070}}, \href
  {http://dx.doi.org/10.1016/S0920-5632(01)01835-7}
  {\path{doi:10.1016/S0920-5632(01)01835-7}}.

\bibitem{Edwards:2005an}
R.~G. Edwards, B.~Joo, A.~D. Kennedy, K.~Orginos, U.~Wenger, {Comparison of
  chiral fermion methods}, PoS LAT2005 (2006) 146.
\newblock \href {http://arxiv.org/abs/hep-lat/0510086}
  {\path{arXiv:hep-lat/0510086}}.

\bibitem{Kennedy:2006ax}
A.~Kennedy, {Algorithms for dynamical fermions}, in: {proceedings of the ILFTN
  workshop `Perspectives in Lattice QCD'}, World Scientific, 2006.
\newblock \href {http://arxiv.org/abs/hep-lat/0607038}
  {\path{arXiv:hep-lat/0607038}}.

\bibitem{Shamir:1993zy}
Y.~Shamir, {Chiral fermions from lattice boundaries}, Nucl. Phys. B406 (1993)
  90--106.
\newblock \href {http://arxiv.org/abs/hep-lat/9303005}
  {\path{arXiv:hep-lat/9303005}}.

\bibitem{Kaplan:1992bt}
D.~B. Kaplan, {A Method for simulating chiral fermions on the lattice}, Phys.
  Lett. B288 (1992) 342--347.
\newblock \href {http://arxiv.org/abs/hep-lat/9206013}
  {\path{arXiv:hep-lat/9206013}}.

\bibitem{Chiu:2002ir}
T.-W. Chiu, {Optimal domain-wall fermions}, Phys. Rev. Lett. 90 (2003) 071601.
\newblock \href {http://arxiv.org/abs/hep-lat/0209153}
  {\path{arXiv:hep-lat/0209153}}.

\bibitem{Cundy:2010uq}
N.~Cundy, A.~Kennedy, A.~Schafer, {A lattice Dirac operator for QCD with light
  dynamical quarks}, Nucl. Phys. B845 (2011) 30--47.
\newblock \href {http://arxiv.org/abs/1010.5629} {\path{arXiv:1010.5629}},
  \href {http://dx.doi.org/10.1016/j.nuclphysb.2010.11.017}
  {\path{doi:10.1016/j.nuclphysb.2010.11.017}}.

\bibitem{Fodor:2003bh}
Z.~Fodor, S.~D. Katz, K.~K. Szabo, {Dynamical overlap fermions, results with
  hybrid Monte-Carlo algorithm}, JHEP 08 (2004) 003.
\newblock \href {http://arxiv.org/abs/hep-lat/0311010}
  {\path{arXiv:hep-lat/0311010}}.

\bibitem{DeGrand:2004nq}
T.~A. DeGrand, S.~Schaefer, {Physics issues in simulations with dynamical
  overlap fermions}, Phys. Rev. D71 (2005) 034507.
\newblock \href {http://arxiv.org/abs/hep-lat/0412005}
  {\path{arXiv:hep-lat/0412005}}.

\bibitem{DeGrand:2005vb}
T.~A. DeGrand, S.~Schaefer, {Chiral properties of two-flavor QCD in small
  volume and at large lattice spacing}, Phys. Rev. D72 (2005) 054503.
\newblock \href {http://arxiv.org/abs/hep-lat/0506021}
  {\path{arXiv:hep-lat/0506021}}, \href
  {http://dx.doi.org/10.1103/PhysRevD.72.054503}
  {\path{doi:10.1103/PhysRevD.72.054503}}.

\bibitem{Cundy:2005pi}
N.~Cundy, et~al., {Numerical methods for the QCD overlap operator. IV: Hybrid
  Monte Carlo}, Comput. Phys. Commun. 180 (2009) 26--54.
\newblock \href {http://arxiv.org/abs/hep-lat/0502007}
  {\path{arXiv:hep-lat/0502007}}, \href
  {http://dx.doi.org/10.1016/j.cpc.2008.08.006}
  {\path{doi:10.1016/j.cpc.2008.08.006}}.

\bibitem{Cundy:2005mr}
N.~Cundy, {Current status of dynamical overlap project}, Nucl. Phys. Proc.
  Suppl. 153 (2006) 54--61.
\newblock \href {http://arxiv.org/abs/hep-lat/0511047}
  {\path{arXiv:hep-lat/0511047}}.

\bibitem{Cundy:2008zc}
N.~Cundy, S.~Krieg, T.~Lippert, A.~Schafer, {Topological tunneling with
  Dynamical overlap fermions}, Comput. Phys. Commun. 180 (2009) 201--208.
\newblock \href {http://arxiv.org/abs/0803.0294} {\path{arXiv:0803.0294}},
  \href {http://dx.doi.org/10.1016/j.cpc.2008.09.010}
  {\path{doi:10.1016/j.cpc.2008.09.010}}.

\bibitem{Cundy:2007la}
N.~Cundy, S.~Krieg, T.~Lippert, A.~Sch{\"a}fer, {Dynamical overlap fermions
  with increased topological tunnelling}, PoS LAT2007 (2007) 030.
\newblock \href {http://arxiv.org/abs/hep-lat:0710.1705}
  {\path{arXiv:hep-lat:0710.1705}}.

\bibitem{Zolotarev}
E.~N. Zolotarev, Zap. Imp. Akad. Nauk St Petersburg 30 (1877) 5.

\bibitem{Sorensen97acceleratingthe}
D.~C. Sorensen, C.~Yang, {Accelerating The Lanczos Algorithm Via Polynomial
  Spectral Transformations}, Tech. rep., Rice University (1997).

\bibitem{IRAM}
R.~B. Lehoucq, D.~C. Sorensen, {Deflation Techniques for an Implicitly
  Restarted Arnoldi Iteration}, SIAM. J. Matrix Anal. and Appl. 17(4) (1996)
  789--821.

\bibitem{Hashimoto:2007vv}
S.~Hashimoto, et~al., {Lattice simulation of 2+1 flavors of overlap light
  quarks}, PoS LAT2007 (2007) 101.
\newblock \href {http://arxiv.org/abs/0710.2730} {\path{arXiv:0710.2730}}.

\bibitem{Arnold:2003sx}
G.~Arnold, N.~Cundy, J.~van~den Eshof, A.~Frommer, S.~Krieg, et~al., {Numerical
  methods for the QCD overlap operator. {II} Optimal Krylov subspace methods},
  Lecture Notes in Computational Science and Engineering 47 (2003) 153--167.
\newblock \href {http://arxiv.org/abs/hep-lat/0311025}
  {\path{arXiv:hep-lat/0311025}}.

\bibitem{Cundy:2004pza}
N.~Cundy, J.~van~den Eshof, A.~Frommer, S.~Krieg, T.~Lippert, et~al.,
  {Numerical methods for the QCD overlap operator. {III}. Nested iterations},
  Comput.Phys.Commun. 165 (2005) 221--242.
\newblock \href {http://arxiv.org/abs/hep-lat/0405003}
  {\path{arXiv:hep-lat/0405003}}, \href
  {http://dx.doi.org/10.1016/j.cpc.2004.10.005}
  {\path{doi:10.1016/j.cpc.2004.10.005}}.

\bibitem{Jagels}
Jagels, Reichel, {A Fast Minimiml Residual Algorithm for Shifted Unitary
  Matrices}, Numerical Linear Algebra with Applications 1(6) (1994) 555.

\bibitem{Cundy:2005mn}
N.~Cundy, S.~Krieg, T.~Lippert, {Improving the dynamical overlap algorithm},
  PoS LAT2005 (2006) 107.
\newblock \href {http://arxiv.org/abs/hep-lat/0511044}
  {\path{arXiv:hep-lat/0511044}}.

\bibitem{Stathopoulos:2007zi}
A.~Stathopoulos, K.~Orginos, {Computing and deflating eigenvalues while solving
  multiple right hand side linear systems in quantum chromodynamics}, SIAM
  J.Sci.Comput. 32 (2010) 439--462.
\newblock \href {http://arxiv.org/abs/0707.0131} {\path{arXiv:0707.0131}}.

\bibitem{Kennedy:2004tk}
A.~Kennedy, {Fast evaluation of Zolotarev coefficients}, Lecture Notes in
  Computational Science and Engineering 47 (2005) 169--189.
\newblock \href {http://arxiv.org/abs/hep-lat/0402038}
  {\path{arXiv:hep-lat/0402038}}.

\bibitem{vandervosrt:1994}
H.~A. van~der Vorst, C.~Vuik, {GMRESR: a family of nested GMRES methods},
  Numer. Linear Algebra Appl. 1~(4) (1994) 369--386.

\bibitem{DuffBeer}
I.~S. Duff, L.~Giraud, J.~Langou, E.~Martin, {Using spectral low rank
  preconditioners for large electromagnetic calculations}, Int. J. Numer. Meth.
  Engng. 62 (2005) 416--434.

\bibitem{Erhel:2000:ACG:354353.354400}
J.~Erhel, F.~Guyomarc'h, \href{http://dx.doi.org/10.1137/S0895479897330194}{{An
  Augmented Conjugate Gradient Method for Solving Consecutive Symmetric
  Positive Definite Linear Systems}}, SIAM J. Matrix Anal. Appl. 21~(4) (2000)
  1279--1299.
\newblock \href {http://dx.doi.org/10.1137/S0895479897330194}
  {\path{doi:10.1137/S0895479897330194}}.
\newline\urlprefix\url{http://dx.doi.org/10.1137/S0895479897330194}

\bibitem{NLA:NLA1680020105}
S.~A. Kharchenko, A.~{Yu. Yeremin},
  \href{http://dx.doi.org/10.1002/nla.1680020105}{{Eigenvalue translation based
  preconditioners for the GMRES(k) method}}, Numerical Linear Algebra with
  Applications 2~(1) (1995) 51--77.
\newblock \href {http://dx.doi.org/10.1002/nla.1680020105}
  {\path{doi:10.1002/nla.1680020105}}.
\newline\urlprefix\url{http://dx.doi.org/10.1002/nla.1680020105}

\bibitem{Giusti:2002sm}
L.~Giusti, C.~Hoelbling, M.~Luscher, H.~Wittig, {Numerical techniques for
  lattice QCD in the epsilon regime}, Comput.Phys.Commun. 153 (2003) 31--51.
\newblock \href {http://arxiv.org/abs/hep-lat/0212012}
  {\path{arXiv:hep-lat/0212012}}, \href
  {http://dx.doi.org/10.1016/S0010-4655(02)00874-3}
  {\path{doi:10.1016/S0010-4655(02)00874-3}}.

\bibitem{Saad:2003:IMS:829576}
Y.~Saad, {Iterative Methods for Sparse Linear Systems}, 2nd Edition, Society
  for Industrial and Applied Mathematics, Philadelphia, PA, USA, 2003.

\bibitem{Chiarappa:2006hz}
T.~Chiarappa, K.~Jansen, K.-I. Nagai, M.~Papinutto, L.~Scorzato, A.~Shindler,
  C.~Urbach, U.~Wenger, I.~Wetzorke, {Iterative methods for overlap and twisted
  mass fermions}, Computational Science \& Discovery 1~(1) (2008) 015001.
\newblock \href {http://arxiv.org/abs/hep-lat/0609023}
  {\path{arXiv:hep-lat/0609023}}.

\bibitem{TILW}
M.~L{\"u}scher, P.~Weisz, {}, Commun Math Phys 97 (1985) 59.

\bibitem{TILW2}
G.~P. Lepage, P.~B. Mackenzie,
  \href{http://link.aps.org/doi/10.1103/PhysRevD.48.2250}{{Viability of lattice
  perturbation theory}}, Phys. Rev. D 48 (1993) 2250--2264.
\newblock \href {http://dx.doi.org/10.1103/PhysRevD.48.2250}
  {\path{doi:10.1103/PhysRevD.48.2250}}.
\newline\urlprefix\url{http://link.aps.org/doi/10.1103/PhysRevD.48.2250}

\bibitem{TILW5}
G.~Curci, P.~Menotti, G.~Paffuti, {}, Phys. Lett. B B130 (1983) 205.

\bibitem{Jacobi-Davidson}
G.~L.~G. Sleijpen, H.~A. van~der Vorst, {A Jacobi-Davidson iteration method for
  linear eigenvalue problems}, SIAM J. Matrix Anal. Appl. 17 (1996) 401--425.

\end{thebibliography}
